\documentclass[%
 reprint,
 superscriptaddress,
nofootinbib,
 amsmath,amssymb,
aps,
prl,
]{revtex4-2}
\usepackage{graphicx}
\usepackage{dcolumn}
\usepackage{bm}
\usepackage{physics}
\usepackage[dvipsnames]{xcolor}
\usepackage{hyperref}
\usepackage{multirow}
\usepackage{mathtools}
\usepackage{verbatim}  
\usepackage[normalem]{ulem}

\newcommand{\affilITMO}{School of Physics and Engineering, ITMO University, St.~Petersburg 197101, Russia}
\usepackage{physics}
\usepackage{bbold}
\usepackage{appendix}

\newcommand{\ve}[1]{\mathbf{#1}}

\renewcommand{\Im}{\operatorname{Im}}

\renewcommand{\dag}{^{\dagger}}
\newcommand{\sig}{\hat{\sigma}}
\newcommand{\sigd}{\hat{\sigma}^{\dag}}

\newcommand{\veps}{\varepsilon}

\definecolor{linkcolor}{HTML}{0176ba}
\definecolor{urlcolor}{HTML}{0176ba} 
\definecolor{citecolor}{HTML}{900020}
\hypersetup{pdfstartview=FitH,  linkcolor=linkcolor,urlcolor=urlcolor, citecolor=citecolor, colorlinks=true}

\begin{document}

\title{Nonradiant multiphoton states in quantum ring oligomers}

\author{N. Ustimenko}
 \affiliation{\affilITMO}

\author{D. Kornovan}%

\affiliation{\affilITMO 
}%

\author{I. Volkov}%

\affiliation{\affilITMO 
}%

\author{A.  Sheremet}
\affiliation{\affilITMO
}%
\author{R. Savelev}

\affiliation{\affilITMO 
}%
\author{M. Petrov}
\email{m.petrov@metalab.ifmo.ru}
\affiliation{\affilITMO
}%


\begin{abstract}
Arrays of coupled dipole emitters support collective single- and multiphoton states that can preserve quantum excitations. One of the crucial characteristics of these states is the lifetime, which is fundamentally limited due to spontaneous emission. Here, we present a mechanism of the external coupling of two states via the radiation continuum, which allows for an increase in the lifetime of both single and double excitations. As an illustrative example, we consider a ringlike ensemble of quantum emitters, demonstrating that upon slight optimization of the structure geometry, one can increase the lifetime of singly and doubly excited states with nonzero orbital momentum by several orders of magnitude. The proposed  mechanism  of  multiphoton excitation  lifetime control has a universal nature and might be applied to a wide class of open quantum systems and quantum ensembles besides the particular geometry considered in this paper. 
\end{abstract}

\maketitle

Assembling quantum emitters in ordered systems allows for enhancing the light-matter interaction~\cite{Chang2018, sheremet_waveguide_2023} which is crucial for quantum information~\cite{Kimble2008,  Hammerer2010}, quantum sensing~\cite{abend_technology_2023,  dubosclard_nondestructive_2021}, and optomechanical~\cite{Meng2018, Jungkind2019, iorsh_waveguide_2020, shahmoon_quantum_2020} applications. Due to advanced trapping techniques, emitters can be assembled into the structured arrays in free space~\cite{barredo_atom-by-atom_2016, Endres2016, Barredo2018, Ebadi2021} or the vicinity of nanophotonic structures~\cite{Vetsch2012,Nayak2018,Goban2015, Corzo2016}, which induces collective effects in single- and multiphoton regimes~\cite{ manzoni_optimization_2018, reitz_cooperative_2021, Prasad2020, liedl_collective_2023, Corzo2019, Ke2019Dec, holzinger_subradiace_2020,Fayard2023,Zhang2023Nov,Cardenas-Lopez2023Jul,Mok2023May,Lohof2023Aug}. However, the reliable manipulation of quantum states requires their stability, which can be easily destroyed by the spontaneous emission inevitably present in open quantum systems. In this context, controlling the lifetime of quantum states remains one of the key problems in modern quantum optics.
 
Fortunately, this problem can be resolved by generating subradiant states characterized by suppressed spontaneous decay. Since the pioneer work of R. Dicke~\cite{Dicke1954}, the emergence of these states has been actively studied both theoretically~\cite{Sheremet2012, Kornovan2016, jen_cooperative_2018, Plankensteiner2019, zhang_theory_2019, Fofanov2021} and experimentally~\cite{pavolini_experimental_1985, devoe_observation_1996, guerin_subradiance_2016, rui_subradiant_2020, Ferioli2021}. Moreover, the spatial ordering of quantum dipole emitters can provide additional control over the lifetime of subradiant states for one-dimensional arrays in free space~\cite{Asenjo-Garcia2017, Zhang2020, Kornovan2021,Fayard2023}, near a waveguide~\cite{sutherland_collective_2016, Asenjo-Garcia2017Mar, Kornovan2019, Ke2019Dec, Zhang2020Feb, Poddubny2020}, for two-dimensional arrays~\cite{facchinetti_storing_2016, bettles_cooperative_2015, ballantine_subradiance-protected_2020}, or for single rings~\cite{Asenjo-Garcia2017}. It has been shown that the radiative decay of large systems can be strongly suppressed, following either a polynomial~$\propto N^{-\alpha}$~\cite{Zhang2020, Kornovan2019} or exponential $\propto e^{-\beta N}$~\cite{Asenjo-Garcia2017} asymptotic dependence on the number of quantum emitters in the array $N$. However, the suppression of radiative decay in smaller structures consisting of several to tens of emitters requires different approaches.
 
In this work, we demonstrate the feasibility of forming singly and doubly excited subradiant eigenstates in finite dipole ensembles based on the mechanism of external coupling, which was initially proposed by H. Friedrich and D. Wintgen~\cite{Friedrich1985a} for open quantum systems. It has recently garnered significant attention in the field of nanophotonics for highly efficient nonlinear generation~\cite{Koshelev2019}, lasing in single semiconductor nanostructures~\cite{Mylnikov2020}, achieving a strong nonlinear response~\cite{ryabov_nonlinear_2022}, and engineering bound states in the continuum of extended periodic structures such as metasurfaces~\cite{Hsu2016, koshelev_bound_2021}. For the first time, we demonstrate that this mechanism can also be extended to form doubly excited subradiant states. Our focus is on concentric rings of two-level dipole emitters, which have already attracted significant attention in quantum optics~\cite{freedhoff_cooperative_1986, cremer_polarization_2020, moreno-cardoner_subradiance-enhanced_2019, holzinger_nanoscale_2020, moreno-cardoner_efficient_2022,Holzinger2023Sep,Cech2023Nov}, owing to their high symmetry and relevance to various natural quantum systems such as organic molecules. The observation of long-lived doubly excited states with nonzero orbital momentum may find utility in quantum information protocols that involve beams with high angular momentum~\cite{Ding2015Feb, mirhosseini_high-dimensional_2015, cozzolino_orbital_2019, jen_cooperative_2018}. The proposed mechanism can be straightforwardly applied to ordered arrays with different symmetries and geometries to extend the lifetime of quantum excitations.
\begin{figure}[t!]
    \centering
    \includegraphics[width=\linewidth]{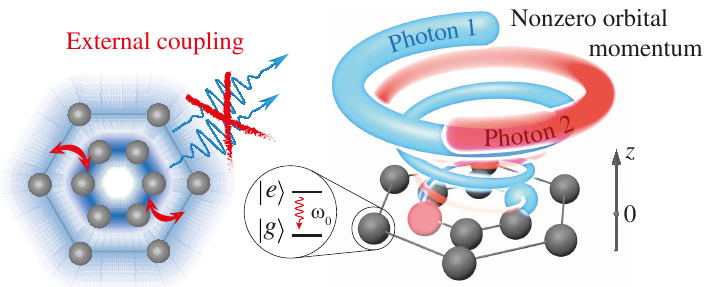}
    \caption{An open system representing a double-ring oligomer of two-level dipole emitters. The doubly excited eigenstates can have nonzero orbital quasimomentum with radiative losses that can be strongly suppressed via the mechanism of external coupling.}
    \label{fig1:main}
\end{figure}

{\it General formalism.}---Let us consider a double-ring ensemble of $N$ two-level dipole emitters shown in Fig.~\ref{fig1:main}. All emitters are located in the $z=0$ plane. We also assume that all emitters have a ground state with $J = 0$ and an excited level with $J' = 1$~\cite{Pellegrino2014Sep, Kemp2020Mar, Antezza2009Sep, Sokolov2019Aug, Fofanov2021, bouscal_systematic_2024}. In such a structure, the lifetime of all three Zeeman sublevels ($J'_z = -1, 0, 1$) is the same $\tau_0 = 1/\gamma_0$ with the decay rate $\gamma_0 = k_0^3 |\mathbf{d}|^2 / (3 \pi \hbar \epsilon_0),$ where $k_0 = \omega_0/c = 2 \pi/\lambda_0$ is the wavenumber in vacuum, $\omega_0$ is the transition frequency, and $\epsilon_0$ is the vacuum permittivity. The transition dipole moment of emitters is fixed to be oriented along the $z$ axis, $\mathbf{d} = |\mathbf{d}| \mathbf{e}_z$, which can be achieved by applying an external magnetic field isolating this transition from the two in-plane ones. The emitters are coupled via free-space electromagnetic modes, and their quantum states are governed by the effective non-Hermitian Hamiltonian (further, the Planck constant is set to be $\hbar = 1$)~\cite{moreno-cardoner_subradiance-enhanced_2019}, 
$\widehat{H}_{\mathrm{eff}} =  - i \dfrac{\gamma_{0}}{2}  \sum\limits_{k = 1}^N  \sigd_k \sig_k + \sum \limits_{k=1}^N \sum\limits_{\substack{l=1, \\ l\neq k}}^N  g(|\mathbf{r}_{kl}|, \omega_0) \sigd_k \sig_l,$ where $\sigd_k$ ($\sig_k$) is the creation (annihilation) operator for excitation on emitter $k$, $\left|\mathbf{r}_{kl}\right| = \left|\mathbf{r}_k - \mathbf{r}_l\right|$ is the relative distance between emitters $k$ and $l$, and the energy of noninteracting system $\omega_0\sum\limits_{k = 1}^N  \sigd_k \sig_k$ has been subtracted. 
The coupling rate between two emitters is defined via the free-space electromagnetic Green's tensor~\cite{Novotny2012}: $g(|\mathbf{r}|, \omega_0) = \left(-3\gamma_0 \pi/k_0 \right) \mathbf{e}_z^{T} \cdot \bm{\mathsf{G}}_0(\mathbf{r}, \omega_0)\cdot \mathbf{e}_z$.
Assumption of the Born-Markov approximation $g(|\mathbf{r}|, \omega) \approx g(|\mathbf{r}|, \omega_0)$ allows to avoid the dispersion of the coupling rate since $\gamma_0 \ll \omega_0$.  

{\it Ring oligomer.}---First, we consider an ensemble of $N_d$ emitters arranged in a ring of radius $R$ with the corresponding separation distance between neighbor emitters $a = 2R\sin(\pi/N_d)$. The eigenvalues of the system can be found by substituting the effective Hamiltonian into Schr\"{o}dinger equation $\widehat{H}_{\mathrm{eff}}\ket{\psi} = \varepsilon \ket{\psi}$ (Sec.~S1 in Supplemental Material~\cite{suppl}).
Each singly excited eigenstate of the ring can be associated with the orbital quasimomentum $m$ due to the discrete rotational symmetry of the system, $\ket{\psi^{(m)}_{\text{ring}}}$. As an illustrative example, we consider $N_d=6$ dipoles, for which $m$ can be equal to $0, \pm 1, \pm 2, 3$ (Sec.~S2 in Supplemental Material~\cite{suppl}). These eigenstates have different parities under the symmetry operations from point group $D_{6h}$ and transform according to the irreducible representations $A_{2u}$, $E_{1g}$, $E_{2u}$, and $B_{1g}$, respectively (Sec.~S4 in Supplemental Material~\cite{suppl}). As a result, the eigenstates with $\pm m$ are degenerate, i.e., $\varepsilon^{(m)} = \varepsilon^{(-m)}$. In the basis of coupled emitters, the eigenstate with a quasimomentum $m$ reads as $\ket{\psi^{(m)}_{\text{ring}}}= \sum\limits_{k=1}^N c^{(m)}_k \hat{\sigma}^{\dagger}_k \ket{g}^{\otimes N_d}$, where $c^{(m)}_k = e^{i m \varphi_k} / \sqrt{N_d}$ is the excitation probability amplitude for emitter $k$ and $\varphi_k = 2 \pi \left(k-1\right) / N_d$ (see Refs.~\cite{freedhoff_cooperative_1986,moreno-cardoner_subradiance-enhanced_2019} and also Sec.~S2 in Supplemental Material~\cite{suppl}).

\begin{figure}[t!]
    \centering
    \includegraphics[scale=0.57]{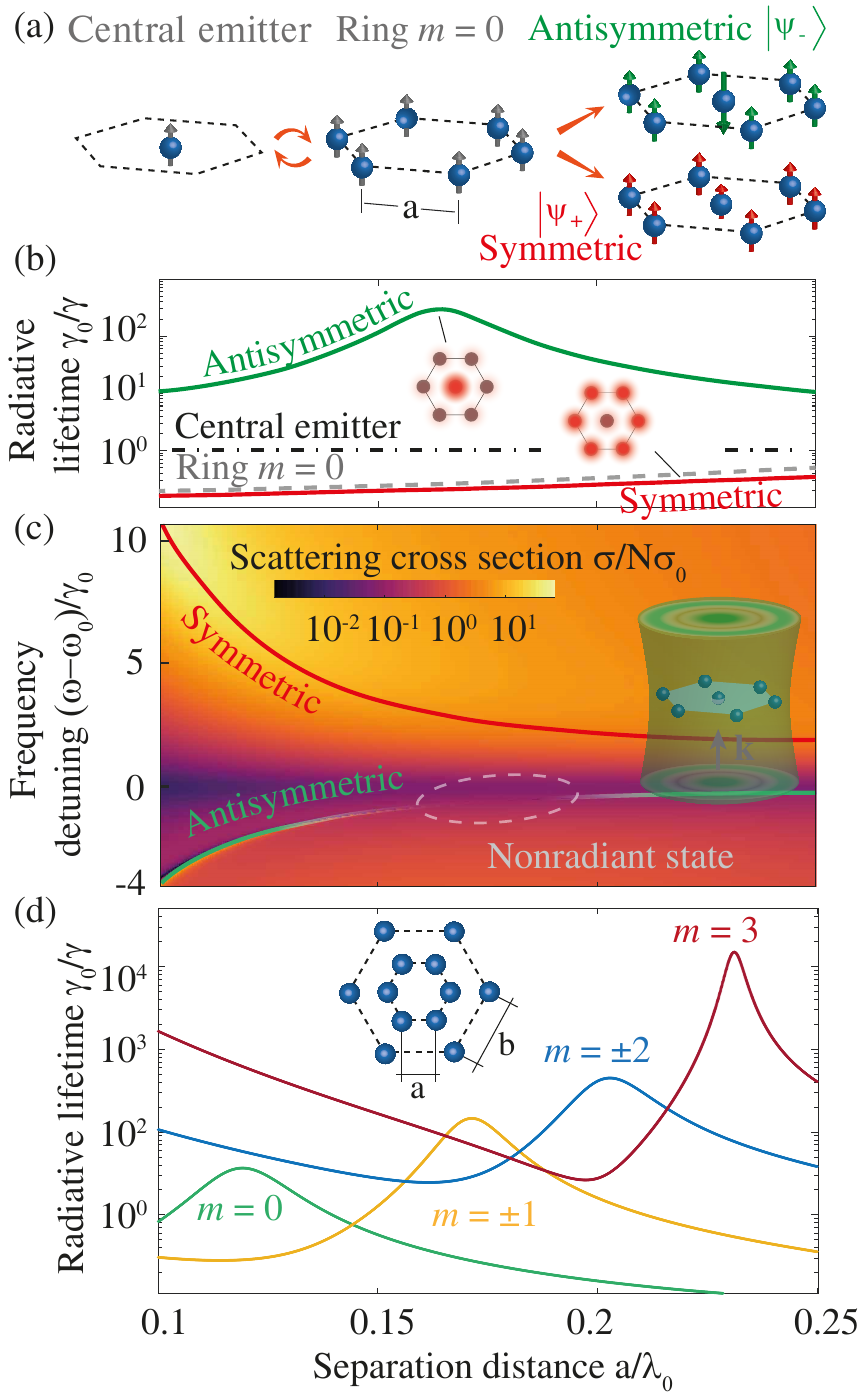}
    \caption{(a) Schematic representation of the hybridization of the ring states with a single emitter state. (b) Lifetime enhancement for the states in panel (a). Inset: The excitation probability is shown in red for the oligomer eigenstates. (c) Scattering cross section (color coded) for the ring oligomer excited with the Bessel beam (inset). The red and green lines represent the eigenfrequencies of the oligomer eigenstates. (d) Lifetime enhancement for antisymmetric eigenstates with different $m$ forming in two rings with $b/a = 2$ (inset).}
    \label{fig:2}
\end{figure}

The radiative losses of a single ring do not exhibit any peculiarities~\cite{Asenjo-Garcia2017,moreno-cardoner_subradiance-enhanced_2019,cremer_polarization_2020} and the collective decay rate of subradiant eigenstates monotonically decreases with the ring size (Sec.~S5 Supplemental Material~\cite{suppl}).  However, the situation can be drastically changed by adding a single emitter at the ring center, forming the \textit{oligomer} ensemble, as shown in Fig.~\ref{fig:2}(a). The external coupling between the ring state with $m=0$ and the central dipole via the radiation continuum leads to the formation of a long-living state. This can be understood by constructing the wave function for the oligomer eigenstates from two contributions,  $\ket{\psi}=c_a\ket{g_{\text{ring}}}\otimes\ket{e_{\text{0}}} + c_b|\psi^{(0)}_{\text{ring}}\rangle\otimes\ket {g_{\text{0}}}$, where $\ket{g_{\text{0}}}$ and $\ket{e_{\text{0}}}$ are the wave functions of the central dipole emitter in the ground and excited states, respectively, and $\ket{g_{\text{ring}}} \equiv \ket{g}^{\otimes N_d}$ is the ground state of the ring. The central emitter can couple only to the ring state with $m=0$ due to symmetry considerations. The Hamiltonian in such a basis can be represented as a sum of the unperturbed $\widehat H_0$ and interaction $\widehat V$ parts~\cite{holzinger_nanoscale_2020},
\begin{gather}
\label{eq:H_1modes}
    \widehat{H} =\widehat H_0+\widehat V=
    \begin{pmatrix}
        {\varepsilon}_0& 0 \\
        0& {\varepsilon}_{\text{ring}}^{(m=0)}
    \end{pmatrix}
    +
    \begin{pmatrix}
       0 & {\varkappa}\\
        {\varkappa} & 0
    \end{pmatrix},
\end{gather}
where ${\varepsilon}_0 = - {i\gamma_0}/{2}$ and ${\varepsilon}_{\text{ring}}^{(m=0)} =  - {i\gamma_0}/{2}+\sum \limits_{k=2}^{N_d}  g(|\mathbf{r}_{1k}|, \omega_0)$ are the energies of the excited central emitter $\ket{e_{0}}$ and the excited ring eigenstate $\ket{\psi^{(0)}_{\text{ring}}}$, respectively, while the coupling rate is given by ${\varkappa} = \sqrt{N_d} g\left(R, \omega_0 \right)$. Although $\varkappa$ is proportional to $\sqrt{N_d}$, the radius of the ring increases and $g$ decreases with $N_d$, leading to the overall decrease of the $\varkappa$. Consequently, the external coupling in such an oligomer is strong enough to induce noticeable lifetime modification only for not very large $N_d$.

As depicted in Fig.~\ref{fig:2}(a), the interaction between subsystems
leads to the appearance of symmetric $\ket{\psi_+}$ and antisymmetric
$\ket{\psi_-}$ hybridized eigenstates with $m = 0$. Their decay rates are defined by the imaginary part of eigenenergies $\gamma_{\pm}= - 2\Im\left(\veps_{\pm}\right)$ where $\veps_{\pm}$ are given in Sec.~S6 in Supplemental Material~\cite{suppl}. The calculations reveal that the antisymmetric eigenstate possesses resonantly increased lifetime by a factor of $\tau_{-}/\tau_0=\gamma_0/\gamma_{-}\approx230$ for the particular separation between emitters $a/\lambda_0 \approx 0.16$ [see Fig.~\ref{fig:2}(b)].  At this point, the phase shift between the probability amplitudes for the central dipole and the ring reaches $\pi$ exactly, while the excitation is primarily located at the central emitter [see the inset in Fig.~\ref{fig:2}(b) and Sec.~S6 in Supplemental Material~\cite{suppl}]. Such a suppression of radiation, caused by external coupling between the ring and the central emitter states, resembles the destructive interference between the radial modes in dielectric cylindrical cavities~\cite{Cao2015a, rybin_high-q_2017}.

The long-living oligomer state can be excited with tightly focused Bessel beams possessing a longitudinal component of the field~\cite{kotlyar_exploiting_2019}. In Fig.~\ref{fig:2}(c), one can see the scattering cross section (SCS) $\sigma$ for the oligomer ensembles of different sizes illuminated by a Bessel beam with orbital angular momentum $\ell=1$ and spin $s=-1$ adding up to the total angular momentum $m = 0$. The SCS  of the oligomer is normalized by $N \sigma_0$ where $N = N_d +1 = 7$ is the total number of emitters in the system, and $\sigma_0$ is the SCS for the central emitter (Sec.~S7 in Supplemental Material~\cite{suppl}). One can observe a resonant enhancement of the SCS for the symmetric (superradiant) eigenstate $\ket{\psi_+}$ and a drastic narrowing of the spectral line corresponding to the antisymmetric (subradiant) eigenstate $\ket{\psi_-}$ when approaching the optimal size condition.

The proposed external coupling mechanism can be also applied to control the lifetime of eigenstates with $m \neq 0$. For instance, in the oligomer ensemble consisting of two concentric rings shown in Fig.~\ref{fig1:main}, the symmetry of singly excited states in the isolated inner and outer rings, or $\ket{\psi^{(m)}_{\text{in-ring}}}$ and $\ket{\psi^{(m)}_{\text{out-ring}}}$, is the same. 
Therefore, the coupling between rings can lead to radiation suppression for arbitrary $m$. The mechanism of such an interaction can be also described within the framework of two coupled states with the wave function of the coupled system given as $\ket{\psi^{(m)}_{\text{two-ring}}} = c_a^{(m)}\ket{\psi^{(m)}_{\text{in-ring}}}\otimes\ket{g_{\text{out-ring}}} + c_b^{(m)}\ket{g_{\text{in-ring}}}\otimes\ket{\psi^{(m)}_{\text{out-ring}}}$. By scaling the oligomer size with the fixed ratio $b/a = 2$, one can reach the regime of resonantly enhanced lifetime for antisymmetric eigenstates with different $m$ as shown in Fig.~\ref{fig:2}(d). Additional details on the superradiant symmetric states are provided in Sec.~S6 in Supplemental Material~\cite{suppl}.

{\it Doubly excited states.}---The mechanism of external coupling can also be exploited to doubly excited states. Doubly excited quantum states form a manifold in the Hilbert space with a dimension of $N(N-1)/2$. Due to the symmetry of the wave function and the Pauli principle, they can be expanded over the wave function basis as $\ket{\Psi}=\sum\limits_{k = 1}^N\sum\limits_{l = k + 1}^Nc_{kl}\hat{ \sigma}^{\dagger}_k\hat{\sigma}_l^{\dagger}\ket{g}^{\otimes N}$. 
Importantly, one can characterize doubly excited eigenstates of ring oligomers with orbital quasimomentum $m$ in a similar manner to the singly excited eigenstates. 
Moreover, a doubly excited eigenstate can be expanded over the products of singly excited eigenstates, $\ket{\Psi^{(m)}} = \sum\limits_{m_1,m_2} v_{m_1, m_2}  \ket{\psi^{(m_1)}}\ket{\psi^{(m_2)}}$, with $v_{m_1, m_2} \neq 0$ if $m_1 + m_2 = m \ (\text{mod} \ N_d)$. This condition for the quasimomentum immediately follows from representation theory. 
For example, a direct product of two wave functions of singly excited states entering $E_{1g}$ ($m_1=\pm1$) and $E_{2u}$ ($m_1=\pm2$) irreducible representations results in the wave function of the doubly excited state that enters one of the three irreducible representations $E_{1g} \otimes E_{2u} = B_{1u} + B_{2u} + E_{1u}$, with a total $m=3$ ($B_{1u}$, $B_{2u}$), or $m = \pm1$ ($E_{1u}$).

\begin{figure}[t!]
    \centering
    \includegraphics[scale=0.57]{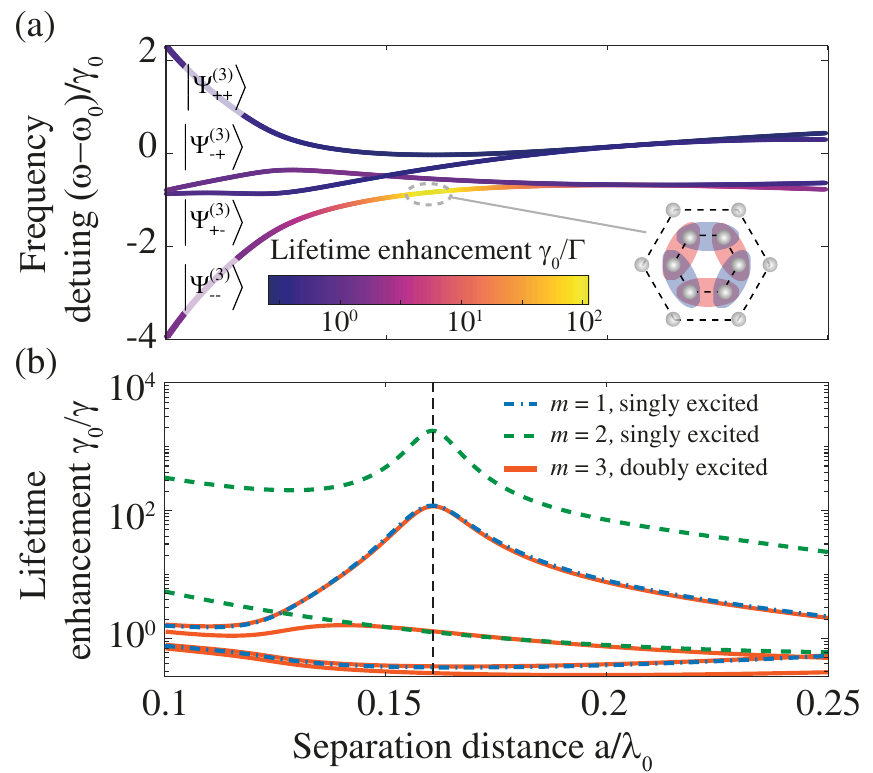}
    \caption{(a) Eigenfrequency curves for eigenstates~\eqref{eq_Psi_B1} where the color corresponds to the lifetime enhancement. The inset schematically shows the largest amplitudes $c_{kl}$ for the nonradiant eigenstate. (b) The lifetime enhancement for singly excited eigenstates with $m = 1$ (blue dashed dotted) and $m = 2$ (green dashed) that form doubly excited eigenstates~\eqref{eq_Psi_B1} with $m = 3$ (orange solid).}
    \label{fig:2exc_state}
\end{figure}

Two rings of $N_d = 6$ emitters support 61 doubly excited eigenstates including ten ones with the largest orbital quasimomentum $m = 3$ (Sec.~S3 in Supplemental Material~\cite{suppl}). From now on, we will pay special attention to the eigenstates $\ket{\Psi^{(3)}}$ with $m = 3$, which also enter the $B_{1u}$ irreducible representation. The indistinguishability of excitations and the Pauli principle imply that the double-ring oligomer supports only four possible doubly excited eigenstates of this type,
\begin{align}
\label{eq_Psi_B1}
\ket{\Psi^{(3)}_{s_1,s_2}} &= \frac{i}{2} \left(\ket{\psi^{(+1)}_{s_1}}\ket{\psi^{(+2)}_{s_2}} + \ket{\psi^{(+2)}_{s_2}}\ket{\psi^{(+1)}_{s_1}}\right. \nonumber \\
        &-  \left. \ket{\psi^{(-1)}_{s_1}}\ket{\psi^{(-2)}_{s_2}} - \ket{\psi^{(-2)}_{s_2}}\ket{\psi^{(-1)}_{s_1}} \right),  
\end{align}
where $s_{1},s_2=\pm$ correspond to symmetric (antisymmetric) singly excited eigenstates of the double-ring oligomer. We note that the other six doubly excited eigenstates with $m=3$ have lower lifetimes and correspond to the $B_{2u}$ symmetry since they contain a direct product of the singly excited eigenstates with $m_1=0$ ($A_{2u}$ representation) and $m_2=3$ ($B_{2u}$ representation).

Doubly excited eigenstates inherit their properties from at least two singly excited eigenstates, and therefore their lifetime can also be controlled with the external coupling mechanism. Indeed, the energy $\mathcal{E}^{(m)} = \omega^{(m)} - i \Gamma^{(m)}/2$ of a doubly excited state $\ket{\Psi^{(m)}}$ can be written as (Sec.~S8 in Supplemental Material~\cite{suppl})
\begin{align}
\label{eq:E2}
\mathcal{E}^{(m)}=\sum_{m_1,m_2}\left|v_{m_1,m_2}\right|^2 \left[\veps^{(m_1)}+\veps^{(m_2)}\right], 
\end{align}
whereas amplitudes $v_{m_1,m_2}$ can be found based on the direct diagonalization of the effective Hamiltonian (Sec.~S1 in Supplemental Material~\cite{suppl}). Hence, the energies of states~\eqref{eq_Psi_B1} are $\mathcal{E}^{(3)}_{s_1,s_2}=\veps^{(+1)}_{s_1}+\veps^{(+2)}_{s_2}$ where $\veps^{(m)}_{\pm}$ are given in Sec.~S6,~\cite{suppl}. Consequently, the radiative decay of states~\eqref{eq_Psi_B1} can also be decomposed into the sum $\Gamma^{(3)}_{s_1,s_2}=\gamma^{(+1)}_{s_1}+\gamma^{(+2)}_{s_2}$. Hence, the suppression of radiative losses for eigenstates~\eqref{eq_Psi_B1} can be achieved for a particular oligomer geometry when both singly excited eigenstates with $m_1 = 1$ and $m_2 = 2$ have the lowest radiative losses. By varying the inner ($a/\lambda_0$) and outer ring ($b/a$ ratio) sizes independently, we find the optimal parameters to be $b/a = 2.2$ and $a/\lambda_0 \approx 0.16$ 
(Sec.~S9 in Supplemental Material~\cite{suppl}).
Indeed, for these parameters,  the fully antisymmetric eigenstate $\ket{\Psi^{(3)}_{--}}$ has a lifetime that is two orders of magnitude larger than that of a single emitter [see Fig. \ref{fig:2exc_state}(a)]. Moreover, this point is characterized by the maximal lifetime of both antisymmetric singly excited eigenstates with $m = 1$ and $m = 2$ [see Fig.~\ref{fig:2exc_state}(b)]. The radiative losses for the antisymmetric state with $m=2$ are much smaller than those for $m=1$, i.e., $\gamma^{(2)}_{-}\ll \gamma^{(1)}_{-}$, therefore, the overall radiative losses for the non-radiant doubly excited eigenstate $\ket{\Psi^{(3)}_{--}}$ are $\Gamma^{(3)}_{--}\approx \gamma_-^{(1)}$ [see Fig.~\ref{fig:2exc_state}(b)]. Additionally, we can emphasize that the form of the nonradiant eigenstate $\ket{\Psi^{(3)}_{--}}$, given by Eq.~\eqref{eq_Psi_B1}, implies that both excitations within this state are predominantly localized on the inner ring [see the inset in Fig.~\ref{fig:2exc_state}(a)], inheriting the properties of $\ket{\psi^{(\pm1)}_{-}}$ and $\ket{\psi^{(\pm2)}_{-}}$ nonradiant eigenstates.

\begin{figure}[t!]
    \centering
    \includegraphics[scale = 0.8]{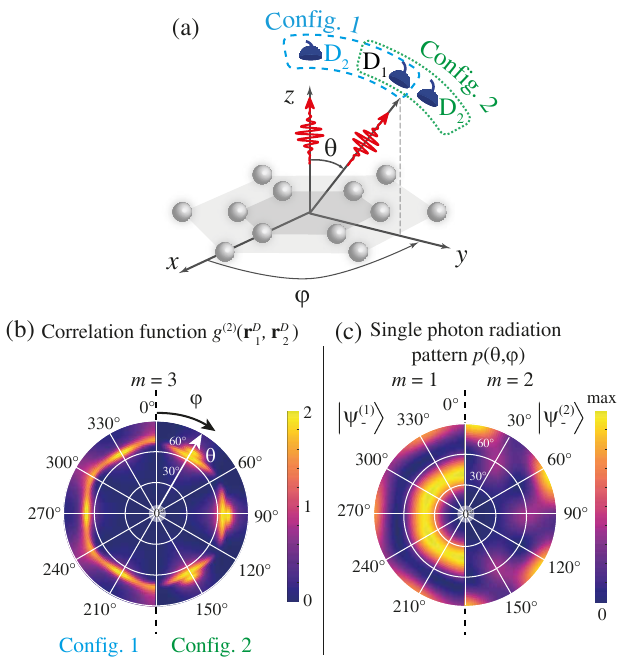}
    \caption{ (a) Configurations for detecting two-photon correlations within doubly excited states. (b) Far-field distribution of $g^{(2)}$ function~\eqref{eq:g2} for the nonradiant eigenstate $\ket{\Psi^{(3)}_{--}}$. (c) Far-field radiation patterns for the singly excited antisymmetric eigenstates $\ket{\psi^{(+1)}_-}$ and $\ket{\psi^{(+2)}_-}$, which form the nonradiant eigenstate $\ket{\Psi^{(3)}_{--}}$. The results in panels (b) and (c) are shown considering the symmetry of distributions. }
    \label{fig:4_g2}
\end{figure}

{\it Photon emission and spatial correlations}.---Finally, the radiative decay of doubly excited states can be characterized by the second-order correlation function, which may be necessary for the future design of potential detection schemes. This function  allows for describing spatial correlations between photons   emitted by a state $\ket{\Psi}$  when detectors $D_1$ and $D_2$ are positioned at coordinates $\ve r^D_1$ and  $\ve r^D_2$, respectively, and reads as~\cite{holzinger_nanoscale_2020,Gulfam2023Jun}:
\begin{multline}
\label{eq:g2}
   g^{(2)}(\ve{r}^D_1, \ve{r}^D_2) = \frac{\sum\limits_{\alpha,\beta}\langle \Psi | \hat E_{\beta,2}\hat E_{\alpha,1} \hat E^{\dagger}_{\alpha,1} \hat E^{\dagger}_{\beta,2} | \Psi \rangle}{\sum\limits_{\alpha,\beta}\langle \Psi | \hat E_{\beta,1}\hat E^{\dagger}_{\beta,1} | \Psi \rangle \langle \Psi | \hat E_{\alpha,2} E^{\dagger}_{\alpha,2} | \Psi \rangle}.
\end{multline}
Here, $\alpha,\beta = x,y,z$ denote the components in the Cartesian coordinate system, and $\hat{E}_{\alpha, 1(2)} \equiv \hat{E}_{\alpha}\left(\mathbf{r}^D_{1(2)}\right)$ is the electric field operator that creates a photon at the detector position $\mathbf{r}^D_{1(2)}$ with the polarization along the $\alpha$ axis. 
The corresponding electric field operator can be expressed via the free-space Green's tensor~\cite{Asenjo-Garcia2017}    $\hat{\mathbf{E}}^{\dagger}(\mathbf{r}) = { k_0^2}/{\epsilon_0} \sum\limits_{k = 1}^N \bm{\mathsf{G}}_0(\mathbf{r} - \mathbf{r}_k, \omega_0) \cdot \mathbf{d}\hat{\sigma}_k$.
Singly excited states can be characterized by the far-field radiation pattern of a single photon  $p(\theta,\varphi)$ (Sec.~S7 in Supplemental Material~\cite{suppl}).

In Fig.~\ref{fig:4_g2}(a), two possible configurations of a two-photon detection scheme are presented: One when a first detector is in the polar position ($\theta^D_1=0$), while a second one scans over the sphere (configuration 1), and another when both detectors are placed in the same point $\ve r^D_1=\ve r^D_2$ (configuration 2). The second-order correlation function~\eqref{eq:g2} of the nonradiant eigenstate $\ket{\Psi_{--}^{(3)}}$ [see Eq.~\eqref{eq_Psi_B1} and Fig.~\ref{fig:2exc_state}]  is shown in Fig.~\ref{fig:4_g2}(b). It exhibits typical hexagonal features with a maximal correlation function at a nodal line around $\theta\approx 60-70^{\circ}$ for both configurations. This behavior can also be explained by radiation patterns for the singly excited eigenstates $\ket{\psi^{(+1)}_-}$ and $\ket{\psi^{(+2)}_-}$ with $m = +1$ and $m = +2$, respectively, shown in Fig.~\ref{fig:4_g2}(c). While maximal emission of the $m=1$ eigenstate is observed around $\theta=35^{\circ}$, the emission of the $m=2$ eigenstate is mainly concentrated around the $\theta=90^{\circ}$ plane, which results in a maximal correlation function at an intermediate polar angle as shown in Fig.~\ref{fig:4_g2}(b).

{\it Discussion and conclusion.} --- As a remark, while the presented result mainly corresponds to linearly polarized $\sigma_z$ transitions, the external coupling mechanism can be also applied for creating subradiant states for circularly polarized $\sigma_{\pm}$ transitions with the dipole moments lying in the plane of the oligomer structure; see Sec.~S10 in Supplemental Material~\cite{suppl}. The recent progress in experimental techniques~\cite{rui_subradiant_2020,jin_two-dimensional_2023} also motivated by a number of fascinating theoretical concepts~\cite{shahmoon_cooperative_2017, Asenjo-Garcia2017, plankensteiner_selective_2015, Cech2023Nov, fedorovich_chirality-driven_2022} proposed for quantum arrays with subwavelength spacing give hope that the small spacing between the atoms required for the observation of the discussed nonradiant states will be reached soon.            

Finally, we have exploited the Friedrich-Wintgen mechanism to demonstrate the formation of subradiant singly and doubly excited eigenstates with a given orbital momentum in the ring oligomers. We have shown that the oligomers can be viewed as two subsystems of emitters supporting states that interact if they possess the same symmetry. The proposed mechanism relies on the destructive interference between the subsystems resulting in the formation of antisymmetric states characterized by suppressed radiative losses for preoptimized oligomer geometry. The suggested approach is not limited to the systems considered in this work and can be applied to control radiative losses of multiphoton states in various open quantum systems. 

\begin{acknowledgments}
   The authors acknowledge Kristina Frizyuk, Andrey Bogdanov, Ivan Iorsh, and Yuri Kivshar for fruitful discussions. The work was supported by the Russian Academic Leadership Program Priority 2030. 
\end{acknowledgments}

\bibliography{ms}

\begin{thebibliography}{89}%
\makeatletter
\providecommand \@ifxundefined [1]{%
 \@ifx{#1\undefined}
}%
\providecommand \@ifnum [1]{%
 \ifnum #1\expandafter \@firstoftwo
 \else \expandafter \@secondoftwo
 \fi
}%
\providecommand \@ifx [1]{%
 \ifx #1\expandafter \@firstoftwo
 \else \expandafter \@secondoftwo
 \fi
}%
\providecommand \natexlab [1]{#1}%
\providecommand \enquote  [1]{``#1''}%
\providecommand \bibnamefont  [1]{#1}%
\providecommand \bibfnamefont [1]{#1}%
\providecommand \citenamefont [1]{#1}%
\providecommand \href@noop [0]{\@secondoftwo}%
\providecommand \href [0]{\begingroup \@sanitize@url \@href}%
\providecommand \@href[1]{\@@startlink{#1}\@@href}%
\providecommand \@@href[1]{\endgroup#1\@@endlink}%
\providecommand \@sanitize@url [0]{\catcode `\\12\catcode `\$12\catcode `\&12\catcode `\#12\catcode `\^12\catcode `\_12\catcode `\%12\relax}%
\providecommand \@@startlink[1]{}%
\providecommand \@@endlink[0]{}%
\providecommand \url  [0]{\begingroup\@sanitize@url \@url }%
\providecommand \@url [1]{\endgroup\@href {#1}{\urlprefix }}%
\providecommand \urlprefix  [0]{URL }%
\providecommand \Eprint [0]{\href }%
\providecommand \doibase [0]{https://doi.org/}%
\providecommand \selectlanguage [0]{\@gobble}%
\providecommand \bibinfo  [0]{\@secondoftwo}%
\providecommand \bibfield  [0]{\@secondoftwo}%
\providecommand \translation [1]{[#1]}%
\providecommand \BibitemOpen [0]{}%
\providecommand \bibitemStop [0]{}%
\providecommand \bibitemNoStop [0]{.\EOS\space}%
\providecommand \EOS [0]{\spacefactor3000\relax}%
\providecommand \BibitemShut  [1]{\csname bibitem#1\endcsname}%
\let\auto@bib@innerbib\@empty
\bibitem [{\citenamefont {Chang}\ \emph {et~al.}(2018)\citenamefont {Chang}, \citenamefont {Douglas}, \citenamefont {Gonz{\ifmmode\acute{a}\else\'{a}\fi}lez-Tudela}, \citenamefont {Hung},\ and\ \citenamefont {Kimble}}]{Chang2018}%
  \BibitemOpen
  \bibfield  {author} {\bibinfo {author} {\bibfnamefont {D.~E.}\ \bibnamefont {Chang}}, \bibinfo {author} {\bibfnamefont {J.~S.}\ \bibnamefont {Douglas}}, \bibinfo {author} {\bibfnamefont {A.}~\bibnamefont {Gonz{\ifmmode\acute{a}\else\'{a}\fi}lez-Tudela}}, \bibinfo {author} {\bibfnamefont {C.-L.}\ \bibnamefont {Hung}},\ and\ \bibinfo {author} {\bibfnamefont {H.~J.}\ \bibnamefont {Kimble}},\ }\bibfield  {title} {\bibinfo {title} {{Colloquium: Quantum matter built from nanoscopic lattices of atoms and photons}},\ }\href {https://doi.org/10.1103/RevModPhys.90.031002} {\bibfield  {journal} {\bibinfo  {journal} {Rev. Mod. Phys.}\ }\textbf {\bibinfo {volume} {90}},\ \bibinfo {pages} {031002} (\bibinfo {year} {2018})}\BibitemShut {NoStop}%
\bibitem [{\citenamefont {Sheremet}\ \emph {et~al.}(2023)\citenamefont {Sheremet}, \citenamefont {Petrov}, \citenamefont {Iorsh}, \citenamefont {Poshakinskiy},\ and\ \citenamefont {Poddubny}}]{sheremet_waveguide_2023}%
  \BibitemOpen
  \bibfield  {author} {\bibinfo {author} {\bibfnamefont {A.~S.}\ \bibnamefont {Sheremet}}, \bibinfo {author} {\bibfnamefont {M.~I.}\ \bibnamefont {Petrov}}, \bibinfo {author} {\bibfnamefont {I.~V.}\ \bibnamefont {Iorsh}}, \bibinfo {author} {\bibfnamefont {A.~V.}\ \bibnamefont {Poshakinskiy}},\ and\ \bibinfo {author} {\bibfnamefont {A.~N.}\ \bibnamefont {Poddubny}},\ }\bibfield  {title} {\bibinfo {title} {{Waveguide quantum electrodynamics: Collective radiance and photon-photon correlations}},\ }\href {https://doi.org/10.1103/RevModPhys.95.015002} {\bibfield  {journal} {\bibinfo  {journal} {Rev. Mod. Phys.}\ }\textbf {\bibinfo {volume} {95}},\ \bibinfo {pages} {015002} (\bibinfo {year} {2023})}\BibitemShut {NoStop}%
\bibitem [{\citenamefont {Kimble}(2008)}]{Kimble2008}%
  \BibitemOpen
  \bibfield  {author} {\bibinfo {author} {\bibfnamefont {H.~J.}\ \bibnamefont {Kimble}},\ }\bibfield  {title} {\bibinfo {title} {{The quantum internet}},\ }\href {https://doi.org/10.1038/nature07127} {\bibfield  {journal} {\bibinfo  {journal} {Nature}\ }\textbf {\bibinfo {volume} {453}},\ \bibinfo {pages} {1023} (\bibinfo {year} {2008})}\BibitemShut {NoStop}%
\bibitem [{\citenamefont {Hammerer}\ \emph {et~al.}(2010)\citenamefont {Hammerer}, \citenamefont {S{\o}rensen},\ and\ \citenamefont {Polzik}}]{Hammerer2010}%
  \BibitemOpen
  \bibfield  {author} {\bibinfo {author} {\bibfnamefont {K.}~\bibnamefont {Hammerer}}, \bibinfo {author} {\bibfnamefont {A.~S.}\ \bibnamefont {S{\o}rensen}},\ and\ \bibinfo {author} {\bibfnamefont {E.~S.}\ \bibnamefont {Polzik}},\ }\bibfield  {title} {\bibinfo {title} {{Quantum interface between light and atomic ensembles}},\ }\href {https://doi.org/10.1103/RevModPhys.82.1041} {\bibfield  {journal} {\bibinfo  {journal} {Rev. Mod. Phys.}\ }\textbf {\bibinfo {volume} {82}},\ \bibinfo {pages} {1041} (\bibinfo {year} {2010})}\BibitemShut {NoStop}%
\bibitem [{\citenamefont {Abend}\ \emph {et~al.}(2023)\citenamefont {Abend}, \citenamefont {Allard}, \citenamefont {Arnold}, \citenamefont {Ban}, \citenamefont {Barry}, \citenamefont {Battelier}, \citenamefont {Bawamia}, \citenamefont {Beaufils}, \citenamefont {Bernon}, \citenamefont {Bertoldi}, \citenamefont {Bonnin}, \citenamefont {Bouyer}, \citenamefont {Bresson}, \citenamefont {Burrow}, \citenamefont {Canuel}, \citenamefont {Desruelle}, \citenamefont {Drougakis}, \citenamefont {Forsberg}, \citenamefont {Gaaloul}, \citenamefont {Gauguet}, \citenamefont {Gersemann}, \citenamefont {Griffin}, \citenamefont {Heine}, \citenamefont {Henderson}, \citenamefont {Herr}, \citenamefont {Kanthak}, \citenamefont {Krutzik}, \citenamefont {Lachmann}, \citenamefont {Lammegger}, \citenamefont {Magnes}, \citenamefont {Mileti}, \citenamefont {Mitchell}, \citenamefont {Mottini}, \citenamefont {Papazoglou}, \citenamefont {Pereira~dos Santos}, \citenamefont {Peters}, \citenamefont {Rasel}, \citenamefont {Riis}, \citenamefont
  {Schubert}, \citenamefont {Seidel}, \citenamefont {Tino}, \citenamefont {Van Den~Bossche}, \citenamefont {von Klitzing}, \citenamefont {Wicht}, \citenamefont {Witkowski}, \citenamefont {Zahzam},\ and\ \citenamefont {Zawada}}]{abend_technology_2023}%
  \BibitemOpen
  \bibfield  {author} {\bibinfo {author} {\bibfnamefont {S.}~\bibnamefont {Abend}}, \bibinfo {author} {\bibfnamefont {B.}~\bibnamefont {Allard}}, \bibinfo {author} {\bibfnamefont {A.~S.}\ \bibnamefont {Arnold}}, \bibinfo {author} {\bibfnamefont {T.}~\bibnamefont {Ban}}, \bibinfo {author} {\bibfnamefont {L.}~\bibnamefont {Barry}}, \bibinfo {author} {\bibfnamefont {B.}~\bibnamefont {Battelier}}, \bibinfo {author} {\bibfnamefont {A.}~\bibnamefont {Bawamia}}, \bibinfo {author} {\bibfnamefont {Q.}~\bibnamefont {Beaufils}}, \bibinfo {author} {\bibfnamefont {S.}~\bibnamefont {Bernon}}, \bibinfo {author} {\bibfnamefont {A.}~\bibnamefont {Bertoldi}}, \bibinfo {author} {\bibfnamefont {A.}~\bibnamefont {Bonnin}}, \bibinfo {author} {\bibfnamefont {P.}~\bibnamefont {Bouyer}}, \bibinfo {author} {\bibfnamefont {A.}~\bibnamefont {Bresson}}, \bibinfo {author} {\bibfnamefont {O.~S.}\ \bibnamefont {Burrow}}, \bibinfo {author} {\bibfnamefont {B.}~\bibnamefont {Canuel}}, \bibinfo {author} {\bibfnamefont {B.}~\bibnamefont
  {Desruelle}}, \bibinfo {author} {\bibfnamefont {G.}~\bibnamefont {Drougakis}}, \bibinfo {author} {\bibfnamefont {R.}~\bibnamefont {Forsberg}}, \bibinfo {author} {\bibfnamefont {N.}~\bibnamefont {Gaaloul}}, \bibinfo {author} {\bibfnamefont {A.}~\bibnamefont {Gauguet}}, \bibinfo {author} {\bibfnamefont {M.}~\bibnamefont {Gersemann}}, \bibinfo {author} {\bibfnamefont {P.~F.}\ \bibnamefont {Griffin}}, \bibinfo {author} {\bibfnamefont {H.}~\bibnamefont {Heine}}, \bibinfo {author} {\bibfnamefont {V.~A.}\ \bibnamefont {Henderson}}, \bibinfo {author} {\bibfnamefont {W.}~\bibnamefont {Herr}}, \bibinfo {author} {\bibfnamefont {S.}~\bibnamefont {Kanthak}}, \bibinfo {author} {\bibfnamefont {M.}~\bibnamefont {Krutzik}}, \bibinfo {author} {\bibfnamefont {M.~D.}\ \bibnamefont {Lachmann}}, \bibinfo {author} {\bibfnamefont {R.}~\bibnamefont {Lammegger}}, \bibinfo {author} {\bibfnamefont {W.}~\bibnamefont {Magnes}}, \bibinfo {author} {\bibfnamefont {G.}~\bibnamefont {Mileti}}, \bibinfo {author} {\bibfnamefont {M.~W.}\
  \bibnamefont {Mitchell}}, \bibinfo {author} {\bibfnamefont {S.}~\bibnamefont {Mottini}}, \bibinfo {author} {\bibfnamefont {D.}~\bibnamefont {Papazoglou}}, \bibinfo {author} {\bibfnamefont {F.}~\bibnamefont {Pereira~dos Santos}}, \bibinfo {author} {\bibfnamefont {A.}~\bibnamefont {Peters}}, \bibinfo {author} {\bibfnamefont {E.}~\bibnamefont {Rasel}}, \bibinfo {author} {\bibfnamefont {E.}~\bibnamefont {Riis}}, \bibinfo {author} {\bibfnamefont {C.}~\bibnamefont {Schubert}}, \bibinfo {author} {\bibfnamefont {S.~T.}\ \bibnamefont {Seidel}}, \bibinfo {author} {\bibfnamefont {G.~M.}\ \bibnamefont {Tino}}, \bibinfo {author} {\bibfnamefont {M.}~\bibnamefont {Van Den~Bossche}}, \bibinfo {author} {\bibfnamefont {W.}~\bibnamefont {von Klitzing}}, \bibinfo {author} {\bibfnamefont {A.}~\bibnamefont {Wicht}}, \bibinfo {author} {\bibfnamefont {M.}~\bibnamefont {Witkowski}}, \bibinfo {author} {\bibfnamefont {N.}~\bibnamefont {Zahzam}},\ and\ \bibinfo {author} {\bibfnamefont {M.}~\bibnamefont {Zawada}},\ }\bibfield  {title}
  {\bibinfo {title} {{Technology roadmap for cold-atoms based quantum inertial sensor in space}},\ }\href {https://doi.org/10.1116/5.0098119} {\bibfield  {journal} {\bibinfo  {journal} {AVS Quantum Sci.}\ }\textbf {\bibinfo {volume} {5}},\ \bibinfo {pages} {019201} (\bibinfo {year} {2023})}\BibitemShut {NoStop}%
\bibitem [{\citenamefont {Dubosclard}\ \emph {et~al.}(2021)\citenamefont {Dubosclard}, \citenamefont {Kim},\ and\ \citenamefont {Garrido~Alzar}}]{dubosclard_nondestructive_2021}%
  \BibitemOpen
  \bibfield  {author} {\bibinfo {author} {\bibfnamefont {W.}~\bibnamefont {Dubosclard}}, \bibinfo {author} {\bibfnamefont {S.}~\bibnamefont {Kim}},\ and\ \bibinfo {author} {\bibfnamefont {C.~L.}\ \bibnamefont {Garrido~Alzar}},\ }\bibfield  {title} {\bibinfo {title} {{Nondestructive microwave detection of a coherent quantum dynamics in cold atoms}},\ }\href {https://doi.org/10.1038/s42005-021-00541-3} {\bibfield  {journal} {\bibinfo  {journal} {Commun. Phys.}\ }\textbf {\bibinfo {volume} {4}},\ \bibinfo {pages} {1} (\bibinfo {year} {2021})}\BibitemShut {NoStop}%
\bibitem [{\citenamefont {Meng}\ \emph {et~al.}(2018)\citenamefont {Meng}, \citenamefont {Dareau}, \citenamefont {Schneeweiss},\ and\ \citenamefont {Rauschenbeutel}}]{Meng2018}%
  \BibitemOpen
  \bibfield  {author} {\bibinfo {author} {\bibfnamefont {Y.}~\bibnamefont {Meng}}, \bibinfo {author} {\bibfnamefont {A.}~\bibnamefont {Dareau}}, \bibinfo {author} {\bibfnamefont {P.}~\bibnamefont {Schneeweiss}},\ and\ \bibinfo {author} {\bibfnamefont {A.}~\bibnamefont {Rauschenbeutel}},\ }\bibfield  {title} {\bibinfo {title} {{Near-Ground-State Cooling of Atoms Optically Trapped 300 nm Away from a Hot Surface}},\ }\href {https://doi.org/10.1103/PhysRevX.8.031054} {\bibfield  {journal} {\bibinfo  {journal} {Phys. Rev. X}\ }\textbf {\bibinfo {volume} {8}},\ \bibinfo {pages} {031054} (\bibinfo {year} {2018})}\BibitemShut {NoStop}%
\bibitem [{\citenamefont {Jungkind}\ \emph {et~al.}(2019)\citenamefont {Jungkind}, \citenamefont {Niedenzu},\ and\ \citenamefont {Ritsch}}]{Jungkind2019}%
  \BibitemOpen
  \bibfield  {author} {\bibinfo {author} {\bibfnamefont {A.}~\bibnamefont {Jungkind}}, \bibinfo {author} {\bibfnamefont {W.}~\bibnamefont {Niedenzu}},\ and\ \bibinfo {author} {\bibfnamefont {H.}~\bibnamefont {Ritsch}},\ }\bibfield  {title} {\bibinfo {title} {{Optomechanical cooling and self-trapping of low field seeking point-like particles}},\ }\href {https://doi.org/10.1088/1361-6455/ab2ab4} {\bibfield  {journal} {\bibinfo  {journal} {J. Phys. B: At. Mol. Opt. Phys.}\ }\textbf {\bibinfo {volume} {52}},\ \bibinfo {pages} {165003} (\bibinfo {year} {2019})}\BibitemShut {NoStop}%
\bibitem [{\citenamefont {Iorsh}\ \emph {et~al.}(2020)\citenamefont {Iorsh}, \citenamefont {Poshakinskiy},\ and\ \citenamefont {Poddubny}}]{iorsh_waveguide_2020}%
  \BibitemOpen
  \bibfield  {author} {\bibinfo {author} {\bibfnamefont {I.}~\bibnamefont {Iorsh}}, \bibinfo {author} {\bibfnamefont {A.}~\bibnamefont {Poshakinskiy}},\ and\ \bibinfo {author} {\bibfnamefont {A.}~\bibnamefont {Poddubny}},\ }\bibfield  {title} {\bibinfo {title} {{Waveguide Quantum Optomechanics: Parity-Time Phase Transitions in Ultrastrong Coupling Regime}},\ }\href {https://doi.org/10.1103/PhysRevLett.125.183601} {\bibfield  {journal} {\bibinfo  {journal} {Phys. Rev. Lett.}\ }\textbf {\bibinfo {volume} {125}},\ \bibinfo {pages} {183601} (\bibinfo {year} {2020})}\BibitemShut {NoStop}%
\bibitem [{\citenamefont {Shahmoon}\ \emph {et~al.}(2020)\citenamefont {Shahmoon}, \citenamefont {Lukin},\ and\ \citenamefont {Yelin}}]{shahmoon_quantum_2020}%
  \BibitemOpen
  \bibfield  {author} {\bibinfo {author} {\bibfnamefont {E.}~\bibnamefont {Shahmoon}}, \bibinfo {author} {\bibfnamefont {M.~D.}\ \bibnamefont {Lukin}},\ and\ \bibinfo {author} {\bibfnamefont {S.~F.}\ \bibnamefont {Yelin}},\ }\bibfield  {title} {\bibinfo {title} {{Quantum optomechanics of a two-dimensional atomic array}},\ }\href {https://doi.org/10.1103/PhysRevA.101.063833} {\bibfield  {journal} {\bibinfo  {journal} {Phys. Rev. A}\ }\textbf {\bibinfo {volume} {101}},\ \bibinfo {pages} {063833} (\bibinfo {year} {2020})}\BibitemShut {NoStop}%
\bibitem [{\citenamefont {Barredo}\ \emph {et~al.}(2016)\citenamefont {Barredo}, \citenamefont {de~L{\ifmmode\acute{e}\else\'{e}\fi}s{\ifmmode\acute{e}\else\'{e}\fi}leuc}, \citenamefont {Lienhard}, \citenamefont {Lahaye},\ and\ \citenamefont {Browaeys}}]{barredo_atom-by-atom_2016}%
  \BibitemOpen
  \bibfield  {author} {\bibinfo {author} {\bibfnamefont {D.}~\bibnamefont {Barredo}}, \bibinfo {author} {\bibfnamefont {S.}~\bibnamefont {de~L{\ifmmode\acute{e}\else\'{e}\fi}s{\ifmmode\acute{e}\else\'{e}\fi}leuc}}, \bibinfo {author} {\bibfnamefont {V.}~\bibnamefont {Lienhard}}, \bibinfo {author} {\bibfnamefont {T.}~\bibnamefont {Lahaye}},\ and\ \bibinfo {author} {\bibfnamefont {A.}~\bibnamefont {Browaeys}},\ }\bibfield  {title} {\bibinfo {title} {{An atom-by-atom assembler of defect-free arbitrary two-dimensional atomic arrays}},\ }\href {https://doi.org/10.1126/science.aah3778} {\bibfield  {journal} {\bibinfo  {journal} {Science}\ }\textbf {\bibinfo {volume} {354}},\ \bibinfo {pages} {1021} (\bibinfo {year} {2016})}\BibitemShut {NoStop}%
\bibitem [{\citenamefont {Endres}\ \emph {et~al.}(2016)\citenamefont {Endres}, \citenamefont {Bernien}, \citenamefont {Keesling}, \citenamefont {Levine}, \citenamefont {Anschuetz}, \citenamefont {Krajenbrink}, \citenamefont {Senko}, \citenamefont {Vuletic}, \citenamefont {Greiner},\ and\ \citenamefont {Lukin}}]{Endres2016}%
  \BibitemOpen
  \bibfield  {author} {\bibinfo {author} {\bibfnamefont {M.}~\bibnamefont {Endres}}, \bibinfo {author} {\bibfnamefont {H.}~\bibnamefont {Bernien}}, \bibinfo {author} {\bibfnamefont {A.}~\bibnamefont {Keesling}}, \bibinfo {author} {\bibfnamefont {H.}~\bibnamefont {Levine}}, \bibinfo {author} {\bibfnamefont {E.~R.}\ \bibnamefont {Anschuetz}}, \bibinfo {author} {\bibfnamefont {A.}~\bibnamefont {Krajenbrink}}, \bibinfo {author} {\bibfnamefont {C.}~\bibnamefont {Senko}}, \bibinfo {author} {\bibfnamefont {V.}~\bibnamefont {Vuletic}}, \bibinfo {author} {\bibfnamefont {M.}~\bibnamefont {Greiner}},\ and\ \bibinfo {author} {\bibfnamefont {M.~D.}\ \bibnamefont {Lukin}},\ }\bibfield  {title} {\bibinfo {title} {{Atom-by-atom assembly of defect-free one-dimensional cold atom arrays}},\ }\href {https://doi.org/10.1126/science.aah3752} {\bibfield  {journal} {\bibinfo  {journal} {Science}\ }\textbf {\bibinfo {volume} {354}},\ \bibinfo {pages} {1024} (\bibinfo {year} {2016})}\BibitemShut {NoStop}%
\bibitem [{\citenamefont {Barredo}\ \emph {et~al.}(2018)\citenamefont {Barredo}, \citenamefont {Lienhard}, \citenamefont {de~L{\ifmmode\acute{e}\else\'{e}\fi}s{\ifmmode\acute{e}\else\'{e}\fi}leuc}, \citenamefont {Lahaye},\ and\ \citenamefont {Browaeys}}]{Barredo2018}%
  \BibitemOpen
  \bibfield  {author} {\bibinfo {author} {\bibfnamefont {D.}~\bibnamefont {Barredo}}, \bibinfo {author} {\bibfnamefont {V.}~\bibnamefont {Lienhard}}, \bibinfo {author} {\bibfnamefont {S.}~\bibnamefont {de~L{\ifmmode\acute{e}\else\'{e}\fi}s{\ifmmode\acute{e}\else\'{e}\fi}leuc}}, \bibinfo {author} {\bibfnamefont {T.}~\bibnamefont {Lahaye}},\ and\ \bibinfo {author} {\bibfnamefont {A.}~\bibnamefont {Browaeys}},\ }\bibfield  {title} {\bibinfo {title} {{Synthetic three-dimensional atomic structures assembled atom by atom}},\ }\href {https://doi.org/10.1038/s41586-018-0450-2} {\bibfield  {journal} {\bibinfo  {journal} {Nature}\ }\textbf {\bibinfo {volume} {561}},\ \bibinfo {pages} {79} (\bibinfo {year} {2018})}\BibitemShut {NoStop}%
\bibitem [{\citenamefont {Ebadi}\ \emph {et~al.}(2021)\citenamefont {Ebadi}, \citenamefont {Wang}, \citenamefont {Levine}, \citenamefont {Keesling}, \citenamefont {Semeghini}, \citenamefont {Omran}, \citenamefont {Bluvstein}, \citenamefont {Samajdar}, \citenamefont {Pichler}, \citenamefont {Ho}, \citenamefont {Choi}, \citenamefont {Sachdev}, \citenamefont {Greiner}, \citenamefont {Vuleti{\ifmmode\acute{c}\else\'{c}\fi}},\ and\ \citenamefont {Lukin}}]{Ebadi2021}%
  \BibitemOpen
  \bibfield  {author} {\bibinfo {author} {\bibfnamefont {S.}~\bibnamefont {Ebadi}}, \bibinfo {author} {\bibfnamefont {T.~T.}\ \bibnamefont {Wang}}, \bibinfo {author} {\bibfnamefont {H.}~\bibnamefont {Levine}}, \bibinfo {author} {\bibfnamefont {A.}~\bibnamefont {Keesling}}, \bibinfo {author} {\bibfnamefont {G.}~\bibnamefont {Semeghini}}, \bibinfo {author} {\bibfnamefont {A.}~\bibnamefont {Omran}}, \bibinfo {author} {\bibfnamefont {D.}~\bibnamefont {Bluvstein}}, \bibinfo {author} {\bibfnamefont {R.}~\bibnamefont {Samajdar}}, \bibinfo {author} {\bibfnamefont {H.}~\bibnamefont {Pichler}}, \bibinfo {author} {\bibfnamefont {W.~W.}\ \bibnamefont {Ho}}, \bibinfo {author} {\bibfnamefont {S.}~\bibnamefont {Choi}}, \bibinfo {author} {\bibfnamefont {S.}~\bibnamefont {Sachdev}}, \bibinfo {author} {\bibfnamefont {M.}~\bibnamefont {Greiner}}, \bibinfo {author} {\bibfnamefont {V.}~\bibnamefont {Vuleti{\ifmmode\acute{c}\else\'{c}\fi}}},\ and\ \bibinfo {author} {\bibfnamefont {M.~D.}\ \bibnamefont {Lukin}},\ }\bibfield
  {title} {\bibinfo {title} {{Quantum phases of matter on a 256-atom programmable quantum simulator}},\ }\href {https://doi.org/10.1038/s41586-021-03582-4} {\bibfield  {journal} {\bibinfo  {journal} {Nature}\ }\textbf {\bibinfo {volume} {595}},\ \bibinfo {pages} {227} (\bibinfo {year} {2021})}\BibitemShut {NoStop}%
\bibitem [{\citenamefont {Vetsch}\ \emph {et~al.}(2012)\citenamefont {Vetsch}, \citenamefont {Dawkins}, \citenamefont {Mitsch}, \citenamefont {Reitz}, \citenamefont {Schneeweiss},\ and\ \citenamefont {Rauschenbeutel}}]{Vetsch2012}%
  \BibitemOpen
  \bibfield  {author} {\bibinfo {author} {\bibfnamefont {E.}~\bibnamefont {Vetsch}}, \bibinfo {author} {\bibfnamefont {S.~T.}\ \bibnamefont {Dawkins}}, \bibinfo {author} {\bibfnamefont {R.}~\bibnamefont {Mitsch}}, \bibinfo {author} {\bibfnamefont {D.}~\bibnamefont {Reitz}}, \bibinfo {author} {\bibfnamefont {P.}~\bibnamefont {Schneeweiss}},\ and\ \bibinfo {author} {\bibfnamefont {A.}~\bibnamefont {Rauschenbeutel}},\ }\bibfield  {title} {\bibinfo {title} {{Nanofiber-Based Optical Trapping of Cold Neutral Atoms}},\ }\href {https://doi.org/10.1109/JSTQE.2012.2196025} {\bibfield  {journal} {\bibinfo  {journal} {IEEE J. Sel. Top. Quantum Electron.}\ }\textbf {\bibinfo {volume} {18}},\ \bibinfo {pages} {1763} (\bibinfo {year} {2012})}\BibitemShut {NoStop}%
\bibitem [{\citenamefont {Nayak}\ \emph {et~al.}(2018)\citenamefont {Nayak}, \citenamefont {Sadgrove}, \citenamefont {Yalla}, \citenamefont {Le~Kien},\ and\ \citenamefont {Hakuta}}]{Nayak2018}%
  \BibitemOpen
  \bibfield  {author} {\bibinfo {author} {\bibfnamefont {K.~P.}\ \bibnamefont {Nayak}}, \bibinfo {author} {\bibfnamefont {M.}~\bibnamefont {Sadgrove}}, \bibinfo {author} {\bibfnamefont {R.}~\bibnamefont {Yalla}}, \bibinfo {author} {\bibfnamefont {F.}~\bibnamefont {Le~Kien}},\ and\ \bibinfo {author} {\bibfnamefont {K.}~\bibnamefont {Hakuta}},\ }\bibfield  {title} {\bibinfo {title} {{Nanofiber quantum photonics}},\ }\href {https://doi.org/10.1088/2040-8986/aac35e} {\bibfield  {journal} {\bibinfo  {journal} {J. Opt.}\ }\textbf {\bibinfo {volume} {20}},\ \bibinfo {pages} {073001} (\bibinfo {year} {2018})}\BibitemShut {NoStop}%
\bibitem [{\citenamefont {Goban}\ \emph {et~al.}(2015)\citenamefont {Goban}, \citenamefont {Hung}, \citenamefont {Hood}, \citenamefont {Yu}, \citenamefont {Muniz}, \citenamefont {Painter},\ and\ \citenamefont {Kimble}}]{Goban2015}%
  \BibitemOpen
  \bibfield  {author} {\bibinfo {author} {\bibfnamefont {A.}~\bibnamefont {Goban}}, \bibinfo {author} {\bibfnamefont {C.-L.}\ \bibnamefont {Hung}}, \bibinfo {author} {\bibfnamefont {J.~D.}\ \bibnamefont {Hood}}, \bibinfo {author} {\bibfnamefont {S.-P.}\ \bibnamefont {Yu}}, \bibinfo {author} {\bibfnamefont {J.~A.}\ \bibnamefont {Muniz}}, \bibinfo {author} {\bibfnamefont {O.}~\bibnamefont {Painter}},\ and\ \bibinfo {author} {\bibfnamefont {H.~J.}\ \bibnamefont {Kimble}},\ }\bibfield  {title} {\bibinfo {title} {{Superradiance for Atoms Trapped along a Photonic Crystal Waveguide}},\ }\href {https://doi.org/10.1103/PhysRevLett.115.063601} {\bibfield  {journal} {\bibinfo  {journal} {Phys. Rev. Lett.}\ }\textbf {\bibinfo {volume} {115}},\ \bibinfo {pages} {063601} (\bibinfo {year} {2015})}\BibitemShut {NoStop}%
\bibitem [{\citenamefont {Corzo}\ \emph {et~al.}(2016)\citenamefont {Corzo}, \citenamefont {Gouraud}, \citenamefont {Chandra}, \citenamefont {Goban}, \citenamefont {Sheremet}, \citenamefont {Kupriyanov},\ and\ \citenamefont {Laurat}}]{Corzo2016}%
  \BibitemOpen
  \bibfield  {author} {\bibinfo {author} {\bibfnamefont {N.~V.}\ \bibnamefont {Corzo}}, \bibinfo {author} {\bibfnamefont {B.}~\bibnamefont {Gouraud}}, \bibinfo {author} {\bibfnamefont {A.}~\bibnamefont {Chandra}}, \bibinfo {author} {\bibfnamefont {A.}~\bibnamefont {Goban}}, \bibinfo {author} {\bibfnamefont {A.~S.}\ \bibnamefont {Sheremet}}, \bibinfo {author} {\bibfnamefont {D.~V.}\ \bibnamefont {Kupriyanov}},\ and\ \bibinfo {author} {\bibfnamefont {J.}~\bibnamefont {Laurat}},\ }\bibfield  {title} {\bibinfo {title} {{Large Bragg Reflection from One-Dimensional Chains of Trapped Atoms Near a Nanoscale Waveguide}},\ }\href {https://doi.org/10.1103/PhysRevLett.117.133603} {\bibfield  {journal} {\bibinfo  {journal} {Phys. Rev. Lett.}\ }\textbf {\bibinfo {volume} {117}},\ \bibinfo {pages} {133603} (\bibinfo {year} {2016})}\BibitemShut {NoStop}%
\bibitem [{\citenamefont {Manzoni}\ \emph {et~al.}(2018)\citenamefont {Manzoni}, \citenamefont {Moreno-Cardoner}, \citenamefont {Asenjo-Garcia}, \citenamefont {Porto}, \citenamefont {Gorshkov},\ and\ \citenamefont {Chang}}]{manzoni_optimization_2018}%
  \BibitemOpen
  \bibfield  {author} {\bibinfo {author} {\bibfnamefont {M.~T.}\ \bibnamefont {Manzoni}}, \bibinfo {author} {\bibfnamefont {M.}~\bibnamefont {Moreno-Cardoner}}, \bibinfo {author} {\bibfnamefont {A.}~\bibnamefont {Asenjo-Garcia}}, \bibinfo {author} {\bibfnamefont {J.~V.}\ \bibnamefont {Porto}}, \bibinfo {author} {\bibfnamefont {A.~V.}\ \bibnamefont {Gorshkov}},\ and\ \bibinfo {author} {\bibfnamefont {D.~E.}\ \bibnamefont {Chang}},\ }\bibfield  {title} {\bibinfo {title} {{Optimization of photon storage fidelity in ordered atomic arrays}},\ }\href {https://doi.org/10.1088/1367-2630/aadb74} {\bibfield  {journal} {\bibinfo  {journal} {New J. Phys.}\ }\textbf {\bibinfo {volume} {20}},\ \bibinfo {pages} {083048} (\bibinfo {year} {2018})}\BibitemShut {NoStop}%
\bibitem [{\citenamefont {Reitz}\ \emph {et~al.}(2022)\citenamefont {Reitz}, \citenamefont {Sommer},\ and\ \citenamefont {Genes}}]{reitz_cooperative_2021}%
  \BibitemOpen
  \bibfield  {author} {\bibinfo {author} {\bibfnamefont {M.}~\bibnamefont {Reitz}}, \bibinfo {author} {\bibfnamefont {C.}~\bibnamefont {Sommer}},\ and\ \bibinfo {author} {\bibfnamefont {C.}~\bibnamefont {Genes}},\ }\bibfield  {title} {\bibinfo {title} {{Cooperative Quantum Phenomena in Light-Matter Platforms}},\ }\href {https://doi.org/10.1103/PRXQuantum.3.010201} {\bibfield  {journal} {\bibinfo  {journal} {PRX Quantum}\ }\textbf {\bibinfo {volume} {3}},\ \bibinfo {pages} {010201} (\bibinfo {year} {2022})}\BibitemShut {NoStop}%
\bibitem [{\citenamefont {Prasad}\ \emph {et~al.}(2020)\citenamefont {Prasad}, \citenamefont {Hinney}, \citenamefont {Mahmoodian}, \citenamefont {Hammerer}, \citenamefont {Rind}, \citenamefont {Schneeweiss}, \citenamefont {S{\o}rensen}, \citenamefont {Volz},\ and\ \citenamefont {Rauschenbeutel}}]{Prasad2020}%
  \BibitemOpen
  \bibfield  {author} {\bibinfo {author} {\bibfnamefont {A.~S.}\ \bibnamefont {Prasad}}, \bibinfo {author} {\bibfnamefont {J.}~\bibnamefont {Hinney}}, \bibinfo {author} {\bibfnamefont {S.}~\bibnamefont {Mahmoodian}}, \bibinfo {author} {\bibfnamefont {K.}~\bibnamefont {Hammerer}}, \bibinfo {author} {\bibfnamefont {S.}~\bibnamefont {Rind}}, \bibinfo {author} {\bibfnamefont {P.}~\bibnamefont {Schneeweiss}}, \bibinfo {author} {\bibfnamefont {A.~S.}\ \bibnamefont {S{\o}rensen}}, \bibinfo {author} {\bibfnamefont {J.}~\bibnamefont {Volz}},\ and\ \bibinfo {author} {\bibfnamefont {A.}~\bibnamefont {Rauschenbeutel}},\ }\bibfield  {title} {\bibinfo {title} {{Correlating photons using the collective nonlinear response of atoms weakly coupled to an optical mode}},\ }\href {https://doi.org/10.1038/s41566-020-0692-z} {\bibfield  {journal} {\bibinfo  {journal} {Nat. Photonics}\ }\textbf {\bibinfo {volume} {14}},\ \bibinfo {pages} {719} (\bibinfo {year} {2020})}\BibitemShut {NoStop}%
\bibitem [{\citenamefont {Liedl}\ \emph {et~al.}(2023)\citenamefont {Liedl}, \citenamefont {Pucher}, \citenamefont {Tebbenjohanns}, \citenamefont {Schneeweiss},\ and\ \citenamefont {Rauschenbeutel}}]{liedl_collective_2023}%
  \BibitemOpen
  \bibfield  {author} {\bibinfo {author} {\bibfnamefont {C.}~\bibnamefont {Liedl}}, \bibinfo {author} {\bibfnamefont {S.}~\bibnamefont {Pucher}}, \bibinfo {author} {\bibfnamefont {F.}~\bibnamefont {Tebbenjohanns}}, \bibinfo {author} {\bibfnamefont {P.}~\bibnamefont {Schneeweiss}},\ and\ \bibinfo {author} {\bibfnamefont {A.}~\bibnamefont {Rauschenbeutel}},\ }\bibfield  {title} {\bibinfo {title} {{Collective Radiation of a Cascaded Quantum System: From Timed Dicke States to Inverted Ensembles}},\ }\href {https://doi.org/10.1103/PhysRevLett.130.163602} {\bibfield  {journal} {\bibinfo  {journal} {Phys. Rev. Lett.}\ }\textbf {\bibinfo {volume} {130}},\ \bibinfo {pages} {163602} (\bibinfo {year} {2023})}\BibitemShut {NoStop}%
\bibitem [{\citenamefont {Corzo}\ \emph {et~al.}(2019)\citenamefont {Corzo}, \citenamefont {Raskop}, \citenamefont {Chandra}, \citenamefont {Sheremet}, \citenamefont {Gouraud},\ and\ \citenamefont {Laurat}}]{Corzo2019}%
  \BibitemOpen
  \bibfield  {author} {\bibinfo {author} {\bibfnamefont {N.~V.}\ \bibnamefont {Corzo}}, \bibinfo {author} {\bibfnamefont {J.}~\bibnamefont {Raskop}}, \bibinfo {author} {\bibfnamefont {A.}~\bibnamefont {Chandra}}, \bibinfo {author} {\bibfnamefont {A.~S.}\ \bibnamefont {Sheremet}}, \bibinfo {author} {\bibfnamefont {B.}~\bibnamefont {Gouraud}},\ and\ \bibinfo {author} {\bibfnamefont {J.}~\bibnamefont {Laurat}},\ }\bibfield  {title} {\bibinfo {title} {{Waveguide-coupled single collective excitation of atomic arrays}},\ }\href {https://doi.org/10.1038/s41586-019-0902-3} {\bibfield  {journal} {\bibinfo  {journal} {Nature}\ }\textbf {\bibinfo {volume} {566}},\ \bibinfo {pages} {359} (\bibinfo {year} {2019})}\BibitemShut {NoStop}%
\bibitem [{\citenamefont {Ke}\ \emph {et~al.}(2019)\citenamefont {Ke}, \citenamefont {Poshakinskiy}, \citenamefont {Lee}, \citenamefont {Kivshar},\ and\ \citenamefont {Poddubny}}]{Ke2019Dec}%
  \BibitemOpen
  \bibfield  {author} {\bibinfo {author} {\bibfnamefont {Y.}~\bibnamefont {Ke}}, \bibinfo {author} {\bibfnamefont {A.~V.}\ \bibnamefont {Poshakinskiy}}, \bibinfo {author} {\bibfnamefont {C.}~\bibnamefont {Lee}}, \bibinfo {author} {\bibfnamefont {Y.~S.}\ \bibnamefont {Kivshar}},\ and\ \bibinfo {author} {\bibfnamefont {A.~N.}\ \bibnamefont {Poddubny}},\ }\bibfield  {title} {\bibinfo {title} {{Inelastic Scattering of Photon Pairs in Qubit Arrays with Subradiant States}},\ }\href {https://doi.org/10.1103/PhysRevLett.123.253601} {\bibfield  {journal} {\bibinfo  {journal} {Phys. Rev. Lett.}\ }\textbf {\bibinfo {volume} {123}},\ \bibinfo {pages} {253601} (\bibinfo {year} {2019})}\BibitemShut {NoStop}%
\bibitem [{\citenamefont {Holzinger}\ \emph {et~al.}(2020{\natexlab{a}})\citenamefont {Holzinger}, \citenamefont {Ostermann},\ and\ \citenamefont {Ritsch}}]{holzinger_subradiace_2020}%
  \BibitemOpen
  \bibfield  {author} {\bibinfo {author} {\bibfnamefont {R.}~\bibnamefont {Holzinger}}, \bibinfo {author} {\bibfnamefont {L.}~\bibnamefont {Ostermann}},\ and\ \bibinfo {author} {\bibfnamefont {H.}~\bibnamefont {Ritsch}},\ }\bibfield  {title} {\bibinfo {title} {{Subradiance in multiply excited states of dipole-coupled V-type atoms}},\ }\href {https://doi.org/10.1209/0295-5075/128/44001} {\bibfield  {journal} {\bibinfo  {journal} {Europhys. Lett.}\ }\textbf {\bibinfo {volume} {128}},\ \bibinfo {pages} {44001} (\bibinfo {year} {2020}{\natexlab{a}})}\BibitemShut {NoStop}%
\bibitem [{\citenamefont {Fayard}\ \emph {et~al.}(2023)\citenamefont {Fayard}, \citenamefont {Ferrier-Barbut}, \citenamefont {Browaeys},\ and\ \citenamefont {Greffet}}]{Fayard2023}%
  \BibitemOpen
  \bibfield  {author} {\bibinfo {author} {\bibfnamefont {N.}~\bibnamefont {Fayard}}, \bibinfo {author} {\bibfnamefont {I.}~\bibnamefont {Ferrier-Barbut}}, \bibinfo {author} {\bibfnamefont {A.}~\bibnamefont {Browaeys}},\ and\ \bibinfo {author} {\bibfnamefont {J.-J.}\ \bibnamefont {Greffet}},\ }\bibfield  {title} {\bibinfo {title} {{Optical control of collective states in one-dimensional ordered atomic chains beyond the linear regime}},\ }\href {https://doi.org/10.1103/PhysRevA.108.023116} {\bibfield  {journal} {\bibinfo  {journal} {Phys. Rev. A}\ }\textbf {\bibinfo {volume} {108}},\ \bibinfo {pages} {023116} (\bibinfo {year} {2023})}\BibitemShut {NoStop}%
\bibitem [{\citenamefont {Zhang}(2023)}]{Zhang2023Nov}%
  \BibitemOpen
  \bibfield  {author} {\bibinfo {author} {\bibfnamefont {Y.-X.}\ \bibnamefont {Zhang}},\ }\bibfield  {title} {\bibinfo {title} {{Zeno Regime of Collective Emission: Non-Markovianity beyond Retardation}},\ }\href {https://doi.org/10.1103/PhysRevLett.131.193603} {\bibfield  {journal} {\bibinfo  {journal} {Phys. Rev. Lett.}\ }\textbf {\bibinfo {volume} {131}},\ \bibinfo {pages} {193603} (\bibinfo {year} {2023})}\BibitemShut {NoStop}%
\bibitem [{\citenamefont {Cardenas-Lopez}\ \emph {et~al.}(2023)\citenamefont {Cardenas-Lopez}, \citenamefont {Masson}, \citenamefont {Zager},\ and\ \citenamefont {Asenjo-Garcia}}]{Cardenas-Lopez2023Jul}%
  \BibitemOpen
  \bibfield  {author} {\bibinfo {author} {\bibfnamefont {S.}~\bibnamefont {Cardenas-Lopez}}, \bibinfo {author} {\bibfnamefont {S.~J.}\ \bibnamefont {Masson}}, \bibinfo {author} {\bibfnamefont {Z.}~\bibnamefont {Zager}},\ and\ \bibinfo {author} {\bibfnamefont {A.}~\bibnamefont {Asenjo-Garcia}},\ }\bibfield  {title} {\bibinfo {title} {{Many-Body Superradiance and Dynamical Mirror Symmetry Breaking in Waveguide QED}},\ }\href {https://doi.org/10.1103/PhysRevLett.131.033605} {\bibfield  {journal} {\bibinfo  {journal} {Phys. Rev. Lett.}\ }\textbf {\bibinfo {volume} {131}},\ \bibinfo {pages} {033605} (\bibinfo {year} {2023})}\BibitemShut {NoStop}%
\bibitem [{\citenamefont {Mok}\ \emph {et~al.}(2023)\citenamefont {Mok}, \citenamefont {Asenjo-Garcia}, \citenamefont {Sum},\ and\ \citenamefont {Kwek}}]{Mok2023May}%
  \BibitemOpen
  \bibfield  {author} {\bibinfo {author} {\bibfnamefont {W.-K.}\ \bibnamefont {Mok}}, \bibinfo {author} {\bibfnamefont {A.}~\bibnamefont {Asenjo-Garcia}}, \bibinfo {author} {\bibfnamefont {T.~C.}\ \bibnamefont {Sum}},\ and\ \bibinfo {author} {\bibfnamefont {L.-C.}\ \bibnamefont {Kwek}},\ }\bibfield  {title} {\bibinfo {title} {{Dicke Superradiance Requires Interactions beyond Nearest Neighbors}},\ }\href {https://doi.org/10.1103/PhysRevLett.130.213605} {\bibfield  {journal} {\bibinfo  {journal} {Phys. Rev. Lett.}\ }\textbf {\bibinfo {volume} {130}},\ \bibinfo {pages} {213605} (\bibinfo {year} {2023})}\BibitemShut {NoStop}%
\bibitem [{\citenamefont {Lohof}\ \emph {et~al.}(2023)\citenamefont {Lohof}, \citenamefont {Schumayer}, \citenamefont {Hutchinson},\ and\ \citenamefont {Gies}}]{Lohof2023Aug}%
  \BibitemOpen
  \bibfield  {author} {\bibinfo {author} {\bibfnamefont {F.}~\bibnamefont {Lohof}}, \bibinfo {author} {\bibfnamefont {D.}~\bibnamefont {Schumayer}}, \bibinfo {author} {\bibfnamefont {D.~A.~W.}\ \bibnamefont {Hutchinson}},\ and\ \bibinfo {author} {\bibfnamefont {C.}~\bibnamefont {Gies}},\ }\bibfield  {title} {\bibinfo {title} {{Signatures of Superradiance as a Witness to Multipartite Entanglement}},\ }\href {https://doi.org/10.1103/PhysRevLett.131.063601} {\bibfield  {journal} {\bibinfo  {journal} {Phys. Rev. Lett.}\ }\textbf {\bibinfo {volume} {131}},\ \bibinfo {pages} {063601} (\bibinfo {year} {2023})}\BibitemShut {NoStop}%
\bibitem [{\citenamefont {Dicke}(1954)}]{Dicke1954}%
  \BibitemOpen
  \bibfield  {author} {\bibinfo {author} {\bibfnamefont {R.~H.}\ \bibnamefont {Dicke}},\ }\bibfield  {title} {\bibinfo {title} {{Coherence in Spontaneous Radiation Processes}},\ }\href {https://doi.org/10.1103/PhysRev.93.99} {\bibfield  {journal} {\bibinfo  {journal} {Phys. Rev.}\ }\textbf {\bibinfo {volume} {93}},\ \bibinfo {pages} {99} (\bibinfo {year} {1954})}\BibitemShut {NoStop}%
\bibitem [{\citenamefont {Sheremet}\ \emph {et~al.}(2012)\citenamefont {Sheremet}, \citenamefont {Manukhova}, \citenamefont {Larionov},\ and\ \citenamefont {Kupriyanov}}]{Sheremet2012}%
  \BibitemOpen
  \bibfield  {author} {\bibinfo {author} {\bibfnamefont {A.~S.}\ \bibnamefont {Sheremet}}, \bibinfo {author} {\bibfnamefont {A.~D.}\ \bibnamefont {Manukhova}}, \bibinfo {author} {\bibfnamefont {N.~V.}\ \bibnamefont {Larionov}},\ and\ \bibinfo {author} {\bibfnamefont {D.~V.}\ \bibnamefont {Kupriyanov}},\ }\bibfield  {title} {\bibinfo {title} {{Cooperative light scattering on an atomic system with degenerate structure of the ground state}},\ }\href {https://doi.org/10.1103/PhysRevA.86.043414} {\bibfield  {journal} {\bibinfo  {journal} {Phys. Rev. A}\ }\textbf {\bibinfo {volume} {86}},\ \bibinfo {pages} {043414} (\bibinfo {year} {2012})}\BibitemShut {NoStop}%
\bibitem [{\citenamefont {Kornovan}\ \emph {et~al.}(2016)\citenamefont {Kornovan}, \citenamefont {Sheremet},\ and\ \citenamefont {Petrov}}]{Kornovan2016}%
  \BibitemOpen
  \bibfield  {author} {\bibinfo {author} {\bibfnamefont {D.~F.}\ \bibnamefont {Kornovan}}, \bibinfo {author} {\bibfnamefont {A.~S.}\ \bibnamefont {Sheremet}},\ and\ \bibinfo {author} {\bibfnamefont {M.~I.}\ \bibnamefont {Petrov}},\ }\bibfield  {title} {\bibinfo {title} {{Collective polaritonic modes in an array of two-level quantum emitters coupled to an optical nanofiber}},\ }\href {https://doi.org/10.1103/PhysRevB.94.245416} {\bibfield  {journal} {\bibinfo  {journal} {Phys. Rev. B}\ }\textbf {\bibinfo {volume} {94}},\ \bibinfo {pages} {245416} (\bibinfo {year} {2016})}\BibitemShut {NoStop}%
\bibitem [{\citenamefont {Jen}\ \emph {et~al.}(2018)\citenamefont {Jen}, \citenamefont {Chang},\ and\ \citenamefont {Chen}}]{jen_cooperative_2018}%
  \BibitemOpen
  \bibfield  {author} {\bibinfo {author} {\bibfnamefont {H.~H.}\ \bibnamefont {Jen}}, \bibinfo {author} {\bibfnamefont {M.-S.}\ \bibnamefont {Chang}},\ and\ \bibinfo {author} {\bibfnamefont {Y.-C.}\ \bibnamefont {Chen}},\ }\bibfield  {title} {\bibinfo {title} {{Cooperative light scattering from helical-phase-imprinted atomic rings}},\ }\href {https://doi.org/10.1038/s41598-018-27888-y} {\bibfield  {journal} {\bibinfo  {journal} {Sci. Rep.}\ }\textbf {\bibinfo {volume} {8}},\ \bibinfo {pages} {9570} (\bibinfo {year} {2018})}\BibitemShut {NoStop}%
\bibitem [{\citenamefont {Plankensteiner}\ \emph {et~al.}(2019)\citenamefont {Plankensteiner}, \citenamefont {Sommer}, \citenamefont {Reitz}, \citenamefont {Ritsch},\ and\ \citenamefont {Genes}}]{Plankensteiner2019}%
  \BibitemOpen
  \bibfield  {author} {\bibinfo {author} {\bibfnamefont {D.}~\bibnamefont {Plankensteiner}}, \bibinfo {author} {\bibfnamefont {C.}~\bibnamefont {Sommer}}, \bibinfo {author} {\bibfnamefont {M.}~\bibnamefont {Reitz}}, \bibinfo {author} {\bibfnamefont {H.}~\bibnamefont {Ritsch}},\ and\ \bibinfo {author} {\bibfnamefont {C.}~\bibnamefont {Genes}},\ }\bibfield  {title} {\bibinfo {title} {{Enhanced collective Purcell effect of coupled quantum emitter systems}},\ }\href {https://doi.org/10.1103/PhysRevA.99.043843} {\bibfield  {journal} {\bibinfo  {journal} {Phys. Rev. A}\ }\textbf {\bibinfo {volume} {99}},\ \bibinfo {pages} {043843} (\bibinfo {year} {2019})}\BibitemShut {NoStop}%
\bibitem [{\citenamefont {Zhang}\ and\ \citenamefont {M{\o}lmer}(2019)}]{zhang_theory_2019}%
  \BibitemOpen
  \bibfield  {author} {\bibinfo {author} {\bibfnamefont {Y.-X.}\ \bibnamefont {Zhang}}\ and\ \bibinfo {author} {\bibfnamefont {K.}~\bibnamefont {M{\o}lmer}},\ }\bibfield  {title} {\bibinfo {title} {{Theory of Subradiant States of a One-Dimensional Two-Level Atom Chain}},\ }\href {https://doi.org/10.1103/PhysRevLett.122.203605} {\bibfield  {journal} {\bibinfo  {journal} {Phys. Rev. Lett.}\ }\textbf {\bibinfo {volume} {122}},\ \bibinfo {pages} {203605} (\bibinfo {year} {2019})}\BibitemShut {NoStop}%
\bibitem [{\citenamefont {Fofanov}\ \emph {et~al.}(2021)\citenamefont {Fofanov}, \citenamefont {Sokolov}, \citenamefont {Kaiser},\ and\ \citenamefont {Guerin}}]{Fofanov2021}%
  \BibitemOpen
  \bibfield  {author} {\bibinfo {author} {\bibfnamefont {Y.~A.}\ \bibnamefont {Fofanov}}, \bibinfo {author} {\bibfnamefont {I.~M.}\ \bibnamefont {Sokolov}}, \bibinfo {author} {\bibfnamefont {R.}~\bibnamefont {Kaiser}},\ and\ \bibinfo {author} {\bibfnamefont {W.}~\bibnamefont {Guerin}},\ }\bibfield  {title} {\bibinfo {title} {{Subradiance in dilute atomic ensembles: Role of pairs and multiple scattering}},\ }\href {https://doi.org/10.1103/PhysRevA.104.023705} {\bibfield  {journal} {\bibinfo  {journal} {Phys. Rev. A}\ }\textbf {\bibinfo {volume} {104}},\ \bibinfo {pages} {023705} (\bibinfo {year} {2021})}\BibitemShut {NoStop}%
\bibitem [{\citenamefont {Pavolini}\ \emph {et~al.}(1985)\citenamefont {Pavolini}, \citenamefont {Crubellier}, \citenamefont {Pillet}, \citenamefont {Cabaret},\ and\ \citenamefont {Liberman}}]{pavolini_experimental_1985}%
  \BibitemOpen
  \bibfield  {author} {\bibinfo {author} {\bibfnamefont {D.}~\bibnamefont {Pavolini}}, \bibinfo {author} {\bibfnamefont {A.}~\bibnamefont {Crubellier}}, \bibinfo {author} {\bibfnamefont {P.}~\bibnamefont {Pillet}}, \bibinfo {author} {\bibfnamefont {L.}~\bibnamefont {Cabaret}},\ and\ \bibinfo {author} {\bibfnamefont {S.}~\bibnamefont {Liberman}},\ }\bibfield  {title} {\bibinfo {title} {{Experimental Evidence for Subradiance}},\ }\href {https://doi.org/10.1103/PhysRevLett.54.1917} {\bibfield  {journal} {\bibinfo  {journal} {Phys. Rev. Lett.}\ }\textbf {\bibinfo {volume} {54}},\ \bibinfo {pages} {1917} (\bibinfo {year} {1985})}\BibitemShut {NoStop}%
\bibitem [{\citenamefont {DeVoe}\ and\ \citenamefont {Brewer}(1996)}]{devoe_observation_1996}%
  \BibitemOpen
  \bibfield  {author} {\bibinfo {author} {\bibfnamefont {R.~G.}\ \bibnamefont {DeVoe}}\ and\ \bibinfo {author} {\bibfnamefont {R.~G.}\ \bibnamefont {Brewer}},\ }\bibfield  {title} {\bibinfo {title} {{Observation of Superradiant and Subradiant Spontaneous Emission of Two Trapped Ions}},\ }\href {https://doi.org/10.1103/PhysRevLett.76.2049} {\bibfield  {journal} {\bibinfo  {journal} {Phys. Rev. Lett.}\ }\textbf {\bibinfo {volume} {76}},\ \bibinfo {pages} {2049} (\bibinfo {year} {1996})}\BibitemShut {NoStop}%
\bibitem [{\citenamefont {Guerin}\ \emph {et~al.}(2016)\citenamefont {Guerin}, \citenamefont {Ara{\ifmmode\acute{u}\else\'{u}\fi}jo},\ and\ \citenamefont {Kaiser}}]{guerin_subradiance_2016}%
  \BibitemOpen
  \bibfield  {author} {\bibinfo {author} {\bibfnamefont {W.}~\bibnamefont {Guerin}}, \bibinfo {author} {\bibfnamefont {M.~O.}\ \bibnamefont {Ara{\ifmmode\acute{u}\else\'{u}\fi}jo}},\ and\ \bibinfo {author} {\bibfnamefont {R.}~\bibnamefont {Kaiser}},\ }\bibfield  {title} {\bibinfo {title} {{Subradiance in a Large Cloud of Cold Atoms}},\ }\href {https://doi.org/10.1103/PhysRevLett.116.083601} {\bibfield  {journal} {\bibinfo  {journal} {Phys. Rev. Lett.}\ }\textbf {\bibinfo {volume} {116}},\ \bibinfo {pages} {083601} (\bibinfo {year} {2016})}\BibitemShut {NoStop}%
\bibitem [{\citenamefont {Rui}\ \emph {et~al.}(2020)\citenamefont {Rui}, \citenamefont {Wei}, \citenamefont {Rubio-Abadal}, \citenamefont {Hollerith}, \citenamefont {Zeiher}, \citenamefont {Stamper-Kurn}, \citenamefont {Gross},\ and\ \citenamefont {Bloch}}]{rui_subradiant_2020}%
  \BibitemOpen
  \bibfield  {author} {\bibinfo {author} {\bibfnamefont {J.}~\bibnamefont {Rui}}, \bibinfo {author} {\bibfnamefont {D.}~\bibnamefont {Wei}}, \bibinfo {author} {\bibfnamefont {A.}~\bibnamefont {Rubio-Abadal}}, \bibinfo {author} {\bibfnamefont {S.}~\bibnamefont {Hollerith}}, \bibinfo {author} {\bibfnamefont {J.}~\bibnamefont {Zeiher}}, \bibinfo {author} {\bibfnamefont {D.~M.}\ \bibnamefont {Stamper-Kurn}}, \bibinfo {author} {\bibfnamefont {C.}~\bibnamefont {Gross}},\ and\ \bibinfo {author} {\bibfnamefont {I.}~\bibnamefont {Bloch}},\ }\bibfield  {title} {\bibinfo {title} {{A subradiant optical mirror formed by a single structured atomic layer}},\ }\href {https://doi.org/10.1038/s41586-020-2463-x} {\bibfield  {journal} {\bibinfo  {journal} {Nature}\ }\textbf {\bibinfo {volume} {583}},\ \bibinfo {pages} {369} (\bibinfo {year} {2020})}\BibitemShut {NoStop}%
\bibitem [{\citenamefont {Ferioli}\ \emph {et~al.}(2021)\citenamefont {Ferioli}, \citenamefont {Glicenstein}, \citenamefont {Henriet}, \citenamefont {Ferrier-Barbut},\ and\ \citenamefont {Browaeys}}]{Ferioli2021}%
  \BibitemOpen
  \bibfield  {author} {\bibinfo {author} {\bibfnamefont {G.}~\bibnamefont {Ferioli}}, \bibinfo {author} {\bibfnamefont {A.}~\bibnamefont {Glicenstein}}, \bibinfo {author} {\bibfnamefont {L.}~\bibnamefont {Henriet}}, \bibinfo {author} {\bibfnamefont {I.}~\bibnamefont {Ferrier-Barbut}},\ and\ \bibinfo {author} {\bibfnamefont {A.}~\bibnamefont {Browaeys}},\ }\bibfield  {title} {\bibinfo {title} {{Storage and release of subradiant excitations in a dense atomic cloud }},\ }\href {https://doi.org/https://doi.org/10.1103/PhysRevX.11.021031} {\bibfield  {journal} {\bibinfo  {journal} {Phys. Rev. X}\ }\textbf {\bibinfo {volume} {11}},\ \bibinfo {pages} {021031} (\bibinfo {year} {2021})}\BibitemShut {NoStop}%
\bibitem [{\citenamefont {Asenjo-Garcia}\ \emph {et~al.}(2017{\natexlab{a}})\citenamefont {Asenjo-Garcia}, \citenamefont {Moreno-Cardoner}, \citenamefont {Albrecht}, \citenamefont {Kimble},\ and\ \citenamefont {Chang}}]{Asenjo-Garcia2017}%
  \BibitemOpen
  \bibfield  {author} {\bibinfo {author} {\bibfnamefont {A.}~\bibnamefont {Asenjo-Garcia}}, \bibinfo {author} {\bibfnamefont {M.}~\bibnamefont {Moreno-Cardoner}}, \bibinfo {author} {\bibfnamefont {A.}~\bibnamefont {Albrecht}}, \bibinfo {author} {\bibfnamefont {H.~J.}\ \bibnamefont {Kimble}},\ and\ \bibinfo {author} {\bibfnamefont {D.~E.}\ \bibnamefont {Chang}},\ }\bibfield  {title} {\bibinfo {title} {{Exponential Improvement in Photon Storage Fidelities Using Subradiance and ``Selective Radiance'' in Atomic Arrays}},\ }\href {https://doi.org/10.1103/PhysRevX.7.031024} {\bibfield  {journal} {\bibinfo  {journal} {Phys. Rev. X}\ }\textbf {\bibinfo {volume} {7}},\ \bibinfo {pages} {031024} (\bibinfo {year} {2017}{\natexlab{a}})}\BibitemShut {NoStop}%
\bibitem [{\citenamefont {Zhang}\ and\ \citenamefont {M{\o}lmer}(2020)}]{Zhang2020}%
  \BibitemOpen
  \bibfield  {author} {\bibinfo {author} {\bibfnamefont {Y.-X.}\ \bibnamefont {Zhang}}\ and\ \bibinfo {author} {\bibfnamefont {K.}~\bibnamefont {M{\o}lmer}},\ }\bibfield  {title} {\bibinfo {title} {{Subradiant Emission from Regular Atomic Arrays: Universal Scaling of Decay Rates from the Generalized Bloch Theorem}},\ }\href {https://doi.org/10.1103/PhysRevLett.125.253601} {\bibfield  {journal} {\bibinfo  {journal} {Phys. Rev. Lett.}\ }\textbf {\bibinfo {volume} {125}},\ \bibinfo {pages} {253601} (\bibinfo {year} {2020})}\BibitemShut {NoStop}%
\bibitem [{\citenamefont {Kornovan}\ \emph {et~al.}(2021)\citenamefont {Kornovan}, \citenamefont {Savelev}, \citenamefont {Kivshar},\ and\ \citenamefont {Petrov}}]{Kornovan2021}%
  \BibitemOpen
  \bibfield  {author} {\bibinfo {author} {\bibfnamefont {D.~F.}\ \bibnamefont {Kornovan}}, \bibinfo {author} {\bibfnamefont {R.~S.}\ \bibnamefont {Savelev}}, \bibinfo {author} {\bibfnamefont {Y.}~\bibnamefont {Kivshar}},\ and\ \bibinfo {author} {\bibfnamefont {M.~I.}\ \bibnamefont {Petrov}},\ }\bibfield  {title} {\bibinfo {title} {{High-$Q$ Localized States in Finite Arrays of Subwavelength Resonators}},\ }\href {https://doi.org/10.1021/acsphotonics.1c01262} {\bibfield  {journal} {\bibinfo  {journal} {ACS Photonics}\ }\textbf {\bibinfo {volume} {8}},\ \bibinfo {pages} {3627} (\bibinfo {year} {2021})}\BibitemShut {NoStop}%
\bibitem [{\citenamefont {Sutherland}\ and\ \citenamefont {Robicheaux}(2016)}]{sutherland_collective_2016}%
  \BibitemOpen
  \bibfield  {author} {\bibinfo {author} {\bibfnamefont {R.~T.}\ \bibnamefont {Sutherland}}\ and\ \bibinfo {author} {\bibfnamefont {F.}~\bibnamefont {Robicheaux}},\ }\bibfield  {title} {\bibinfo {title} {{Collective dipole-dipole interactions in an atomic array}},\ }\href {https://doi.org/10.1103/PhysRevA.94.013847} {\bibfield  {journal} {\bibinfo  {journal} {Phys. Rev. A}\ }\textbf {\bibinfo {volume} {94}},\ \bibinfo {pages} {013847} (\bibinfo {year} {2016})}\BibitemShut {NoStop}%
\bibitem [{\citenamefont {Asenjo-Garcia}\ \emph {et~al.}(2017{\natexlab{b}})\citenamefont {Asenjo-Garcia}, \citenamefont {Hood}, \citenamefont {Chang},\ and\ \citenamefont {Kimble}}]{Asenjo-Garcia2017Mar}%
  \BibitemOpen
  \bibfield  {author} {\bibinfo {author} {\bibfnamefont {A.}~\bibnamefont {Asenjo-Garcia}}, \bibinfo {author} {\bibfnamefont {J.~D.}\ \bibnamefont {Hood}}, \bibinfo {author} {\bibfnamefont {D.~E.}\ \bibnamefont {Chang}},\ and\ \bibinfo {author} {\bibfnamefont {H.~J.}\ \bibnamefont {Kimble}},\ }\bibfield  {title} {\bibinfo {title} {{Atom-light interactions in quasi-one-dimensional nanostructures: A Green's-function perspective}},\ }\href {https://doi.org/10.1103/PhysRevA.95.033818} {\bibfield  {journal} {\bibinfo  {journal} {Phys. Rev. A}\ }\textbf {\bibinfo {volume} {95}},\ \bibinfo {pages} {033818} (\bibinfo {year} {2017}{\natexlab{b}})}\BibitemShut {NoStop}%
\bibitem [{\citenamefont {Kornovan}\ \emph {et~al.}(2019)\citenamefont {Kornovan}, \citenamefont {Corzo}, \citenamefont {Laurat},\ and\ \citenamefont {Sheremet}}]{Kornovan2019}%
  \BibitemOpen
  \bibfield  {author} {\bibinfo {author} {\bibfnamefont {D.~F.}\ \bibnamefont {Kornovan}}, \bibinfo {author} {\bibfnamefont {N.~V.}\ \bibnamefont {Corzo}}, \bibinfo {author} {\bibfnamefont {J.}~\bibnamefont {Laurat}},\ and\ \bibinfo {author} {\bibfnamefont {A.~S.}\ \bibnamefont {Sheremet}},\ }\bibfield  {title} {\bibinfo {title} {{Extremely subradiant states in a periodic one-dimensional atomic array}},\ }\href {https://doi.org/10.1103/PhysRevA.100.063832} {\bibfield  {journal} {\bibinfo  {journal} {Phys. Rev. A}\ }\textbf {\bibinfo {volume} {100}},\ \bibinfo {pages} {063832} (\bibinfo {year} {2019})}\BibitemShut {NoStop}%
\bibitem [{\citenamefont {Zhang}\ \emph {et~al.}(2020)\citenamefont {Zhang}, \citenamefont {Yu},\ and\ \citenamefont {M{\o}lmer}}]{Zhang2020Feb}%
  \BibitemOpen
  \bibfield  {author} {\bibinfo {author} {\bibfnamefont {Y.-X.}\ \bibnamefont {Zhang}}, \bibinfo {author} {\bibfnamefont {C.}~\bibnamefont {Yu}},\ and\ \bibinfo {author} {\bibfnamefont {K.}~\bibnamefont {M{\o}lmer}},\ }\bibfield  {title} {\bibinfo {title} {{Subradiant bound dimer excited states of emitter chains coupled to a one dimensional waveguide}},\ }\href {https://doi.org/10.1103/PhysRevResearch.2.013173} {\bibfield  {journal} {\bibinfo  {journal} {Phys. Rev. Res.}\ }\textbf {\bibinfo {volume} {2}},\ \bibinfo {pages} {013173} (\bibinfo {year} {2020})}\BibitemShut {NoStop}%
\bibitem [{\citenamefont {Poddubny}(2020)}]{Poddubny2020}%
  \BibitemOpen
  \bibfield  {author} {\bibinfo {author} {\bibfnamefont {A.~N.}\ \bibnamefont {Poddubny}},\ }\bibfield  {title} {\bibinfo {title} {{Quasiflat band enabling subradiant two-photon bound states}},\ }\href {https://doi.org/10.1103/PhysRevA.101.043845} {\bibfield  {journal} {\bibinfo  {journal} {Phys. Rev. A}\ }\textbf {\bibinfo {volume} {101}},\ \bibinfo {pages} {043845} (\bibinfo {year} {2020})}\BibitemShut {NoStop}%
\bibitem [{\citenamefont {Facchinetti}\ \emph {et~al.}(2016)\citenamefont {Facchinetti}, \citenamefont {Jenkins},\ and\ \citenamefont {Ruostekoski}}]{facchinetti_storing_2016}%
  \BibitemOpen
  \bibfield  {author} {\bibinfo {author} {\bibfnamefont {G.}~\bibnamefont {Facchinetti}}, \bibinfo {author} {\bibfnamefont {S.~D.}\ \bibnamefont {Jenkins}},\ and\ \bibinfo {author} {\bibfnamefont {J.}~\bibnamefont {Ruostekoski}},\ }\bibfield  {title} {\bibinfo {title} {{Storing Light with Subradiant Correlations in Arrays of Atoms}},\ }\href {https://doi.org/10.1103/PhysRevLett.117.243601} {\bibfield  {journal} {\bibinfo  {journal} {Phys. Rev. Lett.}\ }\textbf {\bibinfo {volume} {117}},\ \bibinfo {pages} {243601} (\bibinfo {year} {2016})}\BibitemShut {NoStop}%
\bibitem [{\citenamefont {Bettles}\ \emph {et~al.}(2015)\citenamefont {Bettles}, \citenamefont {Gardiner},\ and\ \citenamefont {Adams}}]{bettles_cooperative_2015}%
  \BibitemOpen
  \bibfield  {author} {\bibinfo {author} {\bibfnamefont {R.~J.}\ \bibnamefont {Bettles}}, \bibinfo {author} {\bibfnamefont {S.~A.}\ \bibnamefont {Gardiner}},\ and\ \bibinfo {author} {\bibfnamefont {C.~S.}\ \bibnamefont {Adams}},\ }\bibfield  {title} {\bibinfo {title} {{Cooperative ordering in lattices of interacting two-level dipoles}},\ }\href {https://doi.org/10.1103/PhysRevA.92.063822} {\bibfield  {journal} {\bibinfo  {journal} {Phys. Rev. A}\ }\textbf {\bibinfo {volume} {92}},\ \bibinfo {pages} {063822} (\bibinfo {year} {2015})}\BibitemShut {NoStop}%
\bibitem [{\citenamefont {Ballantine}\ and\ \citenamefont {Ruostekoski}(2020)}]{ballantine_subradiance-protected_2020}%
  \BibitemOpen
  \bibfield  {author} {\bibinfo {author} {\bibfnamefont {K.~E.}\ \bibnamefont {Ballantine}}\ and\ \bibinfo {author} {\bibfnamefont {J.}~\bibnamefont {Ruostekoski}},\ }\bibfield  {title} {\bibinfo {title} {{Subradiance-protected excitation spreading in the generation of collimated photon emission from an atomic array}},\ }\href {https://doi.org/10.1103/PhysRevResearch.2.023086} {\bibfield  {journal} {\bibinfo  {journal} {Phys. Rev. Res.}\ }\textbf {\bibinfo {volume} {2}},\ \bibinfo {pages} {023086} (\bibinfo {year} {2020})}\BibitemShut {NoStop}%
\bibitem [{\citenamefont {Friedrich}\ and\ \citenamefont {Wintgen}(1985)}]{Friedrich1985a}%
  \BibitemOpen
  \bibfield  {author} {\bibinfo {author} {\bibfnamefont {H.}~\bibnamefont {Friedrich}}\ and\ \bibinfo {author} {\bibfnamefont {D.}~\bibnamefont {Wintgen}},\ }\bibfield  {title} {\bibinfo {title} {{Interfering resonances and bound states in the continuum}},\ }\href {https://doi.org/10.1103/PhysRevA.32.3231} {\bibfield  {journal} {\bibinfo  {journal} {Phys. Rev. A}\ }\textbf {\bibinfo {volume} {32}},\ \bibinfo {pages} {3231} (\bibinfo {year} {1985})}\BibitemShut {NoStop}%
\bibitem [{\citenamefont {Koshelev}\ \emph {et~al.}(2019)\citenamefont {Koshelev}, \citenamefont {Favraud}, \citenamefont {Bogdanov}, \citenamefont {Kivshar},\ and\ \citenamefont {Fratalocchi}}]{Koshelev2019}%
  \BibitemOpen
  \bibfield  {author} {\bibinfo {author} {\bibfnamefont {K.}~\bibnamefont {Koshelev}}, \bibinfo {author} {\bibfnamefont {G.}~\bibnamefont {Favraud}}, \bibinfo {author} {\bibfnamefont {A.}~\bibnamefont {Bogdanov}}, \bibinfo {author} {\bibfnamefont {Y.}~\bibnamefont {Kivshar}},\ and\ \bibinfo {author} {\bibfnamefont {A.}~\bibnamefont {Fratalocchi}},\ }\bibfield  {title} {\bibinfo {title} {{Nonradiating photonics with resonant dielectric nanostructures}},\ }\href {https://doi.org/10.1515/nanoph-2019-0024} {\bibfield  {journal} {\bibinfo  {journal} {Nanophotonics}\ }\textbf {\bibinfo {volume} {8}},\ \bibinfo {pages} {725} (\bibinfo {year} {2019})}\BibitemShut {NoStop}%
\bibitem [{\citenamefont {Mylnikov}\ \emph {et~al.}(2020)\citenamefont {Mylnikov}, \citenamefont {Ha}, \citenamefont {Pan}, \citenamefont {Valuckas}, \citenamefont {Paniagua-Dom{\ifmmode\acute{\imath}\else\'{\i}\fi}nguez}, \citenamefont {Demir},\ and\ \citenamefont {Kuznetsov}}]{Mylnikov2020}%
  \BibitemOpen
  \bibfield  {author} {\bibinfo {author} {\bibfnamefont {V.}~\bibnamefont {Mylnikov}}, \bibinfo {author} {\bibfnamefont {S.~T.}\ \bibnamefont {Ha}}, \bibinfo {author} {\bibfnamefont {Z.}~\bibnamefont {Pan}}, \bibinfo {author} {\bibfnamefont {V.}~\bibnamefont {Valuckas}}, \bibinfo {author} {\bibfnamefont {R.}~\bibnamefont {Paniagua-Dom{\ifmmode\acute{\imath}\else\'{\i}\fi}nguez}}, \bibinfo {author} {\bibfnamefont {H.~V.}\ \bibnamefont {Demir}},\ and\ \bibinfo {author} {\bibfnamefont {A.~I.}\ \bibnamefont {Kuznetsov}},\ }\bibfield  {title} {\bibinfo {title} {{Lasing Action in Single Subwavelength Particles Supporting Supercavity Modes}},\ }\href {https://doi.org/10.1021/acsnano.0c02730} {\bibfield  {journal} {\bibinfo  {journal} {ACS Nano}\ }\textbf {\bibinfo {volume} {14}},\ \bibinfo {pages} {7338} (\bibinfo {year} {2020})}\BibitemShut {NoStop}%
\bibitem [{\citenamefont {Ryabov}\ \emph {et~al.}(2022)\citenamefont {Ryabov}, \citenamefont {Pashina}, \citenamefont {Zograf}, \citenamefont {Makarov},\ and\ \citenamefont {Petrov}}]{ryabov_nonlinear_2022}%
  \BibitemOpen
  \bibfield  {author} {\bibinfo {author} {\bibfnamefont {D.}~\bibnamefont {Ryabov}}, \bibinfo {author} {\bibfnamefont {O.}~\bibnamefont {Pashina}}, \bibinfo {author} {\bibfnamefont {G.}~\bibnamefont {Zograf}}, \bibinfo {author} {\bibfnamefont {S.}~\bibnamefont {Makarov}},\ and\ \bibinfo {author} {\bibfnamefont {M.}~\bibnamefont {Petrov}},\ }\bibfield  {title} {\bibinfo {title} {{Nonlinear optical heating of all-dielectric super-cavity: efficient light-to-heat conversion through giant thermorefractive bistability}},\ }\href {https://doi.org/10.1515/nanoph-2022-0074} {\bibfield  {journal} {\bibinfo  {journal} {Nanophotonics}\ }\textbf {\bibinfo {volume} {11}},\ \bibinfo {pages} {3981} (\bibinfo {year} {2022})}\BibitemShut {NoStop}%
\bibitem [{\citenamefont {Hsu}\ \emph {et~al.}(2016)\citenamefont {Hsu}, \citenamefont {Zhen}, \citenamefont {Stone}, \citenamefont {Joannopoulos},\ and\ \citenamefont {Solja{\ifmmode\check{c}\else\v{c}\fi}i{\ifmmode\acute{c}\else\'{c}\fi}}}]{Hsu2016}%
  \BibitemOpen
  \bibfield  {author} {\bibinfo {author} {\bibfnamefont {C.~W.}\ \bibnamefont {Hsu}}, \bibinfo {author} {\bibfnamefont {B.}~\bibnamefont {Zhen}}, \bibinfo {author} {\bibfnamefont {A.~D.}\ \bibnamefont {Stone}}, \bibinfo {author} {\bibfnamefont {J.~D.}\ \bibnamefont {Joannopoulos}},\ and\ \bibinfo {author} {\bibfnamefont {M.}~\bibnamefont {Solja{\ifmmode\check{c}\else\v{c}\fi}i{\ifmmode\acute{c}\else\'{c}\fi}}},\ }\bibfield  {title} {\bibinfo {title} {{Bound states in the continuum}},\ }\href {https://doi.org/10.1038/natrevmats.2016.48} {\bibfield  {journal} {\bibinfo  {journal} {Nat. Rev. Mater.}\ }\textbf {\bibinfo {volume} {1}},\ \bibinfo {pages} {1} (\bibinfo {year} {2016})}\BibitemShut {NoStop}%
\bibitem [{\citenamefont {Koshelev}\ \emph {et~al.}(2023)\citenamefont {Koshelev}, \citenamefont {Sadrieva}, \citenamefont {Shcherbakov}, \citenamefont {Kivshar},\ and\ \citenamefont {Bogdanov}}]{koshelev_bound_2021}%
  \BibitemOpen
  \bibfield  {author} {\bibinfo {author} {\bibfnamefont {K.~L.}\ \bibnamefont {Koshelev}}, \bibinfo {author} {\bibfnamefont {Z.~F.}\ \bibnamefont {Sadrieva}}, \bibinfo {author} {\bibfnamefont {A.~A.}\ \bibnamefont {Shcherbakov}}, \bibinfo {author} {\bibfnamefont {{\relax Yu}.~S.}\ \bibnamefont {Kivshar}},\ and\ \bibinfo {author} {\bibfnamefont {A.~A.}\ \bibnamefont {Bogdanov}},\ }\bibfield  {title} {\bibinfo {title} {{Bound states in the continuum in photonic structures}},\ }\href {https://ufn.ru/en/articles/2023/5/c} {\bibfield  {journal} {\bibinfo  {journal} {Phys.-Usp.}\ }\textbf {\bibinfo {volume} {66}},\ \bibinfo {pages} {494} (\bibinfo {year} {2023})}\BibitemShut {NoStop}%
\bibitem [{\citenamefont {Freedhoff}(1986)}]{freedhoff_cooperative_1986}%
  \BibitemOpen
  \bibfield  {author} {\bibinfo {author} {\bibfnamefont {H.~S.}\ \bibnamefont {Freedhoff}},\ }\bibfield  {title} {\bibinfo {title} {{Cooperative single{-}quantum excitations of a closed{-}ring polymer chain}},\ }\href {https://doi.org/10.1063/1.451476} {\bibfield  {journal} {\bibinfo  {journal} {J. Chem. Phys.}\ }\textbf {\bibinfo {volume} {85}},\ \bibinfo {pages} {6110} (\bibinfo {year} {1986})}\BibitemShut {NoStop}%
\bibitem [{\citenamefont {Cremer}\ \emph {et~al.}(2020)\citenamefont {Cremer}, \citenamefont {Plankensteiner}, \citenamefont {Moreno-Cardoner}, \citenamefont {Ostermann},\ and\ \citenamefont {Ritsch}}]{cremer_polarization_2020}%
  \BibitemOpen
  \bibfield  {author} {\bibinfo {author} {\bibfnamefont {J.}~\bibnamefont {Cremer}}, \bibinfo {author} {\bibfnamefont {D.}~\bibnamefont {Plankensteiner}}, \bibinfo {author} {\bibfnamefont {M.}~\bibnamefont {Moreno-Cardoner}}, \bibinfo {author} {\bibfnamefont {L.}~\bibnamefont {Ostermann}},\ and\ \bibinfo {author} {\bibfnamefont {H.}~\bibnamefont {Ritsch}},\ }\bibfield  {title} {\bibinfo {title} {{Polarization control of radiation and energy flow in dipole-coupled nanorings}},\ }\href {https://doi.org/10.1088/1367-2630/aba4d4} {\bibfield  {journal} {\bibinfo  {journal} {New J. Phys.}\ }\textbf {\bibinfo {volume} {22}},\ \bibinfo {pages} {083052} (\bibinfo {year} {2020})}\BibitemShut {NoStop}%
\bibitem [{\citenamefont {Moreno-Cardoner}\ \emph {et~al.}(2019)\citenamefont {Moreno-Cardoner}, \citenamefont {Plankensteiner}, \citenamefont {Ostermann}, \citenamefont {Chang},\ and\ \citenamefont {Ritsch}}]{moreno-cardoner_subradiance-enhanced_2019}%
  \BibitemOpen
  \bibfield  {author} {\bibinfo {author} {\bibfnamefont {M.}~\bibnamefont {Moreno-Cardoner}}, \bibinfo {author} {\bibfnamefont {D.}~\bibnamefont {Plankensteiner}}, \bibinfo {author} {\bibfnamefont {L.}~\bibnamefont {Ostermann}}, \bibinfo {author} {\bibfnamefont {D.~E.}\ \bibnamefont {Chang}},\ and\ \bibinfo {author} {\bibfnamefont {H.}~\bibnamefont {Ritsch}},\ }\bibfield  {title} {\bibinfo {title} {{Subradiance-enhanced excitation transfer between dipole-coupled nanorings of quantum emitters}},\ }\href {https://doi.org/10.1103/PhysRevA.100.023806} {\bibfield  {journal} {\bibinfo  {journal} {Phys. Rev. A}\ }\textbf {\bibinfo {volume} {100}},\ \bibinfo {pages} {023806} (\bibinfo {year} {2019})}\BibitemShut {NoStop}%
\bibitem [{\citenamefont {Holzinger}\ \emph {et~al.}(2020{\natexlab{b}})\citenamefont {Holzinger}, \citenamefont {Plankensteiner}, \citenamefont {Ostermann},\ and\ \citenamefont {Ritsch}}]{holzinger_nanoscale_2020}%
  \BibitemOpen
  \bibfield  {author} {\bibinfo {author} {\bibfnamefont {R.}~\bibnamefont {Holzinger}}, \bibinfo {author} {\bibfnamefont {D.}~\bibnamefont {Plankensteiner}}, \bibinfo {author} {\bibfnamefont {L.}~\bibnamefont {Ostermann}},\ and\ \bibinfo {author} {\bibfnamefont {H.}~\bibnamefont {Ritsch}},\ }\bibfield  {title} {\bibinfo {title} {{Nanoscale Coherent Light Source}},\ }\href {https://doi.org/10.1103/PhysRevLett.124.253603} {\bibfield  {journal} {\bibinfo  {journal} {Phys. Rev. Lett.}\ }\textbf {\bibinfo {volume} {124}},\ \bibinfo {pages} {253603} (\bibinfo {year} {2020}{\natexlab{b}})}\BibitemShut {NoStop}%
\bibitem [{\citenamefont {Moreno-Cardoner}\ \emph {et~al.}(2022)\citenamefont {Moreno-Cardoner}, \citenamefont {Holzinger},\ and\ \citenamefont {Ritsch}}]{moreno-cardoner_efficient_2022}%
  \BibitemOpen
  \bibfield  {author} {\bibinfo {author} {\bibfnamefont {M.}~\bibnamefont {Moreno-Cardoner}}, \bibinfo {author} {\bibfnamefont {R.}~\bibnamefont {Holzinger}},\ and\ \bibinfo {author} {\bibfnamefont {H.}~\bibnamefont {Ritsch}},\ }\bibfield  {title} {\bibinfo {title} {{Efficient nano-photonic antennas based on dark states in quantum emitter rings}},\ }\href {https://doi.org/10.1364/OE.437396} {\bibfield  {journal} {\bibinfo  {journal} {Opt. Express}\ }\textbf {\bibinfo {volume} {30}},\ \bibinfo {pages} {10779} (\bibinfo {year} {2022})}\BibitemShut {NoStop}%
\bibitem [{\citenamefont {Holzinger}\ \emph {et~al.}(2023)\citenamefont {Holzinger}, \citenamefont {Peter}, \citenamefont {Ostermann}, \citenamefont {Ritsch},\ and\ \citenamefont {Yelin}}]{Holzinger2023Sep}%
  \BibitemOpen
  \bibfield  {author} {\bibinfo {author} {\bibfnamefont {R.}~\bibnamefont {Holzinger}}, \bibinfo {author} {\bibfnamefont {J.}~\bibnamefont {Peter}}, \bibinfo {author} {\bibfnamefont {S.}~\bibnamefont {Ostermann}}, \bibinfo {author} {\bibfnamefont {H.}~\bibnamefont {Ritsch}},\ and\ \bibinfo {author} {\bibfnamefont {S.}~\bibnamefont {Yelin}},\ }\bibfield  {title} {\bibinfo {title} {{Harnessing quantum emitter rings for efficient energy transport and trapping}},\ }\bibfield  {journal} {\bibinfo  {journal} {arXiv}\ }\href {https://doi.org/10.48550/arXiv.2309.11376} {10.48550/arXiv.2309.11376} (\bibinfo {year} {2023})\BibitemShut {NoStop}%
\bibitem [{\citenamefont {Cech}\ \emph {et~al.}(2023)\citenamefont {Cech}, \citenamefont {Lesanovsky},\ and\ \citenamefont {Olmos}}]{Cech2023Nov}%
  \BibitemOpen
  \bibfield  {author} {\bibinfo {author} {\bibfnamefont {M.}~\bibnamefont {Cech}}, \bibinfo {author} {\bibfnamefont {I.}~\bibnamefont {Lesanovsky}},\ and\ \bibinfo {author} {\bibfnamefont {B.}~\bibnamefont {Olmos}},\ }\bibfield  {title} {\bibinfo {title} {{Dispersionless subradiant photon storage in one-dimensional emitter chains}},\ }\href {https://doi.org/10.1103/PhysRevA.108.L051702} {\bibfield  {journal} {\bibinfo  {journal} {Phys. Rev. A}\ }\textbf {\bibinfo {volume} {108}},\ \bibinfo {pages} {L051702} (\bibinfo {year} {2023})}\BibitemShut {NoStop}%
\bibitem [{\citenamefont {Ding}\ \emph {et~al.}(2015)\citenamefont {Ding}, \citenamefont {Zhang}, \citenamefont {Zhou}, \citenamefont {Shi}, \citenamefont {Xiang}, \citenamefont {Wang}, \citenamefont {Jiang}, \citenamefont {Shi},\ and\ \citenamefont {Guo}}]{Ding2015Feb}%
  \BibitemOpen
  \bibfield  {author} {\bibinfo {author} {\bibfnamefont {D.-S.}\ \bibnamefont {Ding}}, \bibinfo {author} {\bibfnamefont {W.}~\bibnamefont {Zhang}}, \bibinfo {author} {\bibfnamefont {Z.-Y.}\ \bibnamefont {Zhou}}, \bibinfo {author} {\bibfnamefont {S.}~\bibnamefont {Shi}}, \bibinfo {author} {\bibfnamefont {G.-Y.}\ \bibnamefont {Xiang}}, \bibinfo {author} {\bibfnamefont {X.-S.}\ \bibnamefont {Wang}}, \bibinfo {author} {\bibfnamefont {Y.-K.}\ \bibnamefont {Jiang}}, \bibinfo {author} {\bibfnamefont {B.-S.}\ \bibnamefont {Shi}},\ and\ \bibinfo {author} {\bibfnamefont {G.-C.}\ \bibnamefont {Guo}},\ }\bibfield  {title} {\bibinfo {title} {{Quantum Storage of Orbital Angular Momentum Entanglement in an Atomic Ensemble}},\ }\href {https://doi.org/10.1103/PhysRevLett.114.050502} {\bibfield  {journal} {\bibinfo  {journal} {Phys. Rev. Lett.}\ }\textbf {\bibinfo {volume} {114}},\ \bibinfo {pages} {050502} (\bibinfo {year} {2015})}\BibitemShut {NoStop}%
\bibitem [{\citenamefont {Mirhosseini}\ \emph {et~al.}(2015)\citenamefont {Mirhosseini}, \citenamefont {Maga{\ifmmode\tilde{n}\else\~{n}\fi}a-Loaiza}, \citenamefont {O{'}Sullivan}, \citenamefont {Rodenburg}, \citenamefont {Malik}, \citenamefont {Lavery}, \citenamefont {Padgett}, \citenamefont {Gauthier},\ and\ \citenamefont {Boyd}}]{mirhosseini_high-dimensional_2015}%
  \BibitemOpen
  \bibfield  {author} {\bibinfo {author} {\bibfnamefont {M.}~\bibnamefont {Mirhosseini}}, \bibinfo {author} {\bibfnamefont {O.~S.}\ \bibnamefont {Maga{\ifmmode\tilde{n}\else\~{n}\fi}a-Loaiza}}, \bibinfo {author} {\bibfnamefont {M.~N.}\ \bibnamefont {O{'}Sullivan}}, \bibinfo {author} {\bibfnamefont {B.}~\bibnamefont {Rodenburg}}, \bibinfo {author} {\bibfnamefont {M.}~\bibnamefont {Malik}}, \bibinfo {author} {\bibfnamefont {M.~P.~J.}\ \bibnamefont {Lavery}}, \bibinfo {author} {\bibfnamefont {M.~J.}\ \bibnamefont {Padgett}}, \bibinfo {author} {\bibfnamefont {D.~J.}\ \bibnamefont {Gauthier}},\ and\ \bibinfo {author} {\bibfnamefont {R.~W.}\ \bibnamefont {Boyd}},\ }\bibfield  {title} {\bibinfo {title} {{High-dimensional quantum cryptography with twisted light}},\ }\href {https://doi.org/10.1088/1367-2630/17/3/033033} {\bibfield  {journal} {\bibinfo  {journal} {New J. Phys.}\ }\textbf {\bibinfo {volume} {17}},\ \bibinfo {pages} {033033} (\bibinfo {year} {2015})}\BibitemShut {NoStop}%
\bibitem [{\citenamefont {Cozzolino}\ \emph {et~al.}(2019)\citenamefont {Cozzolino}, \citenamefont {Bacco}, \citenamefont {Da~Lio}, \citenamefont {Ingerslev}, \citenamefont {Ding}, \citenamefont {Dalgaard}, \citenamefont {Kristensen}, \citenamefont {Galili}, \citenamefont {Rottwitt}, \citenamefont {Ramachandran},\ and\ \citenamefont {Oxenl{\o}we}}]{cozzolino_orbital_2019}%
  \BibitemOpen
  \bibfield  {author} {\bibinfo {author} {\bibfnamefont {D.}~\bibnamefont {Cozzolino}}, \bibinfo {author} {\bibfnamefont {D.}~\bibnamefont {Bacco}}, \bibinfo {author} {\bibfnamefont {B.}~\bibnamefont {Da~Lio}}, \bibinfo {author} {\bibfnamefont {K.}~\bibnamefont {Ingerslev}}, \bibinfo {author} {\bibfnamefont {Y.}~\bibnamefont {Ding}}, \bibinfo {author} {\bibfnamefont {K.}~\bibnamefont {Dalgaard}}, \bibinfo {author} {\bibfnamefont {P.}~\bibnamefont {Kristensen}}, \bibinfo {author} {\bibfnamefont {M.}~\bibnamefont {Galili}}, \bibinfo {author} {\bibfnamefont {K.}~\bibnamefont {Rottwitt}}, \bibinfo {author} {\bibfnamefont {S.}~\bibnamefont {Ramachandran}},\ and\ \bibinfo {author} {\bibfnamefont {L.~K.}\ \bibnamefont {Oxenl{\o}we}},\ }\bibfield  {title} {\bibinfo {title} {{Orbital Angular Momentum States Enabling Fiber-based High-dimensional Quantum Communication}},\ }\href {https://doi.org/10.1103/PhysRevApplied.11.064058} {\bibfield  {journal} {\bibinfo  {journal} {Phys. Rev. Appl.}\ }\textbf {\bibinfo {volume}
  {11}},\ \bibinfo {pages} {064058} (\bibinfo {year} {2019})}\BibitemShut {NoStop}%
\bibitem [{\citenamefont {Pellegrino}\ \emph {et~al.}(2014)\citenamefont {Pellegrino}, \citenamefont {Bourgain}, \citenamefont {Jennewein}, \citenamefont {Sortais}, \citenamefont {Browaeys}, \citenamefont {Jenkins},\ and\ \citenamefont {Ruostekoski}}]{Pellegrino2014Sep}%
  \BibitemOpen
  \bibfield  {author} {\bibinfo {author} {\bibfnamefont {J.}~\bibnamefont {Pellegrino}}, \bibinfo {author} {\bibfnamefont {R.}~\bibnamefont {Bourgain}}, \bibinfo {author} {\bibfnamefont {S.}~\bibnamefont {Jennewein}}, \bibinfo {author} {\bibfnamefont {Y.~R.~P.}\ \bibnamefont {Sortais}}, \bibinfo {author} {\bibfnamefont {A.}~\bibnamefont {Browaeys}}, \bibinfo {author} {\bibfnamefont {S.~D.}\ \bibnamefont {Jenkins}},\ and\ \bibinfo {author} {\bibfnamefont {J.}~\bibnamefont {Ruostekoski}},\ }\bibfield  {title} {\bibinfo {title} {{Observation of Suppression of Light Scattering Induced by Dipole-Dipole Interactions in a Cold-Atom Ensemble}},\ }\href {https://doi.org/10.1103/PhysRevLett.113.133602} {\bibfield  {journal} {\bibinfo  {journal} {Phys. Rev. Lett.}\ }\textbf {\bibinfo {volume} {113}},\ \bibinfo {pages} {133602} (\bibinfo {year} {2014})}\BibitemShut {NoStop}%
\bibitem [{\citenamefont {Kemp}\ \emph {et~al.}(2020)\citenamefont {Kemp}, \citenamefont {Roof}, \citenamefont {Havey}, \citenamefont {Sokolov}, \citenamefont {Kupriyanov},\ and\ \citenamefont {Guerin}}]{Kemp2020Mar}%
  \BibitemOpen
  \bibfield  {author} {\bibinfo {author} {\bibfnamefont {K.~J.}\ \bibnamefont {Kemp}}, \bibinfo {author} {\bibfnamefont {S.~J.}\ \bibnamefont {Roof}}, \bibinfo {author} {\bibfnamefont {M.~D.}\ \bibnamefont {Havey}}, \bibinfo {author} {\bibfnamefont {I.~M.}\ \bibnamefont {Sokolov}}, \bibinfo {author} {\bibfnamefont {D.~V.}\ \bibnamefont {Kupriyanov}},\ and\ \bibinfo {author} {\bibfnamefont {W.}~\bibnamefont {Guerin}},\ }\bibfield  {title} {\bibinfo {title} {{Optical-depth scaling of light scattering from a dense and cold atomic $^{87}\mathrm{Rb}$ gas}},\ }\href {https://doi.org/10.1103/PhysRevA.101.033832} {\bibfield  {journal} {\bibinfo  {journal} {Phys. Rev. A}\ }\textbf {\bibinfo {volume} {101}},\ \bibinfo {pages} {033832} (\bibinfo {year} {2020})}\BibitemShut {NoStop}%
\bibitem [{\citenamefont {Antezza}\ and\ \citenamefont {Castin}(2009)}]{Antezza2009Sep}%
  \BibitemOpen
  \bibfield  {author} {\bibinfo {author} {\bibfnamefont {M.}~\bibnamefont {Antezza}}\ and\ \bibinfo {author} {\bibfnamefont {Y.}~\bibnamefont {Castin}},\ }\bibfield  {title} {\bibinfo {title} {{Spectrum of Light in a Quantum Fluctuating Periodic Structure}},\ }\href {https://doi.org/10.1103/PhysRevLett.103.123903} {\bibfield  {journal} {\bibinfo  {journal} {Phys. Rev. Lett.}\ }\textbf {\bibinfo {volume} {103}},\ \bibinfo {pages} {123903} (\bibinfo {year} {2009})}\BibitemShut {NoStop}%
\bibitem [{\citenamefont {Sokolov}\ and\ \citenamefont {Guerin}(2019)}]{Sokolov2019Aug}%
  \BibitemOpen
  \bibfield  {author} {\bibinfo {author} {\bibfnamefont {I.~M.}\ \bibnamefont {Sokolov}}\ and\ \bibinfo {author} {\bibfnamefont {W.}~\bibnamefont {Guerin}},\ }\bibfield  {title} {\bibinfo {title} {{Comparison of three approaches to light scattering by dilute cold atomic ensembles}},\ }\href {https://doi.org/10.1364/JOSAB.36.002030} {\bibfield  {journal} {\bibinfo  {journal} {J. Opt. Soc. Am. B, JOSAB}\ }\textbf {\bibinfo {volume} {36}},\ \bibinfo {pages} {2030} (\bibinfo {year} {2019})}\BibitemShut {NoStop}%
\bibitem [{\citenamefont {Bouscal}\ \emph {et~al.}(2024)\citenamefont {Bouscal}, \citenamefont {Kemiche}, \citenamefont {Mahapatra}, \citenamefont {Fayard}, \citenamefont {Berroir}, \citenamefont {Ray}, \citenamefont {Greffet}, \citenamefont {Raineri}, \citenamefont {Levenson}, \citenamefont {Bencheikh}, \citenamefont {Sauvan}, \citenamefont {Urvoy},\ and\ \citenamefont {Laurat}}]{bouscal_systematic_2024}%
  \BibitemOpen
  \bibfield  {author} {\bibinfo {author} {\bibfnamefont {A.}~\bibnamefont {Bouscal}}, \bibinfo {author} {\bibfnamefont {M.}~\bibnamefont {Kemiche}}, \bibinfo {author} {\bibfnamefont {S.}~\bibnamefont {Mahapatra}}, \bibinfo {author} {\bibfnamefont {N.}~\bibnamefont {Fayard}}, \bibinfo {author} {\bibfnamefont {J.}~\bibnamefont {Berroir}}, \bibinfo {author} {\bibfnamefont {T.}~\bibnamefont {Ray}}, \bibinfo {author} {\bibfnamefont {J.-J.}\ \bibnamefont {Greffet}}, \bibinfo {author} {\bibfnamefont {F.}~\bibnamefont {Raineri}}, \bibinfo {author} {\bibfnamefont {A.}~\bibnamefont {Levenson}}, \bibinfo {author} {\bibfnamefont {K.}~\bibnamefont {Bencheikh}}, \bibinfo {author} {\bibfnamefont {C.}~\bibnamefont {Sauvan}}, \bibinfo {author} {\bibfnamefont {A.}~\bibnamefont {Urvoy}},\ and\ \bibinfo {author} {\bibfnamefont {J.}~\bibnamefont {Laurat}},\ }\bibfield  {title} {\bibinfo {title} {{Systematic design of a robust half-W1 photonic crystal waveguide for interfacing slow light and trapped cold atoms}},\ }\href
  {https://doi.org/10.1088/1367-2630/ad23a4} {\bibfield  {journal} {\bibinfo  {journal} {New J. Phys.}\ }\textbf {\bibinfo {volume} {26}},\ \bibinfo {pages} {023026} (\bibinfo {year} {2024})}\BibitemShut {NoStop}%
\bibitem [{\citenamefont {Novotny}\ and\ \citenamefont {Hecht}(2012)}]{Novotny2012}%
  \BibitemOpen
  \bibfield  {author} {\bibinfo {author} {\bibfnamefont {L.}~\bibnamefont {Novotny}}\ and\ \bibinfo {author} {\bibfnamefont {B.}~\bibnamefont {Hecht}},\ }\href {http://www.amazon.com/Principles-Nano-Optics-Lukas-Novotny/dp/1107005469} {\emph {\bibinfo {title} {Principles of {Nano}-{Optics}}}}\ (\bibinfo  {publisher} {Cambridge University Press},\ \bibinfo {year} {2012})\BibitemShut {NoStop}%
\bibitem [{sup()}]{suppl}%
  \BibitemOpen
  \href@noop {} {}\bibinfo {note} {See Supplemental Material at ... for effective Schr\"{o}dinger equations, the single-excitation spectrum for a single ring, number of doubly excited eigenstates in ring structures, correspondence between $m$ and irreducible representations for $C_{6v}$ symmetry group, numerical calculation of the lifetime for ring eigenstates, details on the hybridized states of the ring oligomers, derivation of the far-field radiation pattern and scattering cross section for a dipole array in the semi-classical picture, the relation between energies of doubly and singly excited eigenstates, numerical optimization of the lifetime for the subradiant doubly excited state, and features of subradiant doubly excited state in a case of circular dipoles, which includes Ref.~\cite{Loudon2006Jul,Wigner1931Jan,Tsimokha2022Apr,Gladyshev2020Aug,Poleva2023Jan}}\BibitemShut {NoStop}%
\bibitem [{\citenamefont {Cao}\ and\ \citenamefont {Wiersig}(2015)}]{Cao2015a}%
  \BibitemOpen
  \bibfield  {author} {\bibinfo {author} {\bibfnamefont {H.}~\bibnamefont {Cao}}\ and\ \bibinfo {author} {\bibfnamefont {J.}~\bibnamefont {Wiersig}},\ }\bibfield  {title} {\bibinfo {title} {{Dielectric microcavities: Model systems for wave chaos and non-Hermitian physics}},\ }\href {https://doi.org/10.1103/RevModPhys.87.61} {\bibfield  {journal} {\bibinfo  {journal} {Rev. Mod. Phys.}\ }\textbf {\bibinfo {volume} {87}},\ \bibinfo {pages} {61} (\bibinfo {year} {2015})}\BibitemShut {NoStop}%
\bibitem [{\citenamefont {Rybin}\ \emph {et~al.}(2017)\citenamefont {Rybin}, \citenamefont {Koshelev}, \citenamefont {Sadrieva}, \citenamefont {Samusev}, \citenamefont {Bogdanov}, \citenamefont {Limonov},\ and\ \citenamefont {Kivshar}}]{rybin_high-q_2017}%
  \BibitemOpen
  \bibfield  {author} {\bibinfo {author} {\bibfnamefont {M.~V.}\ \bibnamefont {Rybin}}, \bibinfo {author} {\bibfnamefont {K.~L.}\ \bibnamefont {Koshelev}}, \bibinfo {author} {\bibfnamefont {Z.~F.}\ \bibnamefont {Sadrieva}}, \bibinfo {author} {\bibfnamefont {K.~B.}\ \bibnamefont {Samusev}}, \bibinfo {author} {\bibfnamefont {A.~A.}\ \bibnamefont {Bogdanov}}, \bibinfo {author} {\bibfnamefont {M.~F.}\ \bibnamefont {Limonov}},\ and\ \bibinfo {author} {\bibfnamefont {Y.~S.}\ \bibnamefont {Kivshar}},\ }\bibfield  {title} {\bibinfo {title} {{High-$Q$ Supercavity Modes in Subwavelength Dielectric Resonators}},\ }\href {https://doi.org/10.1103/PhysRevLett.119.243901} {\bibfield  {journal} {\bibinfo  {journal} {Phys. Rev. Lett.}\ }\textbf {\bibinfo {volume} {119}},\ \bibinfo {pages} {243901} (\bibinfo {year} {2017})}\BibitemShut {NoStop}%
\bibitem [{\citenamefont {Kotlyar}\ \emph {et~al.}(2019)\citenamefont {Kotlyar}, \citenamefont {Nalimov},\ and\ \citenamefont {Stafeev}}]{kotlyar_exploiting_2019}%
  \BibitemOpen
  \bibfield  {author} {\bibinfo {author} {\bibfnamefont {V.~V.}\ \bibnamefont {Kotlyar}}, \bibinfo {author} {\bibfnamefont {A.~G.}\ \bibnamefont {Nalimov}},\ and\ \bibinfo {author} {\bibfnamefont {S.~S.}\ \bibnamefont {Stafeev}},\ }\bibfield  {title} {\bibinfo {title} {{Exploiting the circular polarization of light to obtain a spiral energy flow at the subwavelength focus}},\ }\href {https://doi.org/10.1364/JOSAB.36.002850} {\bibfield  {journal} {\bibinfo  {journal} {J. Opt. Soc. Am. B}\ }\textbf {\bibinfo {volume} {36}},\ \bibinfo {pages} {2850} (\bibinfo {year} {2019})}\BibitemShut {NoStop}%
\bibitem [{\citenamefont {Gulfam}(2023)}]{Gulfam2023Jun}%
  \BibitemOpen
  \bibfield  {author} {\bibinfo {author} {\bibfnamefont {Q.-u.-A.}\ \bibnamefont {Gulfam}},\ }\bibfield  {title} {\bibinfo {title} {{Photon bunching from an equilateral triangle of atoms}},\ }\href {https://www.tandfonline.com/doi/full/10.1080/09500340.2023.2220162} {\bibfield  {journal} {\bibinfo  {journal} {J. Mod. Opt.}\ ,\ \bibinfo {pages} {217}} (\bibinfo {year} {2023})}\BibitemShut {NoStop}%
\bibitem [{\citenamefont {Jin}\ \emph {et~al.}(2023)\citenamefont {Jin}, \citenamefont {Gao}, \citenamefont {Chandrashekara}, \citenamefont {G{\ifmmode\ddot{o}\else\"{o}\fi}lzh{\ifmmode\ddot{a}\else\"{a}\fi}user}, \citenamefont {Sch{\ifmmode\ddot{o}\else\"{o}\fi}ner},\ and\ \citenamefont {Chomaz}}]{jin_two-dimensional_2023}%
  \BibitemOpen
  \bibfield  {author} {\bibinfo {author} {\bibfnamefont {S.}~\bibnamefont {Jin}}, \bibinfo {author} {\bibfnamefont {J.}~\bibnamefont {Gao}}, \bibinfo {author} {\bibfnamefont {K.}~\bibnamefont {Chandrashekara}}, \bibinfo {author} {\bibfnamefont {C.}~\bibnamefont {G{\ifmmode\ddot{o}\else\"{o}\fi}lzh{\ifmmode\ddot{a}\else\"{a}\fi}user}}, \bibinfo {author} {\bibfnamefont {J.}~\bibnamefont {Sch{\ifmmode\ddot{o}\else\"{o}\fi}ner}},\ and\ \bibinfo {author} {\bibfnamefont {L.}~\bibnamefont {Chomaz}},\ }\bibfield  {title} {\bibinfo {title} {{Two-dimensional magneto-optical trap of dysprosium atoms as a compact source for efficient loading of a narrow-line three-dimensional magneto-optical trap}},\ }\href {https://doi.org/10.1103/PhysRevA.108.023719} {\bibfield  {journal} {\bibinfo  {journal} {Phys. Rev. A}\ }\textbf {\bibinfo {volume} {108}},\ \bibinfo {pages} {023719} (\bibinfo {year} {2023})}\BibitemShut {NoStop}%
\bibitem [{\citenamefont {Shahmoon}\ \emph {et~al.}(2017)\citenamefont {Shahmoon}, \citenamefont {Wild}, \citenamefont {Lukin},\ and\ \citenamefont {Yelin}}]{shahmoon_cooperative_2017}%
  \BibitemOpen
  \bibfield  {author} {\bibinfo {author} {\bibfnamefont {E.}~\bibnamefont {Shahmoon}}, \bibinfo {author} {\bibfnamefont {D.~S.}\ \bibnamefont {Wild}}, \bibinfo {author} {\bibfnamefont {M.~D.}\ \bibnamefont {Lukin}},\ and\ \bibinfo {author} {\bibfnamefont {S.~F.}\ \bibnamefont {Yelin}},\ }\bibfield  {title} {\bibinfo {title} {{Cooperative Resonances in Light Scattering from Two-Dimensional Atomic Arrays}},\ }\href {https://doi.org/10.1103/PhysRevLett.118.113601} {\bibfield  {journal} {\bibinfo  {journal} {Phys. Rev. Lett.}\ }\textbf {\bibinfo {volume} {118}},\ \bibinfo {pages} {113601} (\bibinfo {year} {2017})}\BibitemShut {NoStop}%
\bibitem [{\citenamefont {Plankensteiner}\ \emph {et~al.}(2015)\citenamefont {Plankensteiner}, \citenamefont {Ostermann}, \citenamefont {Ritsch},\ and\ \citenamefont {Genes}}]{plankensteiner_selective_2015}%
  \BibitemOpen
  \bibfield  {author} {\bibinfo {author} {\bibfnamefont {D.}~\bibnamefont {Plankensteiner}}, \bibinfo {author} {\bibfnamefont {L.}~\bibnamefont {Ostermann}}, \bibinfo {author} {\bibfnamefont {H.}~\bibnamefont {Ritsch}},\ and\ \bibinfo {author} {\bibfnamefont {C.}~\bibnamefont {Genes}},\ }\bibfield  {title} {\bibinfo {title} {{Selective protected state preparation of coupled dissipative quantum emitters}},\ }\href {https://doi.org/10.1038/srep16231} {\bibfield  {journal} {\bibinfo  {journal} {Sci. Rep.}\ }\textbf {\bibinfo {volume} {5}},\ \bibinfo {pages} {1} (\bibinfo {year} {2015})}\BibitemShut {NoStop}%
\bibitem [{\citenamefont {Fedorovich}\ \emph {et~al.}(2022)\citenamefont {Fedorovich}, \citenamefont {Kornovan}, \citenamefont {Poddubny},\ and\ \citenamefont {Petrov}}]{fedorovich_chirality-driven_2022}%
  \BibitemOpen
  \bibfield  {author} {\bibinfo {author} {\bibfnamefont {G.}~\bibnamefont {Fedorovich}}, \bibinfo {author} {\bibfnamefont {D.}~\bibnamefont {Kornovan}}, \bibinfo {author} {\bibfnamefont {A.}~\bibnamefont {Poddubny}},\ and\ \bibinfo {author} {\bibfnamefont {M.}~\bibnamefont {Petrov}},\ }\bibfield  {title} {\bibinfo {title} {Chirality-driven delocalization in disordered waveguide-coupled quantum arrays},\ }\href {https://doi.org/10.1103/PhysRevA.106.043723} {\bibfield  {journal} {\bibinfo  {journal} {Physical Review A}\ }\textbf {\bibinfo {volume} {106}},\ \bibinfo {pages} {043723} (\bibinfo {year} {2022})}\BibitemShut {NoStop}%
\bibitem [{\citenamefont {Loudon}\ and\ \citenamefont {Barnett}(2006)}]{Loudon2006Jul}%
  \BibitemOpen
  \bibfield  {author} {\bibinfo {author} {\bibfnamefont {R.}~\bibnamefont {Loudon}}\ and\ \bibinfo {author} {\bibfnamefont {S.~M.}\ \bibnamefont {Barnett}},\ }\bibfield  {title} {\bibinfo {title} {{Theory of the linear polarizability of a two-level atom}},\ }\href {https://doi.org/10.1088/0953-4075/39/15/S04} {\bibfield  {journal} {\bibinfo  {journal} {J. Phys. B: At. Mol. Opt. Phys.}\ }\textbf {\bibinfo {volume} {39}},\ \bibinfo {pages} {S555} (\bibinfo {year} {2006})}\BibitemShut {NoStop}%
\bibitem [{\citenamefont {Wigner}(1931)}]{Wigner1931Jan}%
  \BibitemOpen
  \bibfield  {author} {\bibinfo {author} {\bibfnamefont {E.}~\bibnamefont {Wigner}},\ }\href@noop {} {\emph {\bibinfo {title} {{Gruppentheorie und ihre Anwendung auf die Quanten mechanik der Atomspektren}}}}\ (\bibinfo  {publisher} {Braunschweig, Germany: Friedrich Vieweg und Sohn},\ \bibinfo {year} {1931})\ pp.\ \bibinfo {pages} {251--254}\BibitemShut {NoStop}%
\bibitem [{\citenamefont {Tsimokha}\ \emph {et~al.}(2022)\citenamefont {Tsimokha}, \citenamefont {Igoshin}, \citenamefont {Nikitina}, \citenamefont {Toftul}, \citenamefont {Frizyuk},\ and\ \citenamefont {Petrov}}]{Tsimokha2022Apr}%
  \BibitemOpen
  \bibfield  {author} {\bibinfo {author} {\bibfnamefont {M.}~\bibnamefont {Tsimokha}}, \bibinfo {author} {\bibfnamefont {V.}~\bibnamefont {Igoshin}}, \bibinfo {author} {\bibfnamefont {A.}~\bibnamefont {Nikitina}}, \bibinfo {author} {\bibfnamefont {I.}~\bibnamefont {Toftul}}, \bibinfo {author} {\bibfnamefont {K.}~\bibnamefont {Frizyuk}},\ and\ \bibinfo {author} {\bibfnamefont {M.}~\bibnamefont {Petrov}},\ }\bibfield  {title} {\bibinfo {title} {{Acoustic resonators: Symmetry classification and multipolar content of the eigenmodes}},\ }\href {https://doi.org/10.1103/PhysRevB.105.165311} {\bibfield  {journal} {\bibinfo  {journal} {Phys. Rev. B}\ }\textbf {\bibinfo {volume} {105}},\ \bibinfo {pages} {165311} (\bibinfo {year} {2022})}\BibitemShut {NoStop}%
\bibitem [{\citenamefont {Gladyshev}\ \emph {et~al.}(2020)\citenamefont {Gladyshev}, \citenamefont {Frizyuk},\ and\ \citenamefont {Bogdanov}}]{Gladyshev2020Aug}%
  \BibitemOpen
  \bibfield  {author} {\bibinfo {author} {\bibfnamefont {S.}~\bibnamefont {Gladyshev}}, \bibinfo {author} {\bibfnamefont {K.}~\bibnamefont {Frizyuk}},\ and\ \bibinfo {author} {\bibfnamefont {A.}~\bibnamefont {Bogdanov}},\ }\bibfield  {title} {\bibinfo {title} {{Symmetry analysis and multipole classification of eigenmodes in electromagnetic resonators for engineering their optical properties}},\ }\href {https://doi.org/10.1103/PhysRevB.102.075103} {\bibfield  {journal} {\bibinfo  {journal} {Phys. Rev. B}\ }\textbf {\bibinfo {volume} {102}},\ \bibinfo {pages} {075103} (\bibinfo {year} {2020})}\BibitemShut {NoStop}%
\bibitem [{\citenamefont {Poleva}\ \emph {et~al.}(2023)\citenamefont {Poleva}, \citenamefont {Frizyuk}, \citenamefont {Baryshnikova}, \citenamefont {Evlyukhin}, \citenamefont {Petrov},\ and\ \citenamefont {Bogdanov}}]{Poleva2023Jan}%
  \BibitemOpen
  \bibfield  {author} {\bibinfo {author} {\bibfnamefont {M.}~\bibnamefont {Poleva}}, \bibinfo {author} {\bibfnamefont {K.}~\bibnamefont {Frizyuk}}, \bibinfo {author} {\bibfnamefont {K.}~\bibnamefont {Baryshnikova}}, \bibinfo {author} {\bibfnamefont {A.}~\bibnamefont {Evlyukhin}}, \bibinfo {author} {\bibfnamefont {M.}~\bibnamefont {Petrov}},\ and\ \bibinfo {author} {\bibfnamefont {A.}~\bibnamefont {Bogdanov}},\ }\bibfield  {title} {\bibinfo {title} {{Multipolar theory of bianisotropic response of meta-atoms}},\ }\href {https://doi.org/10.1103/PhysRevB.107.L041304} {\bibfield  {journal} {\bibinfo  {journal} {Phys. Rev. B}\ }\textbf {\bibinfo {volume} {107}},\ \bibinfo {pages} {L041304} (\bibinfo {year} {2023})}\BibitemShut {NoStop}%
\end{thebibliography}%


\begin{thebibliography}{11}%
\makeatletter
\providecommand \@ifxundefined [1]{%
 \@ifx{#1\undefined}
}%
\providecommand \@ifnum [1]{%
 \ifnum #1\expandafter \@firstoftwo
 \else \expandafter \@secondoftwo
 \fi
}%
\providecommand \@ifx [1]{%
 \ifx #1\expandafter \@firstoftwo
 \else \expandafter \@secondoftwo
 \fi
}%
\providecommand \natexlab [1]{#1}%
\providecommand \enquote  [1]{``#1''}%
\providecommand \bibnamefont  [1]{#1}%
\providecommand \bibfnamefont [1]{#1}%
\providecommand \citenamefont [1]{#1}%
\providecommand \href@noop [0]{\@secondoftwo}%
\providecommand \href [0]{\begingroup \@sanitize@url \@href}%
\providecommand \@href[1]{\@@startlink{#1}\@@href}%
\providecommand \@@href[1]{\endgroup#1\@@endlink}%
\providecommand \@sanitize@url [0]{\catcode `\\12\catcode `\$12\catcode `\&12\catcode `\#12\catcode `\^12\catcode `\_12\catcode `\%12\relax}%
\providecommand \@@startlink[1]{}%
\providecommand \@@endlink[0]{}%
\providecommand \url  [0]{\begingroup\@sanitize@url \@url }%
\providecommand \@url [1]{\endgroup\@href {#1}{\urlprefix }}%
\providecommand \urlprefix  [0]{URL }%
\providecommand \Eprint [0]{\href }%
\providecommand \doibase [0]{https://doi.org/}%
\providecommand \selectlanguage [0]{\@gobble}%
\providecommand \bibinfo  [0]{\@secondoftwo}%
\providecommand \bibfield  [0]{\@secondoftwo}%
\providecommand \translation [1]{[#1]}%
\providecommand \BibitemOpen [0]{}%
\providecommand \bibitemStop [0]{}%
\providecommand \bibitemNoStop [0]{.\EOS\space}%
\providecommand \EOS [0]{\spacefactor3000\relax}%
\providecommand \BibitemShut  [1]{\csname bibitem#1\endcsname}%
\let\auto@bib@innerbib\@empty
\bibitem [{\citenamefont {Asenjo-Garcia}\ \emph {et~al.}(2017)\citenamefont {Asenjo-Garcia}, \citenamefont {Moreno-Cardoner}, \citenamefont {Albrecht}, \citenamefont {Kimble},\ and\ \citenamefont {Chang}}]{Asenjo-Garcia2017}%
  \BibitemOpen
  \bibfield  {author} {\bibinfo {author} {\bibfnamefont {A.}~\bibnamefont {Asenjo-Garcia}}, \bibinfo {author} {\bibfnamefont {M.}~\bibnamefont {Moreno-Cardoner}}, \bibinfo {author} {\bibfnamefont {A.}~\bibnamefont {Albrecht}}, \bibinfo {author} {\bibfnamefont {H.~J.}\ \bibnamefont {Kimble}},\ and\ \bibinfo {author} {\bibfnamefont {D.~E.}\ \bibnamefont {Chang}},\ }\bibfield  {title} {\bibinfo {title} {{Exponential Improvement in Photon Storage Fidelities Using Subradiance and ``Selective Radiance'' in Atomic Arrays}},\ }\href {https://doi.org/10.1103/PhysRevX.7.031024} {\bibfield  {journal} {\bibinfo  {journal} {Phys. Rev. X}\ }\textbf {\bibinfo {volume} {7}},\ \bibinfo {pages} {031024} (\bibinfo {year} {2017})}\BibitemShut {NoStop}%
\bibitem [{\citenamefont {Novotny}\ and\ \citenamefont {Hecht}(2012)}]{Novotny2012}%
  \BibitemOpen
  \bibfield  {author} {\bibinfo {author} {\bibfnamefont {L.}~\bibnamefont {Novotny}}\ and\ \bibinfo {author} {\bibfnamefont {B.}~\bibnamefont {Hecht}},\ }\href {http://www.amazon.com/Principles-Nano-Optics-Lukas-Novotny/dp/1107005469} {\emph {\bibinfo {title} {Principles of {Nano}-{Optics}}}}\ (\bibinfo  {publisher} {Cambridge University Press},\ \bibinfo {year} {2012})\BibitemShut {NoStop}%
\bibitem [{\citenamefont {Freedhoff}(1986)}]{freedhoff_cooperative_1986}%
  \BibitemOpen
  \bibfield  {author} {\bibinfo {author} {\bibfnamefont {H.~S.}\ \bibnamefont {Freedhoff}},\ }\bibfield  {title} {\bibinfo {title} {{Cooperative single{-}quantum excitations of a closed{-}ring polymer chain}},\ }\href {https://doi.org/10.1063/1.451476} {\bibfield  {journal} {\bibinfo  {journal} {J. Chem. Phys.}\ }\textbf {\bibinfo {volume} {85}},\ \bibinfo {pages} {6110} (\bibinfo {year} {1986})}\BibitemShut {NoStop}%
\bibitem [{\citenamefont {Moreno-Cardoner}\ \emph {et~al.}(2019)\citenamefont {Moreno-Cardoner}, \citenamefont {Plankensteiner}, \citenamefont {Ostermann}, \citenamefont {Chang},\ and\ \citenamefont {Ritsch}}]{moreno-cardoner_subradiance-enhanced_2019}%
  \BibitemOpen
  \bibfield  {author} {\bibinfo {author} {\bibfnamefont {M.}~\bibnamefont {Moreno-Cardoner}}, \bibinfo {author} {\bibfnamefont {D.}~\bibnamefont {Plankensteiner}}, \bibinfo {author} {\bibfnamefont {L.}~\bibnamefont {Ostermann}}, \bibinfo {author} {\bibfnamefont {D.~E.}\ \bibnamefont {Chang}},\ and\ \bibinfo {author} {\bibfnamefont {H.}~\bibnamefont {Ritsch}},\ }\bibfield  {title} {\bibinfo {title} {{Subradiance-enhanced excitation transfer between dipole-coupled nanorings of quantum emitters}},\ }\href {https://doi.org/10.1103/PhysRevA.100.023806} {\bibfield  {journal} {\bibinfo  {journal} {Phys. Rev. A}\ }\textbf {\bibinfo {volume} {100}},\ \bibinfo {pages} {023806} (\bibinfo {year} {2019})}\BibitemShut {NoStop}%
\bibitem [{\citenamefont {Cremer}\ \emph {et~al.}(2020)\citenamefont {Cremer}, \citenamefont {Plankensteiner}, \citenamefont {Moreno-Cardoner}, \citenamefont {Ostermann},\ and\ \citenamefont {Ritsch}}]{cremer_polarization_2020}%
  \BibitemOpen
  \bibfield  {author} {\bibinfo {author} {\bibfnamefont {J.}~\bibnamefont {Cremer}}, \bibinfo {author} {\bibfnamefont {D.}~\bibnamefont {Plankensteiner}}, \bibinfo {author} {\bibfnamefont {M.}~\bibnamefont {Moreno-Cardoner}}, \bibinfo {author} {\bibfnamefont {L.}~\bibnamefont {Ostermann}},\ and\ \bibinfo {author} {\bibfnamefont {H.}~\bibnamefont {Ritsch}},\ }\bibfield  {title} {\bibinfo {title} {{Polarization control of radiation and energy flow in dipole-coupled nanorings}},\ }\href {https://doi.org/10.1088/1367-2630/aba4d4} {\bibfield  {journal} {\bibinfo  {journal} {New J. Phys.}\ }\textbf {\bibinfo {volume} {22}},\ \bibinfo {pages} {083052} (\bibinfo {year} {2020})}\BibitemShut {NoStop}%
\bibitem [{\citenamefont {Wigner}(1931)}]{Wigner1931Jan}%
  \BibitemOpen
  \bibfield  {author} {\bibinfo {author} {\bibfnamefont {E.}~\bibnamefont {Wigner}},\ }\href@noop {} {\emph {\bibinfo {title} {{Gruppentheorie und ihre Anwendung auf die Quanten mechanik der Atomspektren}}}}\ (\bibinfo  {publisher} {Braunschweig, Germany: Friedrich Vieweg und Sohn},\ \bibinfo {year} {1931})\ pp.\ \bibinfo {pages} {251--254}\BibitemShut {NoStop}%
\bibitem [{\citenamefont {Tsimokha}\ \emph {et~al.}(2022)\citenamefont {Tsimokha}, \citenamefont {Igoshin}, \citenamefont {Nikitina}, \citenamefont {Toftul}, \citenamefont {Frizyuk},\ and\ \citenamefont {Petrov}}]{Tsimokha2022Apr}%
  \BibitemOpen
  \bibfield  {author} {\bibinfo {author} {\bibfnamefont {M.}~\bibnamefont {Tsimokha}}, \bibinfo {author} {\bibfnamefont {V.}~\bibnamefont {Igoshin}}, \bibinfo {author} {\bibfnamefont {A.}~\bibnamefont {Nikitina}}, \bibinfo {author} {\bibfnamefont {I.}~\bibnamefont {Toftul}}, \bibinfo {author} {\bibfnamefont {K.}~\bibnamefont {Frizyuk}},\ and\ \bibinfo {author} {\bibfnamefont {M.}~\bibnamefont {Petrov}},\ }\bibfield  {title} {\bibinfo {title} {{Acoustic resonators: Symmetry classification and multipolar content of the eigenmodes}},\ }\href {https://doi.org/10.1103/PhysRevB.105.165311} {\bibfield  {journal} {\bibinfo  {journal} {Phys. Rev. B}\ }\textbf {\bibinfo {volume} {105}},\ \bibinfo {pages} {165311} (\bibinfo {year} {2022})}\BibitemShut {NoStop}%
\bibitem [{\citenamefont {Gladyshev}\ \emph {et~al.}(2020)\citenamefont {Gladyshev}, \citenamefont {Frizyuk},\ and\ \citenamefont {Bogdanov}}]{Gladyshev2020Aug}%
  \BibitemOpen
  \bibfield  {author} {\bibinfo {author} {\bibfnamefont {S.}~\bibnamefont {Gladyshev}}, \bibinfo {author} {\bibfnamefont {K.}~\bibnamefont {Frizyuk}},\ and\ \bibinfo {author} {\bibfnamefont {A.}~\bibnamefont {Bogdanov}},\ }\bibfield  {title} {\bibinfo {title} {{Symmetry analysis and multipole classification of eigenmodes in electromagnetic resonators for engineering their optical properties}},\ }\href {https://doi.org/10.1103/PhysRevB.102.075103} {\bibfield  {journal} {\bibinfo  {journal} {Phys. Rev. B}\ }\textbf {\bibinfo {volume} {102}},\ \bibinfo {pages} {075103} (\bibinfo {year} {2020})}\BibitemShut {NoStop}%
\bibitem [{\citenamefont {Poleva}\ \emph {et~al.}(2023)\citenamefont {Poleva}, \citenamefont {Frizyuk}, \citenamefont {Baryshnikova}, \citenamefont {Evlyukhin}, \citenamefont {Petrov},\ and\ \citenamefont {Bogdanov}}]{Poleva2023Jan}%
  \BibitemOpen
  \bibfield  {author} {\bibinfo {author} {\bibfnamefont {M.}~\bibnamefont {Poleva}}, \bibinfo {author} {\bibfnamefont {K.}~\bibnamefont {Frizyuk}}, \bibinfo {author} {\bibfnamefont {K.}~\bibnamefont {Baryshnikova}}, \bibinfo {author} {\bibfnamefont {A.}~\bibnamefont {Evlyukhin}}, \bibinfo {author} {\bibfnamefont {M.}~\bibnamefont {Petrov}},\ and\ \bibinfo {author} {\bibfnamefont {A.}~\bibnamefont {Bogdanov}},\ }\bibfield  {title} {\bibinfo {title} {{Multipolar theory of bianisotropic response of meta-atoms}},\ }\href {https://doi.org/10.1103/PhysRevB.107.L041304} {\bibfield  {journal} {\bibinfo  {journal} {Phys. Rev. B}\ }\textbf {\bibinfo {volume} {107}},\ \bibinfo {pages} {L041304} (\bibinfo {year} {2023})}\BibitemShut {NoStop}%
\bibitem [{\citenamefont {Loudon}\ and\ \citenamefont {Barnett}(2006)}]{Loudon2006Jul}%
  \BibitemOpen
  \bibfield  {author} {\bibinfo {author} {\bibfnamefont {R.}~\bibnamefont {Loudon}}\ and\ \bibinfo {author} {\bibfnamefont {S.~M.}\ \bibnamefont {Barnett}},\ }\bibfield  {title} {\bibinfo {title} {{Theory of the linear polarizability of a two-level atom}},\ }\href {https://doi.org/10.1088/0953-4075/39/15/S04} {\bibfield  {journal} {\bibinfo  {journal} {J. Phys. B: At. Mol. Opt. Phys.}\ }\textbf {\bibinfo {volume} {39}},\ \bibinfo {pages} {S555} (\bibinfo {year} {2006})}\BibitemShut {NoStop}%
\bibitem [{\citenamefont {Ke}\ \emph {et~al.}(2019)\citenamefont {Ke}, \citenamefont {Poshakinskiy}, \citenamefont {Lee}, \citenamefont {Kivshar},\ and\ \citenamefont {Poddubny}}]{Ke2019Dec}%
  \BibitemOpen
  \bibfield  {author} {\bibinfo {author} {\bibfnamefont {Y.}~\bibnamefont {Ke}}, \bibinfo {author} {\bibfnamefont {A.~V.}\ \bibnamefont {Poshakinskiy}}, \bibinfo {author} {\bibfnamefont {C.}~\bibnamefont {Lee}}, \bibinfo {author} {\bibfnamefont {Y.~S.}\ \bibnamefont {Kivshar}},\ and\ \bibinfo {author} {\bibfnamefont {A.~N.}\ \bibnamefont {Poddubny}},\ }\bibfield  {title} {\bibinfo {title} {{Inelastic Scattering of Photon Pairs in Qubit Arrays with Subradiant States}},\ }\href {https://doi.org/10.1103/PhysRevLett.123.253601} {\bibfield  {journal} {\bibinfo  {journal} {Phys. Rev. Lett.}\ }\textbf {\bibinfo {volume} {123}},\ \bibinfo {pages} {253601} (\bibinfo {year} {2019})}\BibitemShut {NoStop}%
\end{thebibliography}%

\end{document}


\title{SUPPLEMENTAL MATERIAL: \\ Nonradiant multiphoton states in quantum ring oligomers}

\author{Nikita Ustimenko}
 \affiliation{\affilITMO}

\author{Danil Kornovan}%

\affiliation{\affilITMO}%

\author{Ilya Volkov}%

\affiliation{\affilITMO}%

\author{Alexandra  Sheremet}
\affiliation{\affilITMO}%
\author{Roman Savelev}

\affiliation{\affilITMO}%
\author{Mihail Petrov}
 \email{m.petrov@metalab.ifmo.ru}
\affiliation{\affilITMO}
\maketitle

\section{Effective Schr\"{o}dinger equations for singly and doubly excited collective eigenstates}
Quantum states of an array of $N$ two-level dipole emitters in free space placed in the $xy$ plane and oriented along the $z$ axis form a Hilbert space $\mathcal{H}$ and are governed by the following effective Hamiltonian ($\hbar = 1$)~\cite{Asenjo-Garcia2017}:
\begin{gather}
\label{eq_Heff}
    \widehat{H}_{\mathrm{eff}} =  - i \dfrac{\gamma_{0}}{2}  \sum\limits_{k = 1}^N  \sigd_k \sig_k + \sum \limits_{k=1}^N \sum\limits_{\substack{l=1, \\ l\neq k}}^N  g(|\mathbf{r}_{kl}|, \omega_0) \sigd_k \sig_l,
\end{gather}
where $\sigd_k$ ($\sig_k$) is the creation (annihilation) operator for excitation on emitter $k$, $\left|\mathbf{r}_{kl}\right| = \left|\mathbf{r}_k - \mathbf{r}_l\right|$ is the relative distance between emitters $k$ and $l$, $g(|\mathbf{r}|, \omega_0) = \left(-3\gamma_0 \pi/k_0 \right) \mathbf{e}_z^{T} \cdot \bm{\mathsf{G}}_0(\mathbf{r}, \omega_0)\cdot \mathbf{e}_z$ is the coupling rate between transversely oriented emitters $k$ and $l$ under the Born-Markov approximation, $\gamma_0$ is the decay rate of a single emitter, and $k_0$ is the wavenumber in vacuum related to the resonant angular frequency $\omega_0$ and the resonant wavelength $\lambda_0$ of a single emitter as $k_0 = \omega_0/c = 2 \pi /\lambda_0$. Electromagnetic Green's tensor in free space reads as~\cite{Novotny2012}:
\begin{gather}
\label{eq_G0}
  \bm{\mathsf{G}}_0(\mathbf{r}, \omega) = \frac{e^{i k |\mathbf{r}|}}{4 \pi |\mathbf{r}|} \left\{ \left(1 + \frac{i}{k|\mathbf{r}|}- \frac{1}{k^2|\mathbf{r}|^2} \right)\bm{\mathsf{I}} + \left(-1 - \frac{i3}{k|\mathbf{r}|}+ \frac{3}{k^2|\mathbf{r}|^2}\right) \frac{\mathbf{r} \otimes \mathbf{r}}{|\mathbf{r}|^2}\right\},   
\end{gather}
where $k = \omega/c$, $\bm{\mathsf{I}}$ is $3 \times 3$ identity matrix.

One can note that effective Hamiltonian~\eqref{eq_Heff} preserves the total number of excitations in the system. Thus, the whole Hilbert space $\mathcal{H}$ of states of the system can be decomposed into a series of manifolds:
\begin{gather}
    \mathcal{H} = \mathcal{H}_0 + \mathcal{H}_1 + \mathcal{H}_2 + \mathcal{H}_3 + ... + \mathcal{H}_N,
\end{gather}
where $\mathcal{H}_n$ is the manifold corresponding to the case of $n$ excitations in the system with a dimension $N!/\left(n!(N-n)!\right)$. For the last term in the expansion, $n = N$ because $N$ two-level emitters cannot contain more than $N$ excitations.

Let us consider the several cases of different numbers of excitations $n$.
\begin{enumerate}
    \item $n = 0$: There are no excitations in the system, therefore, all $N$ emitters occupy the ground state. Thus, manifold $\mathcal{H}_0$ consists only of one state $\ket{g}^{\otimes N} \equiv \underbrace{\ket{g} \otimes \ket{g} \otimes ... \otimes \ket{g}}_{N \ \text{times}}$ with a zero energy usually referred to in the literature as a ground state of the array;
    \item $n = 1$: In this case, manifold $\mathcal{H}_1$ consists of $N$ singly excited states that have wave function of the following form 
    $\ket{\psi} = \sum\limits_{k=1}^N c_k \hat{\sigma}_k^{\dagger} \ket{g}^{\otimes N}$.
    Let us plug this anzatz into Schr\"{o}dinger equation $\widehat{H}_{\mathrm{eff}}\ket{\psi} = \varepsilon \ket{\psi}$. It yields a set of $N$ linear equations on the single-excitation probability amplitudes:
    \begin{gather}
    \label{eq_1exc}
    \varepsilon \times c_k = - i\frac{\gamma_0}{2}c_k +  \sum\limits_{\substack{l=1, \\ l\neq k}}^N  g(\left|\mathbf{r}_{kl}\right|, \omega_0) c_l,
    \end{gather}
    with the eigenenergy $\varepsilon = \omega - \omega_0 - i \dfrac{\gamma}{2}$. Thus, there are $N$ eigenvalues $\varepsilon$ corresponding to $N$ right eigenvectors written in the form $\left(c_1, c_2, \hdots, c_N \right)^T$. 
    \item $n = 2$: A doubly excited state is governed by the following wave function $\ket{\Psi} = \sum\limits_{k = 1}^N\sum\limits_{l = k + 1}^N c_{kl} \hat{\sigma}_k^{\dagger} \hat{\sigma}_l^{\dagger} \ket{g}^{\otimes N}$.
     The limits in sums are taken because of two following facts. The considered emitters have only two energy levels, hence, two excitations cannot occupy the same emitter and $k \neq l$.  Moreover, the emitters and excitations are identical, hence, a double-excitation probability amplitude is symmetric $c_{kl} = c_{kl}$. Combining these two properties of $c_{kl}$, one can consider only the amplitudes with $l > k$. Thus, for a system of $N$ two-level emitters, a total number of doubly excited eigenstates is $N(N-1)/2$,  which can be calculated by diagonalizing the following system of $N(N-1)/2$ linear equations:
    \begin{gather}
    \label{eq_2exc}
    \mathcal{E} \times c_{kl} = - i \gamma_0 c_{kl} + \sum\limits_{k' \neq k} g(\left|\mathbf{r}_{k'l}\right|,\omega_0) c_{kk'} +  \sum\limits_{k' \neq l} g(\left|\mathbf{r}_{kk'}\right|,\omega_0) c_{k'l} ,
    \end{gather}
    where $\mathcal{E} = \omega - \omega_0 - i \dfrac{\Gamma}{2}$.
\end{enumerate}

\section{Singly excited eigenstates and their spectrum for a ring of dipoles. Orbital quasimomentum}\label{sec:1_ring}

\begin{figure}[h!]
    \centering   \includegraphics{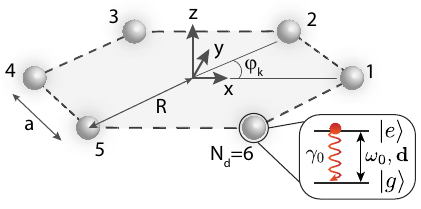}
    \caption{Single ring of $N_d = 6$ two-level dipole emitters placed in the $xy$ plane.}
    \label{fig_one_ring}
\end{figure}

Based on Refs.~\cite{freedhoff_cooperative_1986,moreno-cardoner_subradiance-enhanced_2019,cremer_polarization_2020}, we describe the singly excited eigenstates and their spectrum for a single ring of $N_d$ two-level emitters having a transition dipole moment $\mathbf{d} = \left| \mathbf{d} \right|\mathbf{e}_z$, see Fig.~\ref{fig_one_ring}. In this case, the coordinates of emitters along a circle of radius $R$ read as $\mathbf{r}_k = R \left( \cos \varphi_k, \sin \varphi_k, 0 \right)^T,$
where $k$ runs from 1 to $N_d$ and $\varphi_k = (k - 1) 2 \pi/N_d$ is the angular coordinate of emitter $k$. Thus, the separation distance between neighboring emitters in the ring equals $a = 2 R \sin \left( \pi/ N_d \right)$.

To calculate the single excitation spectrum for the single ring of emitters, we need to use Eq.~\eqref{eq_1exc}. Since the ring of \textit{transversely oriented}, out-of-plane dipoles preserves the rotational symmetry along the $z$ axis corresponding to {$D_{N_d, h}$} symmetry group, the eigenstates of the ring can be described by the orbital quasimomentum $m$. The meaning of $m$ is the following: for the ring eigenstate with a given $m$, the phase shift between the amplitudes of neighboring emitters, $c_{k + 1}^{(m)}$ and $c_k^{(m)}$, equals $\Delta \varphi(m) = 2 \pi m/N_d$. Momenta $m$ and $(m + n N_d)$, where $n$ is an integer, correspond to the same phase shifts $\Delta \varphi(m + n N_d) = \Delta \varphi(m) + 2\pi n$. Therefore, the prefix ``quasi'' in the name of the $m$ indicates that it is determined up to the number of emitters in the ring $N_d$. Hence, one can introduce ``the first Brillouin zone'' for the $m$ as a set $\left\{0, \pm 1, \pm 2, ..., \pm \left(N_d - 2 \right)/2, N_d / 2 \right\}$ for even $N_d$, or $\left\{0, \pm 1, \pm 2, ..., \pm \left(N_d - 2 \right)/2, \pm \left(N_d - 1 \right)/2 \right\}$ for odd $N_d$. Thus, the number of singly excited eigenstates with a given $m$ for a single ring $N_m = 1$. Note that, for even $N_d$, a maximum orbital quasimomentum but with an opposite sign $m = -N_d/2$ corresponds to the same phase shift between emitters of $\Delta \varphi\left(m = \pm N_d /2 \right) = \pi$. For convenience, we choose a positive sign $m = N_d/2$. Note also that a structure consisting of $N_r$ rings with the same number of emitters $N_d$ has $N_r$ singly excited states with each accessible $m$, i.e., $N_m = N_r$.

Thus, the spectrum of a single excitation in the ring can be calculated using the following ansatz for the excitation amplitudes $c_k^{(m)}$: 
\begin{gather}
\label{eq_ansatz_m}
    c_k^{(m)} =  \dfrac{1}{\sqrt{N_d}} e^{i m \varphi_k},
\end{gather}
where the pre-factor $1/\sqrt{N_d}$ is obtained upon the normalization of amplitudes, i.e., $\sum\limits_{k=1}^{N_d} \left| c_k \right|^2 = 1$. Hence,  the singly excited eigenstates of the ring read as
\begin{gather}
\label{1exc_ring_kets}
    \ket{\psi^{(m)}_{\text{ring}}} = \dfrac{1}{\sqrt{N_d}} \sum\limits_{k=1}^{N_d} e^{i m \varphi_k} \hat{\sigma}_k^{\dagger} \ket{g}^{\otimes N_d}.
\end{gather}
After inserting Eq.~\eqref{eq_ansatz_m} in Eq.~\eqref{eq_1exc}, one obtains a set of shifted (by $-\omega_0$) corresponding eigenenergies:
\begin{gather}
\label{eq_1exc_ring_energy}  \varepsilon^{(m)}_{\text{ring}} = - i\frac{\gamma_0}{2} + \sum\limits_{k = 2}^{N_d} g(\left|\mathbf{r}_{1k}\right|, \omega_0) e^{i m \varphi_k}.
\end{gather}
For the ring of $N_d = 6$ emitters, the ring radius is $R = a$ and the dipole sum can be evaluated as
\begin{gather}
\label{eq_dipole_sums_6}
\begin{gathered}
     \sum\limits_{k = 2}^{6} g(\left|\mathbf{r}_{1k}\right|, \omega_0) e^{i m \varphi_k} = 2 \left[g\left(a,\omega_0\right) \cos\left( \frac{\pi m}{3}\right) \right. \\ \left. + g\left(\sqrt{3} a,\omega_0\right) \cos\left( m \frac{2\pi}{3}\right) + \frac{1}{2}g\left(2 a,\omega_0\right) \cos\left( \pi m\right) \right].
\end{gathered}
\end{gather}
One can notice that dipole sums~\eqref{eq_dipole_sums_6} are symmetric with respect to a sign of the $m$ that can be shown for any number of emitters $N_d$. Thus, the exchange of the quasimomentum sign $m \leftrightarrow -m$ preserves the energy $\varepsilon^{(m)} = \varepsilon^{(-m)}$. Hence, the ring eigenstates are doubly degenerate except for two cases. The state with $m = 0$ is nondegenerate for any $N_d$, and the state with $m = N_d/2$ also has no degeneracy for even $N_d$.   

\section{Number of doubly excited eigenstates in multiring structures}
\begin{figure}[h!]
    \centering
    \includegraphics[scale = 0.5]{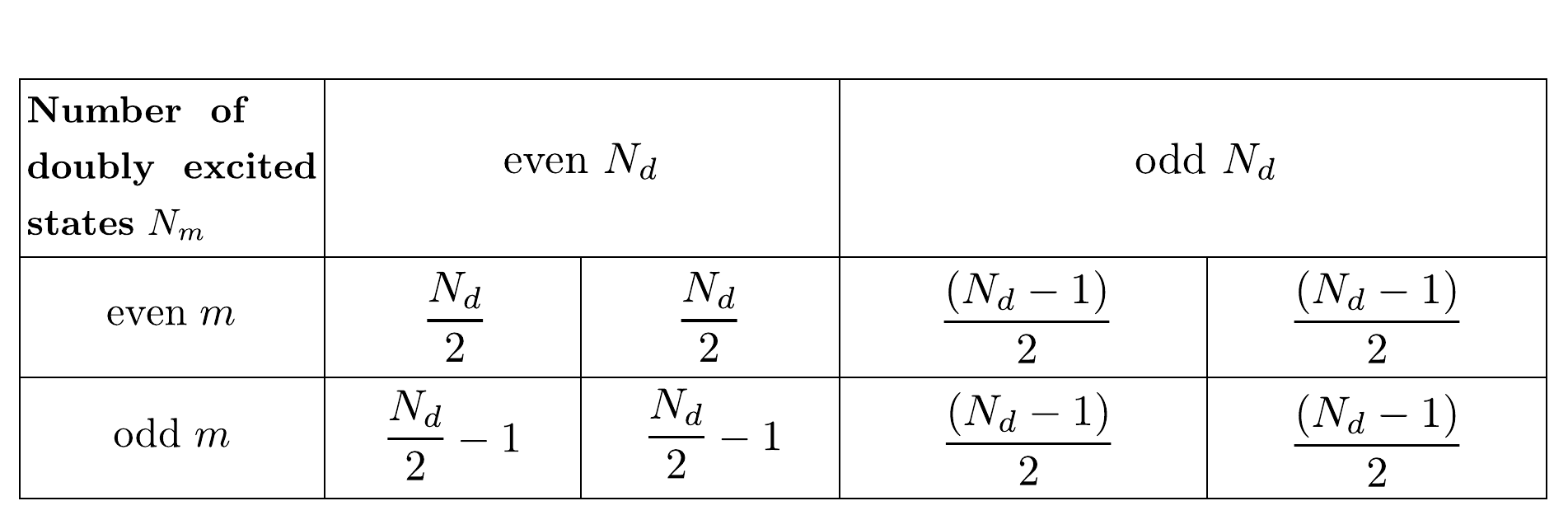}
    \caption{Number of doubly excited eigenstates $N_m$ in a single ring of $N_d$ emitters $N_m$ for a given value of orbital quasimomentum $m$ depending on the parity of $m$ (rows) and $N_d$ (columns).}
    \label{fig_number_of_m_states}
\end{figure}

Considering the rotational symmetry of the ring of transverse dipole emitters, we identify the number of doubly excited eigenstates with a given quasimomentum $m$ and summarize the results in Fig.~\ref{fig_number_of_m_states}. 
Using Fig.~\ref{fig_number_of_m_states}, one can also find that $N_r$ rings with the same number of emitters $N_d$ support the following number of doubly excited eigenstates with a given $m$. For \textit{even} $N_d$, $N_m$ is
\begin{gather}
\label{eq_Nm_even}
    N_m = 
    \begin{cases}
     N_r \times \dfrac{N_d}{2} + \dfrac{N_r(N_r - 1)}{2} \times N_d& \text{even} \  m, \\
     N_r \times \left(\dfrac{N_d}{2} - 1 \right) + \dfrac{N_r(N_r - 1)}{2} \times N_d& \text{odd} \ m.
    \end{cases}
\end{gather}
For \textit{odd} $N_d$:
\begin{gather}
\label{eq_Nm_odd}
    N_m = 
     N_r \times \dfrac{(N_d - 1)}{2} + \dfrac{N_r(N_r - 1)}{2} \times N_d.
\end{gather}

\section{Correspondence between singly excited eigenstates of the ring and irreducible representations of $C_{6v}$ symmetry group}\label{sec:momentum}

\begin{figure}[h!]
    \centering
    \includegraphics[scale = 0.72]{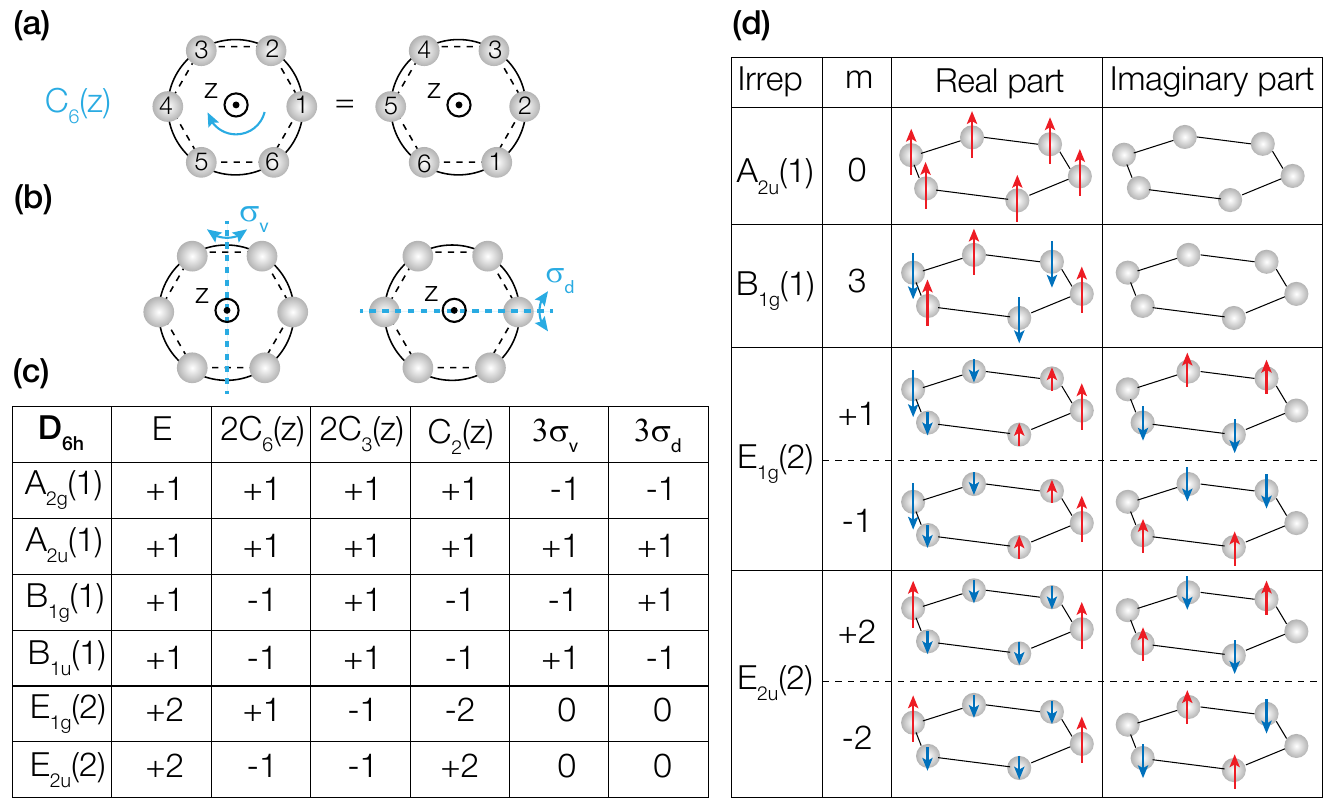}
    \caption{(a) Altering of emitter indices after rotating the ring around the $z$ axis by angle $2 \pi/N_d$. (b) Reflection of the ring with respect to the axis passing between atoms (left) and through atoms (right). (c) Fragment of the table of characters for {$D_{6h}$} symmetry group. The columns contain the symmetry transformations, the rows correspond to the irreducible representations (irreps) with their dimensions in the brackets. The cells show the traces (characters) of irreducible representations for a given transformation. {The indices $g$ and $u$ indicate the even and odd irreps relative to the point reflection (inversion) across the coordinate system origin (from German ``gerade'' and ``ungerade'').} (d) Correspondence of the orbital quasimomentum $m$ of singly excited states~\eqref{1exc_ring_kets} to the irreps of symmetry group for the ring of six emitters {$D_{6h}$}. The arrows show the real part $\propto \cos(m\varphi)$ and the imaginary part $\propto \sin(m\varphi)$ of the states. For clarity, the values with the same magnitude, but with different signs, are highlighted by different colors.}
    \label{fig:charachters}
\end{figure}

Let us look at how states~\eqref{1exc_ring_kets} are transformed under the rotation of the ring in the $xy$ plane by $\varphi' = 2 \pi / N_d$ angle around the $z$ axis  as shown in Fig.~\ref{fig:charachters}(a). One can associate this transformation with an operator of rotation $\widehat{R}(\varphi')$. Note that such a rotation does the following permutation of indices for the excitation amplitudes: $k \to k' = k + 1 \ (\text{mod} \ N_d)$, see Fig.~\ref{fig:charachters}(a). Hence, matrix of $\widehat{R}(\varphi')$ for the basis $\{ \hat{\sigma}_k^{\dagger} \ket{g}^{\otimes N_d} \}_{k = 1}^{N_d}$ has the following entries: $R_{kl}(\varphi') = \delta_{l,k'}$ for $k,l \in \{1..N_d\}$, where $\delta_{l,k'}$ is the Kronecker delta. From Eq.~\eqref{eq_ansatz_m}, it follows that $c_{k + 1} = e^{i m \varphi'} c_k$. Hence, one can write the action of $\widehat{R}(\varphi')$ operator on eigenstates~\eqref{1exc_ring_kets} as follows
$\widehat{R}(\varphi') \ket{\psi^{(m)}_{\text{ring}}} = e^{i m \varphi'} \ket{\psi^{(m)}_{\text{ring}}}.$ Thus, the rotation operator $\widehat{R}$ and effective Hamiltonian $\widehat{H}_{\mathrm{eff}}$ are diagonalizable in the same basis~\eqref{1exc_ring_kets} with the eigenvalue sets $\left\{ e^{i m \varphi'}\right\}$ and $\left\{ \varepsilon^{(m)}\right\}$, respectively. This is reflected in a fact that the operators commute $\widehat{H}_{\mathrm{eff}} \widehat{R} = \widehat{R}\widehat{H}_{\mathrm{eff}}$.

Note that the ring rotation by $\varphi' = 2 \pi / N_d$ angle is included in transformations from symmetry group of the ring of transverse emitters {$D_{N_d,h}$}. For example, symmetry group for the ring of six emitters {$D_{6h}$} consists, among other things, of the rotation by $\pi/3$ angle around the $z$ axis, denoted as $C_6(z)$ in Fig.~\ref{fig:charachters}(a), the rotation $C_3(z)$ by $2\pi/3$ angle, the rotation $C_2(z)$ by $\pi$ angle, and two mirror reflections $\sigma_v$ and $\sigma_d$ relative to the $y$- and $x$-axes, respectively, as shown in Fig.~\ref{fig:charachters}(b). {Note that all considered singly excited eigenstates are anti-symmetric with respect to mirror reflection $\sigma_h(xy)$ across the horizontal $xy$ plane.} All operators corresponding to these transformations can be diagonalized in the same basis $\ket{\psi^{(m)}}$ of eigenstates of $\widehat{H}_{\mathrm{eff}}$. 
Therefore, $\widehat{H}_{\mathrm{eff}}$ commutes with all transformations of the symmetry group for the ring ensembles. Here, we can notice a deep connection of physics with group theory formulated by so-called Wigner's theorem~\cite{Wigner1931Jan}. In simple words, this theorem can be formulated for the Schr\"{o}dinger equation as follows. If a Hamiltonian remains invariant under transformations from the symmetry group, the eigenstates of such a Hamiltonian, which are solutions of the Schr\"{o}dinger equation, are transformed by irreducible representations of the symmetry group.

Let us classify the singly excited eigenstates of six emitters per ring, corresponding to {$D_{6h}$} symmetry group, using the symmetry analysis as in Refs.~\cite{Tsimokha2022Apr,Gladyshev2020Aug,Poleva2023Jan}. Fig.~\ref{fig:charachters}(c) presents a character table for {$D_{6h}$} group. The irreducible representations are essentially the matrices acting on a column of functions $\left( \cos(m \varphi), \sin(m \varphi) \right)^T$ under the symmetry transformations, and the character is a trace of such a matrix. The excitation amplitudes of ring eigenstates~\eqref{1exc_ring_kets} have real part $\propto \cos(m \varphi_k)$ and imaginary part $\propto \sin(m \varphi_k)$, where $\varphi_k = (k - 1) \pi / 3$, as shown in Fig.~\ref{fig:charachters}(d). It is obvious that the imaginary part is zero for $m = 0$ and $m = 3$, therefore, these states should enter one-dimensional representations. For them, we just need to multiply a given state by the corresponding character to obtain a transformed state. The state with $m = 0$ has a uniform distribution of amplitudes, therefore, it remains the same under the symmetry transformations and enters irreducible representation {$A_{2u}$}, see Fig.~\ref{fig:charachters}(c). The state with $m = 3$ is symmetric for reflection $\sigma_d$ and antisymmetric for rotation $C_6(z)$, therefore, it enters {$B_{1g}$} irreducible representation, see Fig.~\ref{fig:charachters}(c). The states with $m = \pm 1$ and $m = \pm 2$ have both nonzero real and imaginary parts. Hence, they enter two-dimensional representations, {$E_{1g}$} or {$E_{2u}$}, and they are double degenerate [cf. Eq.~\eqref{eq_1exc_ring_energy}]. For these states, the real and imaginary parts have opposite symmetry with respect to the reflections, therefore, they have zeros at corresponding cells in Fig.~\ref{fig:charachters}(c). Hence, let us consider rotation $C_2(z)$. Irreducible representations {$E_{1g}$} and {$E_{2u}$} are $2 \times 2$ \textit{diagonal} matrices for this rotation, therefore, negative character (trace) ``-2'' corresponds to both anti-symmetric real and imaginary parts, while positive character ``+2'' is for both symmetric real and imaginary parts. The real and imaginary parts of states with $m = \pm 1$ (resp. $m = \pm 2$) are both anti-symmetric (resp. symmetric) with respect to this rotation, therefore, the states with $m = \pm 1$ (resp. $m = \pm 2$) enter {$E_{1g}$} (resp. {$E_{2u}$}) irreducible representation. 

Thus, singly excited states of the ring of six emitters enter irreducible representations {$A_{2u}$, $E_{1g}$, $E_{2u}$, and $B_{1g}$}. Note that doubly excited states can enter other irreducible representations [cf. states given by Eq.~(2) in the main text].

\section{Lifetime for the eigenstates of the ring}\label{sec:subradiance_1ring}
In this section, we present the decay rates for singly and doubly excited eigenstates of the ring shown in Fig.~\ref{fig_one_ring}.  We recall that the obtained normalized and shifted complex eigenvalue of the eigenstate with $m$ is numerically evaluated as $\varepsilon^{(m)}/\gamma_0 = \left( \omega^{(m)} - \omega_0\right)/\gamma_0  - i \gamma^{(m)}/\left(2 \gamma_0\right)$, where the real part stands for the resonant frequency detuning and the imaginary part is for the normalized collective decay rate. Note that ratio $\gamma/\gamma_0$ represents the inverse radiative lifetime enhancement. 

\begin{figure}[h!]
    \centering
    \includegraphics[scale=0.55]{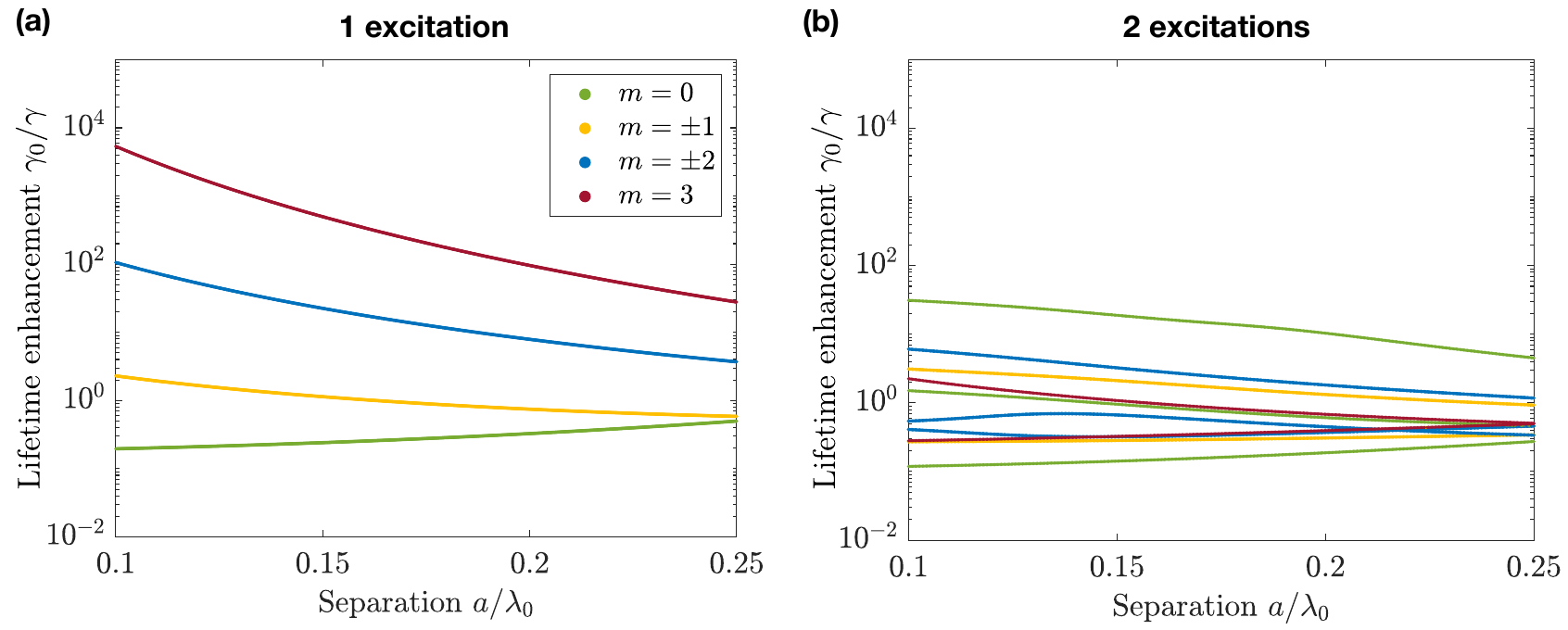}
    \caption{Radiative lifetime enhancement on a logarithmic scale for (a) singly excited and (b) doubly excited eigenstates of the ring of $N_d = 6$ emitters as a function of the separation distance between emitters normalized by the resonant wavelength of a single emitter. The curves in both panels are colored depending on the value of orbital quasimomentum $m$.}
    \label{fig_1ring_Q_vs_a}
\end{figure}

Figs.~\ref{fig_1ring_Q_vs_a}(a) and~\ref{fig_1ring_Q_vs_a}(b) show the radiative lifetime enhancement for singly and doubly excited eigenstates, respectively, as functions of normalized separation between emitters $a/\lambda_0$. The eigenenergies for singly excited states are calculated using Eq.~\eqref{eq_1exc_ring_energy} with analytically derived dipole sum~\eqref{eq_dipole_sums_6} whereas the spectrum of doubly excited states is calculated using Eq.~\eqref{eq_2exc}. In Fig.~\ref{fig_1ring_Q_vs_a}(a) one can see that the decay rates for all eigenstates monotonically change with the separation distance according to the results of Refs.~\cite{Asenjo-Garcia2017,moreno-cardoner_subradiance-enhanced_2019,cremer_polarization_2020}. Moreover, one can note that $\gamma_0/\gamma^{(m)} \lesssim 1$ for $m = 0$, $\gamma_0/\gamma^{(m)} \sim 1$ for $m = \pm 1$, and $\gamma_0/\gamma^{(m)} \gg 1$ for $m = \pm2$ and $m = 3$ for the considered range of separations $a/\lambda_0 \lesssim 0.25$. In contrast, the most subradiant doubly excited state in Fig.~\ref{fig_1ring_Q_vs_a}(b) has $m = 0$ since it can be considered as a direct product of the most subradiant singly excited state having $m = 3$ with itself [cf. Eq.~(3) in the main text].

\section{Hybridized singly excited states in the ring oligomers}
We consider the formation of subradiant states in oligomers viewed as two open subsystems, labeled A and B, interacting via the external coupling mechanism. Then the wave function for a hybridized singly excited eigenstate of an oligomer consists of two contributions $\ket{\psi} = c_a \ket{e_a} \otimes \ket{g_b} + c_b \ket{g_a} \otimes \ket{e_b}$, where $\ket{e_{a(b)}}$ is the excited state of system A (B) with the energy $\varepsilon_{a(b)}, \ket{g_{a(b)}}$ is the ground state of system A (B), and $\left| c_a \right|^2 + \left| c_b \right|^2 = 1$. The effective Hamiltonian of the coupling between these states reads as $2 \times 2$ matrix, which is a generalization of Eq.~(1) in the main text:
\begin{gather}
\label{H_2modes}
    \widehat{H} = 
    \begin{pmatrix}
        \varepsilon_{a} & 0 \\
        0 & \varepsilon_{b}
    \end{pmatrix}
    +
    \begin{pmatrix}
        0 & \varkappa \\
        \varkappa & 0
    \end{pmatrix},
\end{gather}
where $\varkappa$ is the coupling rate between the states of subsystems A and B. The energies for the hybridized eigenstates can be easily found as
\begin{gather}
\label{eq_1exc_epsiloN_dm}
    \varepsilon_{\pm} = \frac{1}{2} \left[ \varepsilon_a + \varepsilon_b \pm \sqrt{\left(\varepsilon_a - \varepsilon_b \right)^2 + 4 \varkappa^2 }\right]. 
\end{gather}
In the main text, we refer the index ``+'' to symmetric states and the index ``-'' to antisymmetric states.
The corresponding eigenstates $\ket{\psi_{\pm}}$ are
\begin{gather}
\label{eq_c}
    \begin{pmatrix}
        c_{a, \pm} \\
        c_{b, \pm}
    \end{pmatrix}
    = \dfrac{1}{\sqrt{1 + \left|\eta_{\pm} \right|^2}}
\left(\begin{array}{cccc}
1\\
\eta_{\pm}
\end{array}\right),
\end{gather}
where
\begin{gather}
    \eta_{\pm} = \frac{1}{2 \varkappa} \left[ \varepsilon_b - \varepsilon_a \pm  \sqrt{\left( \varepsilon_a - \varepsilon_b \right)^2 + 4  \varkappa^2}\right].
\end{gather}

\subsection{The ring with a central emitter}\label{sec:m_0}
First, we analyze the formation of a long-living state with $m = 0$ in the oligomer being the ring of $N_d$ emitters (subsystem A), enumerated from 1 to $N_d$, with a central one, labeled with the ``0'' index (subsystem B). The states of the subsystems read as $\ket{e_a} = \ket{e_{\text{0}}}$, $\ket{g_a} = \ket{g_{\text{0}}}$, $\ket{e_b} = \ket{\psi^{(0)}_{\text{ring}}}$ given by Eq.~\eqref{1exc_ring_kets}, and $\ket{g_b} = \ket {g_{\text{ring}}} = \ket{g}^{\otimes N_d}$. For these states, the parameters in Eq.~\eqref{H_2modes} are ${\varepsilon}_a = - {i\gamma_0}/{2}$, ${\varepsilon}_{b} =  - {i\gamma_0}/{2}+\tilde{\Sigma}$, with $\tilde{\Sigma} = \sum \limits_{k=2}^{N_d}  g(|\mathbf{r}_{1k}|, \omega_0)$, and ${\varkappa} = \sqrt{N_d} g\left(R, \omega_0 \right)$. After some algebra the energies of the hybridized states with $m = 0$, given by Eq.~\eqref{eq_1exc_epsiloN_dm}, can be expressed as
\begin{gather}
\label{eps_pm_oligomer}
    \varepsilon_{\pm} = - i\gamma_0/2 + \tilde{\Sigma}/2 \pm \sqrt{\tilde{\Sigma}^2/4 + N_d \left[g\left(R, \omega_0 \right) \right]^2}.
\end{gather}
The corresponding $\eta_{\pm}$ are 
\begin{gather}
\label{eq_eta_pm_oligomer}
    \eta_{\pm} = \frac{1}{g(R, \omega_0)} \left[\tilde{\Sigma}/2 \pm  \sqrt{\tilde{\Sigma}^2/4 + N_d \left[g\left(R, \omega_0 \right) \right]^2}\right].
\end{gather}

\begin{figure}[h!]
    \centering
    \includegraphics[scale=0.9]{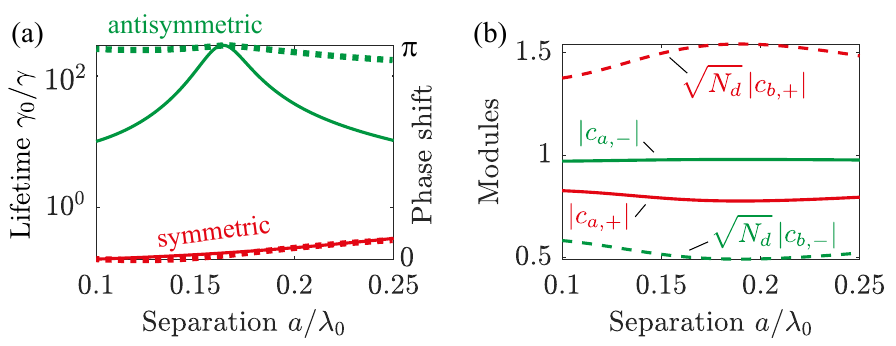}
    \caption{(a) (Left) Radiative lifetime on a logarithmic scale (solid lines) for the symmetric and antisymmetric states of the oligomer consisting of the ring ($N_d = 6$) with the central emitter. The lifetimes are calculated using Eq.~\eqref{eps_pm_oligomer}. (Right) Phase shift (markers) between excitation amplitudes $c_{a,\pm}$ and $c_{b,\pm}$ for these states. (b) The absolute values of excitation amplitudes $c_{a,\pm}$  (solid lines) and $\sqrt{N_d}c_{b,\pm}$ (dashed lines) for these states which are calculated using Eqs.~\eqref{eq_c} and~\eqref{eq_eta_pm_oligomer}.}
    \label{fig_oligomer_Q_vs_a}
\end{figure}

Fig.~\ref{fig_oligomer_Q_vs_a}(a) shows simultaneously the normalized lifetime of the symmetric and antisymmetric hybridized states $\ket{\psi_{\pm}}$ and the phase shift between the excitation amplitudes of the ring and the central emitter within these states. One can see that a maximal lifetime of the antisymmetric state is observed for oligomer size $a/\lambda_0 \approx 0.16$ when the phase shift reaches $\pi$ exactly. This can be intuitively explained by the fact that the overall radiative losses of a dipole array are proportional to a square of the mean dipole moment, $\gamma \propto \left| \langle \mathbf{d} \rangle \right|^2$, which can be approximately written for the oligomer as $\langle \mathbf{d} \rangle \approx \left( c_{a,-} + \sqrt{N_d}c_{b,-}\right)\mathbf{d}$ and it is reduced compared to that of a single dipole at $a/\lambda_0 \approx 0.16$. Thus, the anti-symmetric state is a subradiant one for the considered range of separations $a/\lambda_0 \lesssim 0.25$ while the symmetric state is a superradiant one. Moreover, the lifetime for the anti-symmetric state tends to infinity in the Dicke limit ($a/\lambda_0 \to 0$), while the lifetime for the symmetric state approaches $1/N$ where $N = N_d + 1$.  
Additionally, Fig.~\ref{fig_oligomer_Q_vs_a}(b) demonstrates that an excitation predominantly occupies the inner subsystem (central emitter) within the antisymmetric state, while an excitation prefers to occupy the outer subsystem (ring) within the symmetric state. A similar situation takes place for the subradiant singly excited and doubly excited states in the double-ring structure. 

\begin{figure}[h!]
    \centering
    \includegraphics[scale=0.55]{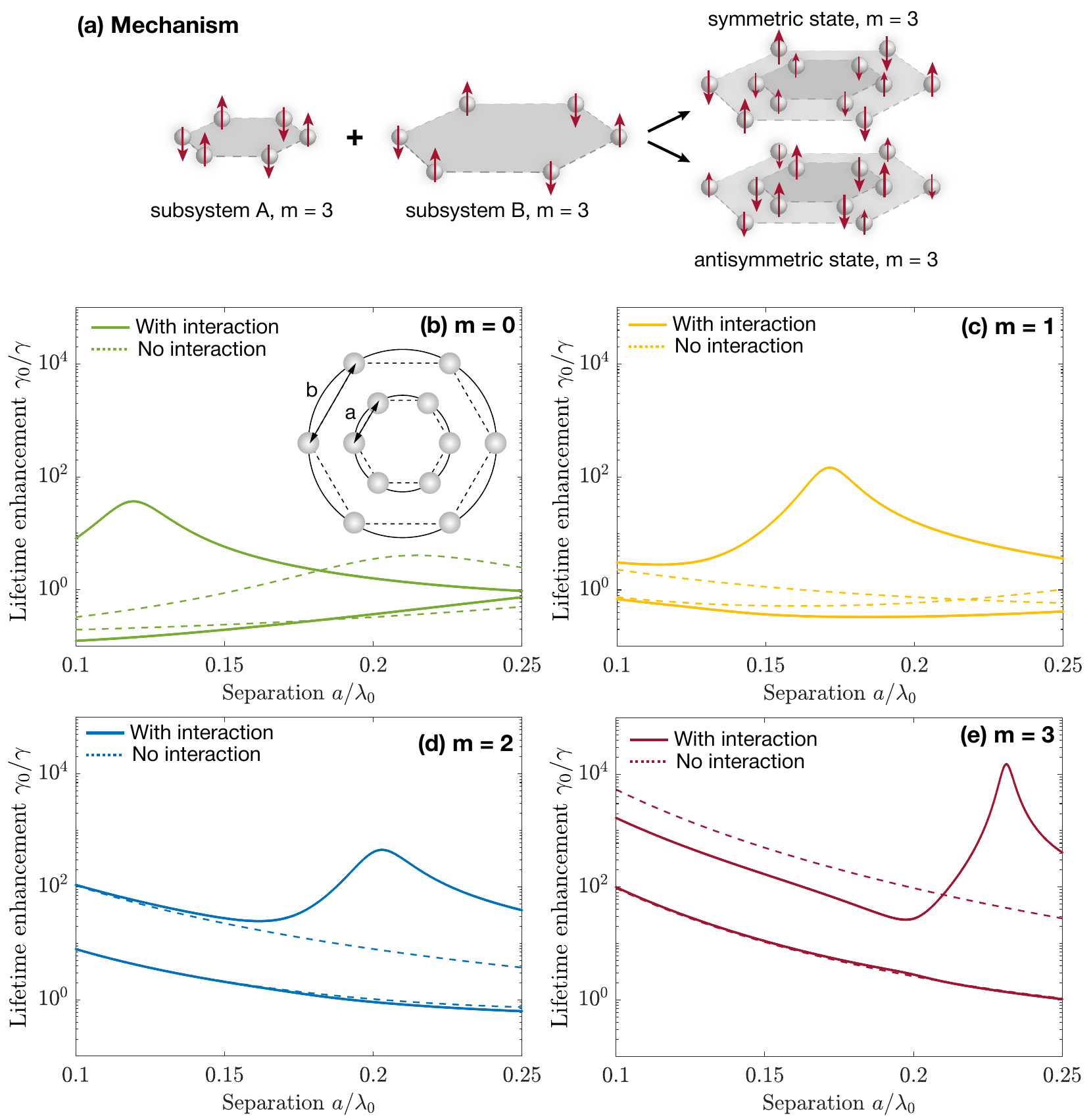}
    \caption{ (a) Hybridization of the ring states possessing orbital quasimomentum $m = 3$. (b-d) Lifetime on a logarithmic scale for the states with (b) $m = 0$, (c) $m = \pm 1$, (d) $m = \pm 2$, and (e) $m = 3$ in the noninteracting (dashed) and coupled (solid) rings with the fixed size ratio $b/a = 2$. The inset in panel (b) shows a two-ring oligomer.}
    \label{fig_2rings_Q_vs_a}
\end{figure}

\subsection{Double-ring oligomer}
Further, we consider the formation of subradiant states with arbitrary $m$ in two concentric rings with the same number of emitters $N_d$ labeled as the inner (``in'') and outer (``out'') ones. The emitters are enumerated from 1 to $N_d$ within the first ring and from $(N_d + 1)$ to $2N_d$ within the second one. The initial wave functions of the system are $\ket{e_a} = \ket{\psi^{(m)}_{\text{in-ring}}}$, $\ket{g_a} = \ket{e_{\text{in-ring}}}$, $\ket{e_b} = \ket{\psi^{(m)}_{\text{out-ring}}}$, and $\ket{g_b} = \ket{g_{\text{out-ring}}}$. The corresponding eigenenergies are $\varepsilon_a = \varepsilon^{(m)}_{\text{in-ring}}$ and $\varepsilon_b = \varepsilon^{(m)}_{\text{out-ring}}$ given by Eq.~\eqref{eq_1exc_ring_energy}. The coupling rate between singly excited rings can be computed as
\begin{gather}
    \varkappa_{\text{ring-ring}} = \sum\limits_{k = N_d + 1}^{2N_d} g(\left|\mathbf{r}_{1k}\right|, \omega_0) e^{i m \varphi_k}.
\end{gather}
Note that the ring states can interact only if they have the same $m$ which leads to the formation of symmetric and antisymmetric states within two rings for each accessible $m$ as shown in Fig.~\ref{fig_2rings_Q_vs_a}(a) for $m = 3$. Figs.~\ref{fig_2rings_Q_vs_a}(b-d) demonstrate lifetimes for the singly excited eigenstates for two rings of $N_d = 6$ emitters.

\section{Far-field radiation of dipole arrays in semi-classical approach}
The classical power radiated in the far field wave zone $r \gg \lambda_0$ can be written through the time-averaged Poynting vector $\langle \mathbf{S} \rangle$ as $P = \int\limits_{4\pi} r^2 \langle \mathbf{S} \rangle \cdot \mathbf{n} \ \mathrm{d} \Omega$.
Here the integration is carried out over the total solid angle $\int\limits_{4\pi}\mathrm{d} \Omega = \int\limits_0^{2\pi} \mathrm{d} \varphi \int\limits_0^{\pi} \mathrm{d} \theta \sin \theta$ for a spherical surface of radius $r \gg \lambda$ with the center at the origin of Cartesian coordinate system, $\mathbf{n} = \left( \cos \theta \cos \varphi,  \cos \theta \sin \varphi, \sin \theta \right)$ is a unit vector normal to this spherical surface. The spherical angles $\theta$ and $\varphi$ are defined according to Fig.~4(a) of the main text. Let us introduce $p(\theta, \varphi) = r^2 \langle \mathbf{S} \rangle \cdot \mathbf{n}$ as the power radiated into a unite solid angle $\mathrm{d} \Omega$ along $\mathbf{n}$. $p(\theta, \varphi)$ determines a \textit{radiation pattern} of the system and depends on a radial component of $\langle \mathbf{S} \rangle = \dfrac{1}{2} \Re\left(\mathbf{E}_{\mathrm{FF}} \times \mathbf{H}^*_{\mathrm{FF}} \right)$ where only far-field terms of generated electric $\mathbf{E}_{\mathrm{FF}}$ and magnetic $\mathbf{H}_{\mathrm{FF}}$ fields contribute to the outgoing radiation. For an electric dipole, $\mathbf{H}_{\mathrm{FF}} = c \epsilon_0 \mathbf{n} \times \mathbf{E}_{\mathrm{FF}}$~\cite{Novotny2012}. Moreover,  $\mathbf{E}_{\mathrm{FF}}$, $\mathbf{H}_{\mathrm{FF}}$, and $\mathbf{n}$ are orthogonal to each other in the far field wave zone, then $\langle \mathbf{S} \rangle = \dfrac{1}{2} c \epsilon_0 \left| \mathbf{E}_{\mathrm{FF}}\right|^2 \mathbf{n}$ is always oriented along $\mathbf{n}$.

The classical electric field generated by a collection of electric dipoles in free space can be written via electromagnetic Green's tensor~\eqref{eq_G0} as
$\mathbf{E}(\mathbf{r}, \omega_0) = k_0^2 / \epsilon_0 \sum\limits_{k = 1}^N \bm{\mathsf{G}}_0 \left( \mathbf{r} - \mathbf{r}_k, \omega_0\right) \mathbf{d}_k,$
where $\{ \mathbf{r}_k \}_{k = 1}^N$ is a set of coordinates of scatterers with dipole moments $\{ \mathbf{d}_k \}_{k = 1}^N$~\cite{Novotny2012}. 
To calculate $\mathbf{E}_{\mathrm{FF}}$, we write down the far field part of electromagnetic Green's tensor~\eqref{eq_G0}:
$\bm{\mathsf{G}}_0^{\mathrm{FF}}(\mathbf{r} - \mathbf{r}_k, \omega_0) =  \dfrac{e^{i k_0 r - i k_0 \mathbf{n} \cdot \mathbf{r}_k}}{4 \pi r} \left(\bm{\mathsf{I}} - \mathbf{n} \otimes \mathbf{n} \right),$ where $\mathbf{n} \equiv \mathbf{r}/r$ and we have used that $|\mathbf{r} - \mathbf{r}_k| \approx r - \mathbf{r} \cdot \mathbf{r}_k/r = r -  \mathbf{n} \cdot \mathbf{r}_k$. Hence, the electric far field can be expressed as $\mathbf{E}_{\mathrm{FF}} = \dfrac{e^{i k_0 r}}{r} \mathbf{f}(\theta, \varphi)$. 

For dipole moments oscillating in the $z$-direction $\mathbf{d}_k = \left| \mathbf{d}_k\right| \mathbf{e}_z$, the scattering amplitude is given as $\mathbf{f}(\theta, \varphi) = -  \dfrac{\omega_0^2}{4 \pi c^2 \epsilon_0}\sin \theta \sum\limits_{k = 1}^N e^{-i k_0 \mathbf{n} \cdot \mathbf{r}_k}  \left|\mathbf{d}_k \right| \mathbf{e}_z$. Thus, the radiation pattern is completely determined by the scattering amplitude $p(\theta, \varphi) = \dfrac{1}{2} c \epsilon_0 \left| \mathbf{f}(\theta, \varphi)\right|^2$ and does not depend on the sphere radius $r$. For the obtained $\mathbf{f}(\theta, \varphi)$, the radiation pattern of electric dipoles oscillating in the $z$-direction is finally written as
\begin{gather}
\label{eq_p}
  p(\theta, \varphi) = \frac{\omega_0^4 }{32 \pi^2 c^3 \epsilon_0} \sin^2 \theta \left| \sum\limits_{k = 1}^N e^{-i k_0 \mathbf{n} \cdot \mathbf{r}_k} \left| \mathbf{d}_k\right|\mathbf{e}_z\right|^2. 
\end{gather}
Hence, the total radiated power for such dipoles reads as
\begin{gather}
\label{eq_P}
    P = \frac{\omega_0^4 }{32 \pi^2 c^3 \epsilon_0} \int\limits_0^{2\pi} \mathrm{d} \varphi \int\limits_0^{\pi} \mathrm{d} \theta \sin^3 \theta \left| \sum\limits_{k = 1}^N e^{-i k_0 \mathbf{n} \cdot \mathbf{r}_k} \left| \mathbf{d}_k\right|\mathbf{e}_z\right|^2.
\end{gather}


\subsection{Scattering cross section}
Let us assume that the array of emitters is illuminated by an external electric field $\mathbf{E}_{\mathrm{inc}}(\mathbf{r})$ at a frequency $\omega$. In this case, the distribution of dipole moments $\mathbf{d}_k$ is governed by the coupled dipole equation describing the electromagnetic interaction of dipoles~\cite{Novotny2012}:
\begin{gather}
\label{eq_cde}
    \mathbf{d}_k = \widehat{\alpha}(\omega) \mathbf{E}_{\mathrm{inc}}(\mathbf{r}_k) + \widehat{\alpha}(\omega) \frac{k_0^2}{\epsilon_0} \sum\limits_{\substack{l=1, \\ l\neq k}}^N\bm{\mathsf{G}}_0(\mathbf{r} - \mathbf{r}_l, \omega_0) \mathbf{d}_l. 
\end{gather}
For the considered emitters, the electric dipole polarizability tensor  is given as
\begin{gather}
    \widehat{\alpha}(\omega) = 
    \begin{pmatrix}
        0 & 0 & 0 \\
        0 & 0 & 0 \\
        0 & 0 & \alpha(\omega)
    \end{pmatrix},
\end{gather}
where the Lorentz polarizability in the proximity of the resonant frequency $\omega_0$ can be approximated as $\alpha(\omega) = - \dfrac{6 \pi}{k_0^3} \dfrac{ \gamma_0/2}{\left(\omega - \omega_0\right) + i\gamma_0/2}$~\cite{Loudon2006Jul}. Such polarizability tensor leads to the excitation of the dipole moments only in the $z$-direction, therefore, we can use Eqs.~\eqref{eq_p} and~\eqref{eq_P} for calculating the scattered power. 

From Eq.~\eqref{eq_cde}, one can obtain self-consistent dipole moments for the emitters within the incident field and further insert them into Eq.~\eqref{eq_P}. Since the radiated power is proportional to the scattering cross section, one can define normalized scattering cross section as:
\begin{gather}
    \frac{\sigma(\omega)}{N \sigma_0(\omega)} = \frac{3}{8 \pi N \left| \alpha(\omega) E_{\mathrm{inc},z}(\mathbf{r} = 0)\right|^2} \int\limits_0^{2\pi} \mathrm{d} \varphi \int\limits_0^{\pi} \mathrm{d} \theta \sin^3 \theta \left| \sum\limits_{k = 1}^N e^{-i k_0 \mathbf{n} \cdot \mathbf{r}_k}\left| \mathbf{d}_k\right|\mathbf{e}_z\right|^2,
\end{gather}
where $\sigma_0(\omega)$ is the scattering cross section for a single emitter placed at $\mathbf{r} = 0$.

\subsection{Radiation pattern for a singly excited eigenstate}
In a semi-classical approach, we can assign a classical dipole moment to quantum emitter $k$ by a simple rule $\mathbf{d}_k = \mathbf{d} \times c_k$, where $c_k$ is the excitation amplitude within a given singly excited eigenstate, and $\mathbf{d}$ is the emitter transition dipole moment. Then, Eq.~\eqref{eq_p} can be simplified to the following one:
\begin{gather}
\label{eq_radiatioN_dattern}
   p(\theta, \varphi) = P_0 \times \frac{3}{32 \pi } \sin^2 \theta \left| \sum\limits_{k = 1}^N e^{-i k_0 \mathbf{n} \cdot \mathbf{r}_k} c_k\right|^2, 
\end{gather}
where $P_0 = \hbar \omega_0 \times \gamma_0$ is the classical total radiated power by a single electric dipole.

\section{Relationship between energies of doubly and singly excited states}
In this section, we derive Eq.~(3) in the main text. Considering the Pauli principle and spatial symmetry of the wave function, doubly excited eigenstates in a system of two-level ($\chi \to \infty$) quantum emitters can be described by the effective double-excitation Hamiltonian~\cite{Ke2019Dec}:
\begin{gather}
\label{eq_H2}
    \widehat{H}^{(2)} = \widehat{H}^{(1)} \otimes \hat{I} + \hat{I} \otimes \widehat{H}^{(1)} + \frac{\chi}{2} \sum\limits_{k=1}^N \hat{\sigma}^{\dagger}_k \hat{\sigma}^{\dagger}_k \hat{\sigma}_k \hat{\sigma}_k.
\end{gather}
Here $\hat{I}$ is the identity operator, while $\widehat{H}^{(1)}$ represents the single-excitation Hamiltonian with the sets of right (ket) and left (bra) singly excited eigenstates corresponding to the same set of eigenvalues $\varepsilon$: $\widehat{H}^{(1)} \ket{\psi_R} = \varepsilon \ket{\psi_R}$ and $\bra{\psi_L} \widehat{H}^{(1)}  = \bra{\psi_L} \varepsilon$, respectively. Similarly, one can introduce the right and left doubly excited eigenstates: $\widehat{H}^{(2)} \ket{\Psi_R} = \mathcal{E} \ket{\Psi_R}$ and $\bra{\Psi_L} \widehat{H}^{(2)}  = \bra{\Psi_L} \mathcal{E}$, respectively.
Note that, considering the Pauli principle $\hat{\sigma}_k \hat{\sigma}_k \ket{\Psi_R} = 0$ and $\bra{\Psi_L} \hat{\sigma}^{\dagger}_k \hat{\sigma}^{\dagger}_k = 0$, we can exclude the nonlinear term from Eq.~\eqref{eq_H2}.

The eigenstates of ring structures can be characterized by the orbital quasimomentum $m$. Moreover, the wave function of right and left doubly excited eigenstates with $m$ can be expanded over the direct products of right and left singly excited eigenstates with $m_1$ and $m_2$ as: 
$\ket{\Psi_R^{(m)}} = \sum\limits_{m_1,m_2} v^R_{m_1,m_2} \ket{\psi_R^{(m_1)}} \ket{\psi_R^{(m_2)}},$ and $
\bra{\Psi_L^{(m)}} = \sum\limits_{m_1,m_2} v^L_{m_1,m_2} \bra{\psi_L^{(m_2)}} \bra{\psi_L^{(m_1)}}$, respectively.
The orthogonality of wave functions $\braket{\Psi_L^{(m')} }{\Psi_R^{(m)}} = \delta_{m',m}$ and $\braket{\psi_L^{(m')} }{\psi_R^{(m)}} = \delta_{m',m}$,
where $\delta_{m',m}$ is the Kronecker delta, implies that $v^L_{m_1,m_2} = \left( v^R_{m_1,m_2} \right)^*$, where $*$ is the complex conjugation. Hence, we can denote $v^R_{m_1,m_2} \equiv v_{m_1,m_2}$ and $v^L_{m_1,m_2} \equiv v^*_{m_1,m_2}$. The energy of the doubly excited state with $m$ is given as
\begin{align}
    \begin{aligned}
        &\mathcal{E}^{(m)} = \\
        &\mel{\Psi_L^{(m)}}{\widehat{H}^{(2)}}{\Psi_R^{(m)}} = \\ &\sum_{m_1,m_2} v^*_{m_1,m_2}  v_{m_1,m_2}  \bra{\psi_L^{(m_2)}} \mel{\psi_L^{(m_1)} }{\widehat{H}^{(1)} \otimes \hat{I} + \hat{I} \otimes \widehat{H}^{(1)}}{\psi_R^{(m_1)}}\ket{\psi_R^{(m_2)}} = \\
        &\sum_{m_1,m_2} \left| v_{m_1,m_2} \right|^2  \left[ \varepsilon^{(m_1)} + \varepsilon^{(m_2)} \right] \underbrace{\bra{\psi_L^{(m_2)}} \braket{\psi_L^{(m_1)} }{\psi_R^{(m_1)}}\ket{\psi_R^{(m_2)}}}_{=1} = \\
        &\sum_{m_1,m_2} \left| v_{m_1,m_2} \right|^2  \left[ \varepsilon^{(m_1)} + \varepsilon^{(m_2)} \right],
    \end{aligned}
\end{align}
where we have used Eq.~\eqref{eq_H2} and the orthogonality of wave functions. Finally, we obtain Eq.~(3) in the main text.

\section{Optimization of the lifetime for the subradiant doubly excited state}
\begin{figure}[h!]
    \centering
    \includegraphics[scale=0.6]{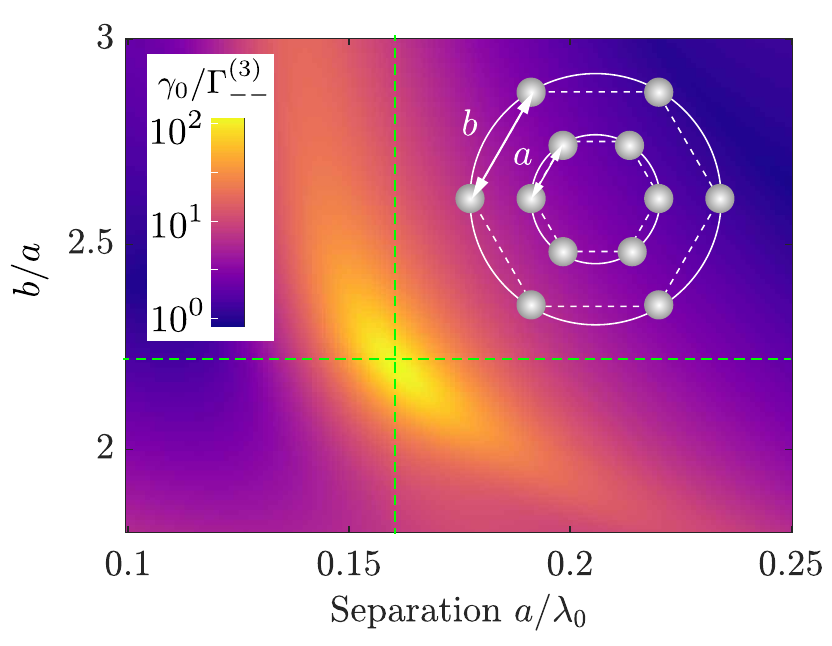}
    \caption{Lifetime enhancement $\gamma_0/\Gamma^{(3)}_{--}$ on a logarithmic scale for the most subradiant doubly excited state with orbital quasimomentum $m = 3$.The inset shows the double-ring oligomer with depicted $a$ and $b$.}
    \label{fig_2exc_2rings_optimization}
\end{figure}
Fig.~\eqref{fig_2exc_2rings_optimization} shows the lifetime for the most subradiant doubly excited state in the double-ring oligomer $\ket{\Psi^{(3)}_{--}}$, given by Eq.~(2) in the main text, as a function of the inner ($a/\lambda_0$) and outer ($b/a$) ring sizes. The collective decay rate of the state $\Gamma^{(3)}_{--}$ is also given in the main text. One can observe a strong enhancement of the normalized lifetime for parameters $a/\lambda_0 \approx 0.16$ and $b/a = 2.2$.

\section{Subradiant doubly excited state in a case of circular dipoles}

\begin{figure}[h!]
    \centering
    \includegraphics[scale=0.55]{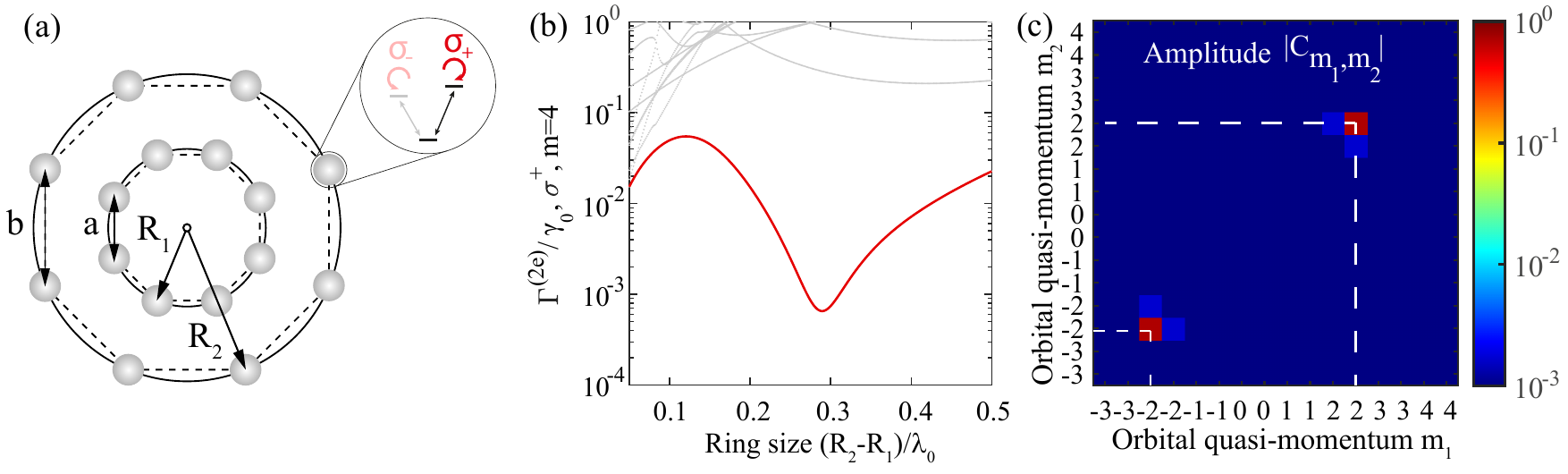}
    \caption{(a)Double-ring oligomer composed of two concentric rings with $N=8$ atoms characterized by a single circular dipole transition. Atoms in the rings are positioned at distances $a$ and $b$. (b) Radiative decay rates for doubly-excited states in the oligomer with $m=4$. The most subradiant state is marked by red. (c) Expansion of this state in terms of singly excited states with different $m_1$ and $m_2$.}
    \label{fig_2rings_circular}
\end{figure}

Here, we would like to demonstrate that the appearance of doubly excited nonradiant states can be also observed for circularly polarized dipoles ($\sigma_{\pm}$ transitions) which can be accessed by exciting maximally stretched atoms. We considered circularly polarized transition dipole moments rotating in the $xy$ plane of the structure as shown in Fig.~\ref{fig_2rings_circular}(a). Our calculations show that one can find nonradiant states in such systems, and one of the examples is shown below for two rings with $N=8$ atoms on each, while the distance between the neighbors on the inner ring is equal to $a = \lambda_0/10$.

One can easily see from Fig.~\ref{fig_2rings_circular}(b) that there is a significant reduction of emission rate for one of the doubly excited states with orbital quasimomentum $m=4$ when the difference between the inner and outer rings radii is $ (R_2-R_1)/\lambda_0 \approx 0.3$.  From the expansion of this doubly excited state in terms of singly excited eigenstates presented in Fig.~\ref{fig_2rings_circular}(c), one can also see that this state has the following form: $\ket{\Psi^{(4)}} \approx \frac{1}{\sqrt{2}} \left(\ket{\psi^{(-2)}_1}\ket{\psi^{(-2)}_2} - \ket{\psi^{(2)}_1}\ket{\psi^{(2)}_2}\right)$, similar to already discussed in the main text. Here, for circular in-plane dipoles, the main mechanism of radiative loss suppression is again related to external coupling between the states.   However, we found that on average it is much easier to find such geometries for linearly polarized $z$-oriented dipoles than for any other polarization because a $z$-dipole immediately suppresses emission in the $+z$, $-z$ direction due to its emission pattern.

\bibliography{supplement}